\documentclass[showpacs,amsmath,amssymb,prd,floatfix,preprintnumbers,nofootinbib]{revtex4-2}
\usepackage{graphicx}
\usepackage{bm}
\usepackage{amsmath}
\usepackage{color}
\usepackage[dvipsnames]{xcolor}
\usepackage{fontawesome5}
\usepackage{hyperref}
\usepackage{comment}
\usepackage{multirow}
\usepackage{url}

\hypersetup{
  bookmarksnumbered=true,
  colorlinks=true,
  linkcolor=orange,
  citecolor=cyan,
  urlcolor=gray
}

\makeatletter
\newcommand{\github}[1]{%
   \href{#1}{\faGithubSquare}%
}
\makeatother

\newcommand{\bc}{\bm{c}}
\newcommand{\bd}{\bm{d}}

\newcommand{\bk}{\bm{k}}

\newcommand{\bn}{\bm{n}}

\newcommand{\bp}{\bm{p}}
\newcommand{\bq}{\bm{q}}
\newcommand{\bbr}{\bm{r}}

\newcommand{\bu}{\bm{u}}
\newcommand{\bv}{\bm{v}}

\newcommand{\bx}{\bm{x}}

\newcommand{\bJ}{\bm{J}}
\newcommand{\bK}{\bm{K}}
\newcommand{\bL}{\bm{L}}

\newcommand{\bP}{\bm{P}}

\newcommand{\btheta}{\bm{\theta}}

\newcommand{\bDelta}{\bm{\Delta}}

\newcommand{\bPsi}{\bm{\Psi}}

\newcommand{\lin}{\mathrm{lin}}

\newcommand{\D}{\mathcal{D}}

\newcommand{\F}{\mathcal{F}}
\newcommand{\G}{\mathcal{G}}

\newcommand{\I}{\mathcal{I}}
\newcommand{\J}{\mathcal{J}}
\renewcommand{\L}{\mathcal{L}}

\newcommand{\N}{\mathcal{N}}
\renewcommand{\O}{\mathcal{O}}

\newcommand{\Q}{\mathcal{Q}}

\newcommand{\U}{\mathcal{U}}

\newcommand{\bin}{\mathrm{bin}}
\newcommand{\dD}{(2\pi)^3 \delta_{\mathrm{D}}}

\def\avrg#1{\left\langle #1 \right\rangle}

\def\avrgc#1{\left\langle #1 \right\rangle_\textrm{c}}

\newcommand{\simgt}{\lower.5ex\hbox{$\; \buildrel > \over \sim \;$}}
\newcommand{\simlt}{\lower.5ex\hbox{$\; \buildrel < \over \sim \;$}}

\newmuskip\pFqmuskip

\newcommand*\pFq[6][8]{%
  \begingroup 
  \pFqmuskip=#1mu\relax
  \mathchardef\normalcomma=\mathcode`,
  \mathcode`\,=\string"8000
  \begingroup\lccode`\~=`\,
  \lowercase{\endgroup\let~}\pFqcomma
  {}_{#2}F_{#3}{\left(\genfrac..{0pt}{}{#4}{#5};#6\right)}%
  \endgroup
}
\newcommand{\pFqcomma}{{\normalcomma}\mskip\pFqmuskip}

\begin{document}

\title[]{Fast and accurate differentiable code of the galaxy power spectrum based on Eulerian and Lagrangian one-loop perturbation theories}

\author{Yosuke Kobayashi${}^{1}$}
\email{yosuke.kobayashi@cc.kyoto-su.ac.jp}
\author{Kazuyuki Akitsu${}^{2}$}
\email{kakitsu@ias.edu}
\affiliation{
${}^{1}$ Department of Astrophysics and Atmospheric Sciences, Faculty of Science, Kyoto Sangyo University, Kyoto 603-8555, Japan\\
${}^{2}$ Theory Center, Institute of Particle and Nuclear Studies, High Energy Accelerator Research Organization (KEK), Tsukuba, Ibaraki 305-0801, Japan
}

\begin{abstract}
This paper presents a JAX-Python implementation of the galaxy power spectrum up to one-loop order in the effective field theory of large-scale structure, based on both the Eulerian and Lagrangian perturbation theories.
The code uses JAX\footnote{\url{https://jax.readthedocs.io/en/latest/}} to enable automatic differentiation with respect to parameters such as cosmology, galaxy bias, or counterterm coefficients.
This facilitates applications such as parameter inference using the gradient of the posterior, \textit{e.g.}, Hamiltonian Monte Carlo sampling.
The code can compute both the EPT and LPT power spectra in $\O(1)$ milliseconds on a GPU, after just-in-time compilation.
We also identify the various sources of numerical errors associated with discretization, and design the numerical implementation to minimize them,
achieving relative errors of order $0.01\%$ at scales $k \lesssim 1 \, h \, {\rm Mpc}^{-1}$, compared with brute-force numerical integration.
Notably, this is the first auto-differentiable implementation of the one-loop galaxy power spectrum based on LPT.
The natural resummation of the long-wavelength displacements in LPT further makes the code well suited for applications involving features in the power spectrum beyond the standard BAO feature, including primordial features.
Our code \texttt{elf} is publicly available at the repository \github{https://github.com/archaeo-pteryx/elf}.
\end{abstract}

\maketitle

\hypersetup{linkcolor=gray}
\tableofcontents
\hypersetup{linkcolor=orange}

\section{Introduction}
\label{sec:intro}

Spatial distributions of luminous objects such as galaxies or quasars over the large-scale structure of the Universe are one of the fundamental sources of cosmological information.
The ongoing or upcoming large-scale structure surveys, such as
Dark Energy Spectroscopic Instrument (DESI; \cite{DESI:2016fyo}),
\textit{Euclid} satellite \cite{EUCLID:2011zbd},
Spectro-Photometer for the History of the Universe, Epoch of Reionization, and Ices Explorer (SPHEREx; \cite{Dore:2014cca}),
Subaru Prime Focus Spectrograph (PFS; \cite{PFSTeam:2012fqu}),
Legacy Survey of Space and Time (LSST; \cite{LSST:2008ijt}),
Nancy Grace Roman Space Telescope \cite{Spergel:2015sza},
DESI-II \cite{DESI:2022lza},
and MegaMapper \cite{Schlegel:2022vrv}
are aiming at constraining the cosmological model with unprecedented precision from the galaxy distributions through the signals of galaxy clustering or galaxy-galaxy weak lensing.

The formation of the large-scale structure is a nonlinear physical process whose exact analytic solution is not available.
Hence, to model the large-scale structure observables, cosmologists have been utilizing either perturbation theory or $N$-body simulations.
The perturbation theory of large-scale structure is based on the fluid equations that govern the evolution of the cosmological matter fluid in an expanding universe.
Since the traditional standard perturbation theory (see \cite{Bernardeau:2001qr} for a review) has several drawbacks \cite{Pajer:2013jj}, such as that higher-order solutions are plagued by small-scale modes that cannot be treated as small perturbations, so-called the effective field theory (EFT) of large-scale structure has been developed over the past decade, for both Eulerian (\textit{e.g.}, \cite{Baumann:2010tm,Carrasco:2012cv,Pajer:2013jj}) and Lagrangian \cite{Porto:2013qua,Vlah:2015zda} approaches of perturbation theory.
This is a large-scale effective theory of the cosmological fluid constructed by integrating out the unreliable small-scale modes and accounting for their impact in terms of the large-scale modes with the coefficients that cannot be determined by the effective theory itself.

Due to the complicated baryonic physics, we do not have an \textit{ab initio} understanding of how the luminous objects that we observe in the surveys are distributed in the underlying invisible dark matter, \textit{i.e.}, galaxy bias.
In perturbation theory, the galaxy bias is modeled perturbatively in terms of a linear combination of the scalar quantities on which the observed galaxy number densities can depend, constructed from the gravitational tidal field or the velocity shear field that are compatible with the boost symmetry of the system \cite{McDonald:2009dh,Assassi:2014fva,Senatore:2014eva} (see also a review \cite{Desjacques:2016bnm}).
The coefficients of this linear combination, \textit{i.e.}, galaxy bias parameters, are also regarded as effective parameters that are not predicted from first principles.

Consequently, when we try to measure the cosmological parameters from survey data, we simultaneously marginalize over many effective nuisance parameters that characterize possible dependences of the spatial distributions of galaxies or quasars on the underlying matter field at the lowest orders of density perturbations.
This implies that we have to perform the parameter inference on high-dimensional parameter spaces, which is hard to tackle using a standard Markov chain Monte Carlo (MCMC) sampling (\textit{e.g.}, \cite{Speagle:2019ffr} for a recent review) or nested sampling.
Recent analyses of large-scale structure surveys aim to extract cosmological information from galaxy clustering or galaxy-galaxy lensing data, combining multiple redshift bins and galaxy subsamples.

In this paper, we present a differentiable code for the one-loop EFT power spectrum, implemented in JAX, which includes both Eulerian perturbation theory (EPT) and Lagrangian perturbation theory (LPT) in redshift space.
This code is compatible with just-in-time (JIT) compilation and enables accelerated computation on a GPU device.
Because of the automatic differentiation of JAX, our code enables the computation of the derivatives of the one-loop power spectrum with respect to the model parameters such as galaxy bias or other EFT parameters.
Since our code is independent of the implementation of the input linear matter power spectrum, users can also compute the derivatives with respect to the cosmological parameters by combining our code with differentiable codes or emulators of the linear power spectrum, such as \texttt{DISCO-EB} \cite{Hahn:2023nvb} and \texttt{CosmoPower-JAX} \cite{Piras:2023aub}.
The differentiability allows us to use gradient-based sampling algorithms, such as Hamiltonian Monte Carlo (see, \textit{e.g.}, \cite{Neal:2011mrf}) or its variants.
By combining gradient-based inference algorithms with accelerated computation using a GPU, the JAX code enables efficient parameter inference in high-dimensional 
parameter spaces.
So far, such differentiable programming compatible with GPUs has been applied to implement the one-loop EFT power spectrum by a few works, such as \texttt{Effort.jl} \cite{Bonici:2025ltp} and \texttt{PyBird-JAX} \cite{Reeves:2025bxn}, both of which are based on a neural network-based emulation of the Eulerian theory.
This work, for the first time, presents a differentiable code for both Eulerian and Lagrangian theories without emulation. 

The rest of this paper is organized as follows.
In Sec.~\ref{sec:theory_model} we describe the theory models of the power spectrum implemented in the code.
Section~\ref{sec:implementation} explains the implementation details of the important parts, such as one-loop integrals, infrared resummation, and Fourier-space angular integrals involving the exponential factor in LPT.
We show the accuracy of our implementation of the one-loop power spectrum as well as the validation against other public codes and the benchmarking result in Sec.~\ref{sec:code_performance}.
In Sec.~\ref{sec:inference}, we investigate the performance of our code when applied to parameter inference using Hamiltonian Monte Carlo sampling, and finally, we conclude in Sec.~\ref{sec:conclusion}.
Throughout the paper, we will use the shorthand notation for vector summations, \textit{e.g.}, $\bk_{1 \cdots n} \equiv \bk_{1} + \cdots + \bk_{n}$, and for 3D integrals, \textit{e.g.}, $\int_{\bp} \equiv \int \frac{\mathrm{d}^{3}p}{(2 \pi)^{3}}$ for a Fourier-space vector $\bp$ or $\int_{\bq} \equiv \int \mathrm{d}^{3}q$ for a position coordinate vector $\bq$.

\section{Theory model}
\label{sec:theory_model}

This section describes the one-loop power spectrum of biased tracers we implemented in our code.
We describe the power spectrum based on Eulerian (Sec.~\ref{sec:model_ept}) and Lagrangian (Sec.~\ref{sec:model_lpt}) perturbation theory.
One of the main features of this work is that we expand the galaxy bias in terms of the Galileon operators \cite{Nicolis:2008in,Chan:2012jj,Assassi:2014fva} for both the EPT and LPT models.
To the best of our knowledge, the LPT power spectrum using Galileon operators has not been implemented in a public code, though in practice a mapping of the bias operator set can be realized by a mapping of the bias parameters.

\subsection{Eulerian perturbation theory}
\label{sec:model_ept}

The EPT power spectrum we implement is basically the same as that in the public code \texttt{CLASS-PT} \cite{Chudaykin:2020aoj}.
We follow the bias expansion of \textit{e.g.}, \cite{Simonovic:2017mhp,Chudaykin:2020aoj},
\begin{align}
    \delta_{\mathrm{g}}(\bx) &= b_1 \delta(\bx) + \frac{b_2}{2} \delta^2(\bx) + b_{\G_2} \G_2(\bx) + \frac{b_3}{6} \delta^3(\bx) + b_{\G_3} \G_3(\bx) + b_{\delta \G_2} \delta(\bx) \G_2(\bx) + b_{\Gamma_3} \Gamma_3(\bx) + b_{\nabla^{2} \delta} \nabla^2 \delta(\bx),
    \label{eq:bias_expansion_ept}
\end{align}
where $\delta$ denotes the matter density fluctuation, and we adopt the Galileon tidal operators $\G_2, \G_3$, and $\Gamma_3$ defined as \cite{Assassi:2014fva}
\begin{align}
    \G_{2}(\Phi_\mathrm{g}) &\equiv (\nabla_i \nabla_j \Phi_\mathrm{g})^2 - (\nabla^2 \Phi_\mathrm{g})^2, \\
    \G_{3}(\Phi_\mathrm{g}) &\equiv - \nabla_i \nabla_j \Phi_\mathrm{g} \nabla^j \nabla_k \Phi_\mathrm{g} \nabla^k \nabla^i \Phi_\mathrm{g} - \frac{1}{2} (\nabla^2 \Phi_\mathrm{g})^3 + \frac{3}{2} (\nabla_i \nabla_j \Phi_\mathrm{g})^2 \nabla^2 \Phi_\mathrm{g}, \\
    \Gamma_{3}(\Phi_\mathrm{g}, \Phi_\mathrm{v}) &\equiv \G_{2}(\Phi_\mathrm{g}) - \G_{2}(\Phi_\mathrm{v}),
\end{align}
where $\Phi_\mathrm{g}$ and $\Phi_\mathrm{v}$ are the rescaled gravitational and velocity potentials.
The bias operators are associated with coefficients called bias parameters, which are $\{ b_1, b_2, b_{\G_2}, b_3, b_{\G_3}, b_{\delta \G_2}, b_{\Gamma_3}, b_{\nabla^{2} \delta} \}$ in our expansion Eq.~(\ref{eq:bias_expansion_ept}).
Among them, at one-loop order of the power spectrum, the $\G_3$ contribution is zero, and $\delta^3$ and $\delta \G_2$ contributions vanish after the renormalization of the linear bias \cite{McDonald:2006mx,McDonald:2009dh,Assassi:2014fva}, and the higher-derivative bias term $\nabla^{2} \delta$ is degenerate with the EFT counterterm described in Sec.~\ref{sec:ctr_and_stoch}.

In spectroscopic surveys, a galaxy's observed redshift is dilated due to the Doppler effect on the galaxy's light.
The redshift-space distortions are also incorporated up to the third order of perturbations (see \textit{e.g.} \cite{Desjacques:2018pfv}):
\begin{align}
    \delta_\mathrm{g}^{\mathrm{(s)}} &= \delta_\mathrm{g} - \nabla_\parallel \left[ u_{\parallel} (1+ \delta_\mathrm{g} ) \right] + \frac{1}{2} \nabla_\parallel^2 \left[ u_{\parallel}^2 (1+ \delta_\mathrm{g} ) \right] - \frac{1}{6} \nabla_\parallel^3 (u_{\parallel}^3),
\end{align}
in configuration space. 
Here $\bu \equiv \bv / (aH)$ and the subscript $\parallel$ denotes the vector component along the line-of-sight direction.
Or equivalently, we have the expression in Fourier space (\textit{e.g.} \cite{Scoccimarro:1999ed,Matsubara:2007wj,Senatore:2014vja}),
\begin{align}
    \delta_\mathrm{g}^{\mathrm{(s)}}(\bk) 
    &= \delta_\mathrm{g}(\bk) - i k_{\parallel} u_{\parallel}(\bk) - i k_{\parallel} [ u_{\parallel} \delta_\mathrm{g} ](\bk) -\frac{1}{2} k_{\parallel}^2 [ u_{\parallel}^2 ](\bk) - \frac{1}{2} k_{\parallel}^2 [ u_{\parallel}^2 \delta_\mathrm{g} ](\bk) + \frac{1}{6} i k_{\parallel}^3 [ u_{\parallel}^3 ](\bk).
    \label{eq:rsd_to_3rd_order_fourier}
\end{align}
In this work, we ignore the velocity bias, \textit{i.e.}, assume a galaxy's velocity is the same as that of the surrounding matter, which is expanded perturbatively in terms of the velocity divergence, $\theta \equiv \nabla \cdot \bv$.
Note that Eq.~(\ref{eq:rsd_to_3rd_order_fourier}) includes the composite operators, \textit{i.e.}, products of multiple operators evaluated at the same position.
These terms have contributions from unreliable UV modes, and hence we need to add counterterms \cite{Senatore:2014via}, as described in Sec.~\ref{sec:ctr_and_stoch}.
Hereafter, we drop the superscript $\mathrm{(s)}$ and use $\delta_{\mathrm{g}}$ to denote the galaxy density in redshift space.

By combining the expansions of the galaxy bias and the redshift-space distortions described above, the galaxy number density fluctuations in redshift space, before adding the EFT counterterms and the stochasticity contributions, are expressed by the perturbative expansion 
\begin{align}
    \delta_\mathrm{g}(\bk) &= \sum_{n=1}^{\infty} \delta_{\mathrm{g}}^{(n)}(\bk) = \sum_{n=1}^{\infty} \int_{\bk_1,\cdots,\bk_n} (2\pi)^3 \delta_\mathrm{D}(\bk-\bk_{1 \cdots n}) Z_{n}(\bk_1, \cdots, \bk_n) \delta^{(1)}(\bk_1) \cdots \delta^{(1)}(\bk_n),
    \label{eq:galaxy_density_perturbations_redshift_space_EPT}
\end{align}
where the explicit expressions of the PT kernels up to third order, which are required to calculate the one-loop power spectrum, are written as
\begin{align}
    Z_{1}(\bk_1) &= b_1 + f \mu_1^2, \\
    Z_{2}(\bk_1,\bk_2) &= b_1 F_2(\bk_1,\bk_2) + \frac{b_2}{2} + b_{\G_2} K_{\G_2}(\bk_1,\bk_2) + f \mu_{12}^2 G_2(\bk_1,\bk_2) + \frac{f \mu_{12} k_{12}}{2} b_1 \left( \frac{\mu_1}{k_1} + \frac{\mu_2}{k_2} \right) + \frac{(f \mu_{12} k_{12})^2}{2} \frac{\mu_1}{k_1} \frac{\mu_2}{k_2}, \\
    Z_{3}(\bk_1,\bk_2,\bk_3) &= b_1 F_3(\bk_1,\bk_2,\bk_3) + b_2 F_2(\bk_2,\bk_3) + 2 b_{\G_2} K_{\G_2}(\bk_1,\bk_2+\bk_3) F_2(\bk_2,\bk_3) + \frac{b_3}{6} + b_{\delta \G_2} K_{\G_2}(\bk_2,\bk_3) \nonumber\\
    &\,\, + b_{\G_3} K_{\G_3}(\bk_1,\bk_2,\bk_3) + 2 b_{\Gamma_3} K_{\G_2}(\bk_1,\bk_2+\bk_3) [F_2(\bk_2,\bk_3) - G_2(\bk_2,\bk_3)] + f \mu_{123}^2 G_3(\bk_1,\bk_2,\bk_3) \nonumber\\
    &\,\, + (f \mu_{123} k_{123}) b_1 \frac{\mu_{23}}{k_{23}} G_2(\bk_2,\bk_3) + (f \mu_{123} k_{123}) \frac{\mu_1}{k_1} \left[ b_1 F_2(\bk_2,\bk_3) + \frac{b_2}{2} + b_{\G_2} K_{\G_2}(\bk_2,\bk_3) \right] \nonumber\\
    &\,\, + (f \mu_{123} k_{123})^2 \frac{\mu_1}{k_1} \frac{\mu_{23}}{k_{23}} G_2(\bk_2,\bk_3) + \frac{(f \mu_{123} k_{123})^2}{2} b_1 \frac{\mu_2}{k_2} \frac{\mu_3}{k_3} + \frac{(f \mu_{123} k_{123})^3}{6} \frac{\mu_1}{k_1} \frac{\mu_2}{k_2}  \frac{\mu_3}{k_3},
    \label{eq:Z3_kernel}
\end{align}
In the above, the $F_{n}$ and $G_{n}$ are the $n$th-order kernels for matter density fluctuation and velocity divergence (see \textit{e.g.}, a review \cite{Bernardeau:2001qr}), we have defined $\mu_i \equiv (\bk_i \cdot \hat{\bn}) / k_i$ and $\mu_{1 \cdots n} \equiv (\bk_{1 \cdots n} \cdot \hat{\bn}) / k_{1 \cdots n}$, with $\hat{\bn}$ being the unit vector along the line-of-sight.
We also have defined the Fourier-space kernels of $\G_{2}$ and $\G_{3}$,
\begin{align}
    K_{\G_{2}}(\bk_{1}, \bk_{2}) &= \frac{ (\bk_{1} \cdot \bk_{2})^{2} }{ k_{1}^{2} k_{2}^{2} } - 1, \label{eq:KG2} \\
    K_{\G_{3}}(\bk_{1}, \bk_{2}, \bk_{3}) &= \frac{1}{2} \left[ \frac{(\bk_1 \cdot \bk_2)^2}{k_1^2 k_2^2} + \frac{(\bk_2 \cdot \bk_3)^2}{k_2^2 k_3^2} + \frac{(\bk_3 \cdot \bk_1)^2}{k_3^2 k_1^2} \right] - \frac{(\bk_1 \cdot \bk_2) (\bk_2 \cdot \bk_3) (\bk_3 \cdot \bk_1)}{k_1^2 k_2^2 k_3^2} - \frac{1}{2}, \label{eq:KG3}
\end{align}
and note that the third-order kernel (\ref{eq:Z3_kernel}) needs to be symmetrized.

The one-loop EPT power spectrum, including the EFT counterterms and the stochasticity terms, is written as follows:
\begin{align}
    P^{\mathrm{EPT}}(\bk) &= P^{\mathrm{EPT}}_{\mathrm{tree}}(\bk) + P^{\mathrm{EPT}}_{\text{1-loop}}(\bk) + P_{\mathrm{ctr}}(\bk) + P_{\mathrm{stoch}}(\bk).
\end{align}
Using the above expansion (\ref{eq:galaxy_density_perturbations_redshift_space_EPT}), the tree-level and one-loop order parts of the galaxy power spectrum are written as
\begin{align}
    P^{\mathrm{EPT}}_{\mathrm{tree}}(\bk) &=  \left[ Z_{1}(\bk) \right]^2 P_{\lin}(k), \label{eq:pkmu_ept_tree} \\
    P^{\mathrm{EPT}}_{\text{1-loop}}(\bk) &= 2 \int_{\bp} \left[ Z_{2}(\bp,\bk-\bp) \right]^2 P_{\lin}(p) P_{\lin}(|\bk-\bp|) + 6 Z_{1}(\bk) P_{\lin}(k) \int_{\bp} Z_{3}(\bk,\bp,-\bp) P_{\lin}(p),
    \label{eq:pkmu_ept_1loop}
\end{align}
where $P_{\lin}(k)$ is the linear matter power spectrum.
Note that $\{b_3, b_{\G_3}, b_{\delta \G_2} \}$ do not appear in the one-loop power spectrum after the renormalization, and also the higher-derivative term is absorbed into the counterterm in the matter one-loop power spectrum.
The counterterm and the stochasticity term are described in Sec.~\ref{sec:ctr_and_stoch}.

\subsubsection{Infrared resummation}
\label{sec:ir_resum}

The displacements from long-wavelength modes affect features in the clustering, such as the BAO \cite{Senatore:2014via,Baldauf:2015xfa}.
A naive EPT calculation fully expands the effect of the long-wavelength displacements, hence fails to accurately capture the nonlinear damping of the oscillatory features.
This is contrastive with the LPT, which keeps the IR displacements exponentiated as described in Sec.~\ref{sec:model_lpt}.
In fact, if one Taylor expands the exponential factor in the LPT, it reduces to the EPT result (see Appendix~\ref{sec:relation_btw_ept_lpt} for an explicit calculation on the one-loop matter power spectrum).
Thus, in the EPT we resum the IR modes of the displacements to retain the nonlinear BAO damping, under the approximation that the IR displacements affect only the wiggle feature.
Following \cite{Ivanov:2018gjr}, we calculate the IR-resummed one-loop EPT power spectrum as
\begin{align}
    P(\bk) &= \left[ Z_{1}(\bk) \right]^{2} \left\{ P_{\mathrm{nw}}(k) + e^{- k^{2} \Sigma_{\mathrm{s}}^{2}(\mu) } P_{\mathrm{w}}(k) \left[ 1 + k^{2} \Sigma_{\mathrm{s}}^{2}(\mu) \right] \right\} \nonumber\\
    &\,\, + P_{\text{1-loop}}[P_{\mathrm{nw}}](\bk) + e^{- k^{2} \Sigma_{\mathrm{s}}^{2}(\mu) } P_{\text{1-loop}}[P_{\mathrm{w}}](\bk),
    \label{eq:ir_resummed_pkmu}
\end{align}
where $P_{\mathrm{nw}}$ and $P_{\mathrm{w}}$ are the no-wiggle and wiggle parts of the linear matter power spectrum, \textit{i.e.},
\begin{align}
    P_{\lin}(k) = P_{\mathrm{nw}}(k) + P_{\mathrm{w}}(k).
    \label{eq:pk_wiggle_split}
\end{align}
This split is not unique, and we describe the splitting methods in Sec.~\ref{sec:ir_resum_impl}.
In Eq.~(\ref{eq:ir_resummed_pkmu}), we apply the damping factor $\Sigma_{\mathrm{s}}^{2}(\mu)$ on the wiggle part of the one-loop contribution. 
$P_{\text{1-loop}}[P_{\mathrm{nw}}](\bk)$ is the no-wiggle part of the one-loop term, \textit{i.e.}, the one-loop term computed using $P_{\mathrm{nw}}$ as the linear power spectrum, and we denote the remaining wiggle part by $P_{\text{1-loop}}[P_{\mathrm{w}}](\bk) \equiv P_{\text{1-loop}}[P_{\lin}](\bk) - P_{\text{1-loop}}[P_{\mathrm{nw}}](\bk)$.
Note that $P_{\text{1-loop}}$ in the above should be understood to include the counterterm as well.
The damping factor in redshift space is \cite{Ivanov:2018gjr}
\begin{align}
    \Sigma^{2}_{\mathrm{s}}(\mu) &= [1 + f (2 + f) \mu^{2}] \Sigma^{2} + f^{2} \mu^{2} (\mu^{2} - 1) \delta \Sigma^{2}.
    \label{eq:damping_redshift_space}
\end{align}
with the two separate pieces written as
\begin{align}
    \Sigma^{2} &= \int_{0}^{\Lambda_\mathrm{IR}} \frac{\mathrm{d}k}{6 \pi^{2}} P_{\mathrm{nw}}(k) [ 1 - j_{0}(k r_{\mathrm{BAO}}) + 2 j_{2}(k r_{\mathrm{BAO}}) ], \\
    \delta \Sigma^{2} &= \int_{0}^{\Lambda_\mathrm{IR}} \frac{\mathrm{d}k}{2 \pi^{2}} P_{\mathrm{nw}}(k) j_{2}(k r_{\mathrm{BAO}}).
\end{align}
$\Sigma^{2}$ is the damping factor that arises in real space \cite{Baldauf:2015xfa,Blas:2016sfa}, and $\delta \Sigma^{2}$ arises due to the redshift-space distortions.
In the above, we set the BAO scale $r_{\mathrm{BAO}} = 110 \, h^{-1} \, \mathrm{Mpc}$ and $\Lambda_\mathrm{IR} = 0.2 \, h \, \mathrm{Mpc}^{-1}$ as the default.
While the redshift-space damping described above was first explicitly shown by \cite{Ivanov:2018gjr} in the context of the time-sliced perturbation theory \cite{Blas:2015qsi}, it can also be derived from the argument in \cite{Baldauf:2015xfa} and in relation to LPT, as described in Appendix~\ref{sec:ir_resum_redshift_space}.

\subsection{Lagrangian perturbation theory}
\label{sec:model_lpt}

In LPT, we focus on the displacement $\bPsi$ that relates the Eulerian coordinates $\bx$ and the corresponding Lagrangian coordinates $\bq$,
\begin{align}
    \bx &= \bq + \bPsi(\bq),
    \label{eq:displacement_btw_q_and_x}
\end{align}
and expand it into a perturbative series rather than expanding the Eulerian fields such as the density contrast, as done in EPT.
In Fourier space, the displacement field $\bPsi$ is expanded as
\begin{align}
    \bPsi(\bk) &= \sum_{n=1}^{\infty} \bPsi^{(n)}(\bk) = \sum_{n=1}^{\infty} \int_{\bk_{1}, \cdots, \bk_{n}} \dD(\bk-\bk_{1 \cdots n}) i \bL_{n}(\bk_{1}, \cdots, \bk_{n}) \delta^{(1)}(\bk_{1}) \cdots \delta^{(1)}(\bk_{n}),
\end{align}
where the recursion relations are known for the kernels for each order $\bL_{n}$ \cite{Rampf:2012up,Matsubara:2015ipa}.
The one-loop power spectrum requires the displacement expanded up to third order.
Corresponding to the coordinate transformation (\ref{eq:displacement_btw_q_and_x}), the galaxy number density fluctuations $\delta_{\mathrm{g}}(\bx)$ at an Eulerian position $\bx$ is related to the Lagrangian field as   
\begin{align}
    1 + \delta_{\mathrm{g}}(\bx) = \int_{\bq} F(\bq) \delta_{\mathrm{D}}(\bx - \bq - \bPsi(\bq)) 
    \text{ and } \dD(\bk) + \delta_{\mathrm{g}}(\bk) = \int_{\bq} F(\bq) e^{-i \bk \cdot [\bq + \bPsi(\bq)]},
    \label{eq:eulerian_to_lagrangian_mapping}
\end{align}
where the function $F(\bq) = 1 + \delta_{\mathrm{g}}^{\mathrm{L}}(\bq)$ describes how the galaxy density field at late time is expressed in the corresponding initial Lagrangian coordinates $\bq$.
The case of matter distributions implies that $F(\bq) = 1$.
By taking the two-point correlation of $\delta_{\mathrm{g}}(\bk)$, we obtain the galaxy power spectrum as
\begin{align}
    P(\bk) &= \int_{\bq} e^{i \bk \cdot \bq} \avrg{e^{i \bk \cdot \bDelta} F(\bq_1) F(\bq_2)}_{\bq = \bq_{2} - \bq_{1}}, 
    \label{eq:pk_bias_lpt}
\end{align}
where we have defined the difference of the displacements between two Lagrangian positions, $\bDelta \equiv \bPsi(\bq_2) - \bPsi(\bq_1)$.
In the Lagrangian bias expansion, we include the following set of operators up to third order constructed from the initial linear field,
\begin{align}
    F(\bq) 
    &= 1 + b_{1}^{\mathrm{L}} \delta(\bq) + \frac{b_{2}^{\mathrm{L}}}{2} \left[ \delta^{2}(\bq) - \sigma^{2} \right] + b_{\G_{2}}^{\mathrm{L}} \G_{2}(\bq) + \frac{b_{3}^{\mathrm{L}}}{6} \left[ \delta^{3}(\bq) - 3 \sigma^{2} \delta(\bq) \right] \nonumber\\
    &\,\, + b_{\G_{3}}^{\mathrm{L}} \G_{3}(\bq) + b_{\delta \G_{2}}^{\mathrm{L}} \left[ \delta(\bq) \G_{2}(\bq) + \frac{4}{3} \sigma^{2} \delta(\bq) \right] + b_{\Gamma_{3}}^{\mathrm{L}} \Gamma_{3}(\bq) + b_{\nabla^{2} \delta}^{\mathrm{L}} \nabla^{2} \delta(\bq),
    \label{eq:lagrangian_bias_expansion}
\end{align}
where we have employed the renormalized operators with the trivial zero-lag pieces being subtracted.
Using the bias expansion (\ref{eq:lagrangian_bias_expansion}), one can rewrite Eq.~(\ref{eq:pk_bias_lpt}) as \cite{Vlah:2018ygt}
\begin{align}
    P(\bk) 
    &= \int_{\bq} e^{i \bk \cdot \bq} \left. \D_{1} \D_{2} Z(\bJ_{1}, \bJ_{2} ; \bk, \bq) \right|_{\bJ_{1}, \bJ_{2} = 0},
\end{align}
in terms of the generating function for the displacement difference and Lagrangian bias operators,
\begin{align}
    Z(\bJ_{1}, \bJ_{2} ; \bk, \bq) &\equiv \avrg{ e^{i \bk \cdot \bDelta + \sum_{O} J_{O,1} O(\bq_{1}) + \sum_{O} J_{O,2} O(\bq_{2}) } }_{\bq = \bq_{2} - \bq_{1}} \text{ where } O \in \{ \delta, \G_{2}, \G_{3}, \Gamma_{3}, \nabla^{2} \delta \},
    \label{eq:lpt_generating_function}
\end{align}
and the derivative operator acting on it corresponding to the bias expansion (\ref{eq:lagrangian_bias_expansion}),
\begin{align}
    \D_{a} &\equiv 1 + b_{1}^{\mathrm{L}} \partial_{J_{\delta, a}} + \frac{b_{2}^{\mathrm{L}}}{2} (\partial_{J_{\delta, a}}^{2} - \sigma^{2}) + b_{\G_{2}}^{\mathrm{L}} \partial_{J_{\G_{2}, a}} + \frac{b_{3}^{\mathrm{L}}}{6} (\partial_{J_{\delta, a}}^{3} - 3 \sigma^{2} \partial_{J_{\delta, a}}) \nonumber\\
    &\,\, + b_{\G_{3}}^{\mathrm{L}} \partial_{J_{\G_{3}, a}} + b_{\delta \G_{2}}^{\mathrm{L}} \left( \partial_{J_{\delta, a}} \partial_{J_{\G_{2}, a}} + \frac{4}{3} \sigma^{2} \partial_{J_{\delta, a}} \right) + b_{\Gamma_{3}}^{\mathrm{L}} \partial_{J_{\Gamma_{3}, a}} + b_{\nabla^{2} \delta}^{\mathrm{L}} \partial_{J_{\nabla^{2} \delta, a}}.
\end{align}
To move on to perturbation theory, we perform the cumulant expansion of Eq.~(\ref{eq:lpt_generating_function}) and truncate it at the desired order; in the case of the one-loop power spectrum, the cumulant expansion is up to the fourth order of the linear density $\delta^{(1)}$.
This expansion generates the cumulants such as $A_{ij} \equiv \avrgc{\Delta_{i} \Delta_{j}}$ and $W_{ijk} \equiv \avrgc{\Delta_{i} \Delta_{j} \Delta_{k}}$, which exist in the matter clustering, or the terms such as $U^{10}_{i} \equiv \avrgc{ \Delta_{i} \delta }$, which arise from galaxy biasing.
We defer the details of these terms to Appendix~\ref{sec:lpt_formulas}.
The lowest-order cumulant of the displacement difference, $A_{ij}$, has the linear and the one-loop order terms,
\begin{align}
    A_{ij}(\bq) = A_{ij}^{\lin}(\bq) + A_{ij}^{\text{1-loop}}(\bq).
\end{align}
We further separate $A_{ij}^{\lin}$ into the IR part $A_{ij}^{<}$ and the UV part $A_{ij}^{>}$ by imposing an IR cut-off scale $k_{\mathrm{IR}}$, and keep only the IR part exponentiated \cite{Chen:2020zjt}, which corresponds to resumming the long-wavelength displacements that affect features like the BAO.
Following \cite{Chen:2020zjt}, we set $k_{\mathrm{IR}} = 0.2 \, h \, \mathrm{Mpc}^{-1}$ by default.
The other contributions are Taylor-expanded and are truncated at the order of $\O(P_{\lin}^2)$. 
Thus, the one-loop LPT power spectrum is
\begin{align}
    P^{\mathrm{LPT}}(\bk) &= P^{\mathrm{LPT}}_{\mathrm{tree}}(\bk) + P^{\mathrm{LPT}}_{\text{1-loop}}(\bk) + P^{\mathrm{LPT}}_{\mathrm{ctr}}(\bk) + P_{\mathrm{stoch}}(\bk).
    \label{eq:ps_1loop_lpt}
\end{align}
In the above, we denote the terms of $\O(P_{\lin})$ as the ``tree-level'' terms
\footnote{
These terms are approximately equivalent to the IR-resummed tree-level EPT in the saddle-point approximation \cite{Vlah:2015zda,Chen:2024pyp}.
Splitting $A_{ij}^{<}$ into the no-wiggle and wiggle parts, $A_{ij}^{\mathrm{nw},<}$ and $A_{ij}^{\mathrm{w},<}$, and regarding $A_{ij}^{>}$ as consisting only of the no-wiggle contribution, we can separate $P^{\mathrm{LPT}}_{\mathrm{tree}}$ into the no-wiggle and wiggle parts, 
$P^{\mathrm{LPT,nw}}_{\mathrm{tree}}(\bk) = \int_{\bq} e^{i \bk \cdot \bq} e^{ - \frac{1}{2} k_{i} k_{j} A_{ij}^{\mathrm{nw},<} } \left\{ 1 - \frac{1}{2} k_{i} k_{j} A_{ij}^{\mathrm{nw},>} + 2 i b_{1}^{\mathrm{L}} k_{i} U^{\mathrm{nw}}_{i} + (b_{1}^{\mathrm{L}})^{2} \xi^{\mathrm{nw}} \right\}$ 
and 
$P^{\mathrm{LPT,w}}_{\mathrm{tree}}(\bk) = \int_{\bq} e^{i \bk \cdot \bq} e^{ - \frac{1}{2} k_{i} k_{j} A_{ij}^{\mathrm{nw},<} } \left \{ - \frac{1}{2} k_{i} k_{j} A_{ij}^{\mathrm{w},<} + 2 i b_{1}^{\mathrm{L}} k_{i} U^{\mathrm{w}}_{i} + (b_{1}^{\mathrm{L}})^{2} \xi^{\mathrm{w}} \right \}$.
Expanding the exponential factor in the no-wiggle contribution and keeping terms up to linear order, the first line reduces to $\dD(\bk) + Z_{1}^{2}(\bk) P_{\mathrm{nw}}(k)$. 
On the other hand, for the wiggle contribution, the exponential factor can be evaluated using the saddle-point approximation for $k r_{\mathrm{BAO}} \gg 1$, yielding 
$P^{\mathrm{LPT,w}}_{\mathrm{tree}}(\bk) \simeq e^{-k^{2}\Sigma^{2}(\mu)} Z_{1}^{2}(\bk) P_{\mathrm{w}}(k)$.
}
and the terms of $\O(P_{\lin}^{2})$ as the one-loop terms, \textit{i.e.},
\begin{align}
    P^{\mathrm{LPT}}_{\mathrm{tree}}(\bk) &= \int_{\bq} e^{i \bk \cdot \bq} e^{- \frac{1}{2} k_{i} k_{j} A_{ij}^{<}} \left\{ 1 - \frac{1}{2} k_{i} k_{j} A_{ij}^{>} + 2 i b_{1}^{\mathrm{L}} k_{i} U_{i}^{\lin} + (b_{1}^{\mathrm{L}})^{2} \xi_{\lin} \right\},
    \label{eq:pkmu_lpt_tree} \\
    P^{\mathrm{LPT}}_{\text{1-loop}}(\bk) 
    &= \int_{\bq} e^{i \bk \cdot \bq} e^{- \frac{1}{2} k_{i} k_{j} A_{ij}^{<}} \left\{ - \frac{1}{2} k_{i} k_{j} A^{\text{1-loop}}_{ij} - \frac{i}{6} k_{i} k_{j} k_{k} W_{ijk} + \frac{1}{8} k_{i} k_{j} k_{k} k_{l} A^{>}_{ij} A^{>}_{kl} \right. \nonumber\\ 
    &\,\, \left. + 2 i b_{1}^{\mathrm{L}} \left( k_{i} U^{(3)}_{i} - \frac{1}{2} k_{i} k_{j} A_{ij}^{>} k_{k} U^{\lin}_{k} \right) - \frac{1}{2} (b_{1}^{\mathrm{L}})^2 k_{i} k_{j} A_{ij}^{>}
    \xi_{\lin} + i b_{2}^{\mathrm{L}} k_{i} U^{20}_{i} + i (b_{1}^{\mathrm{L}})^2 k_{i} U^{11}_{i} - b_{1}^{\mathrm{L}} k_{i} k_{j} A^{10}_{ij} \right. \nonumber\\
    &\,\, \left. - ( (b_{1}^{\mathrm{L}})^2 + b_{2}^{\mathrm{L}} ) k_{i} k_{j} U^{\lin}_{i} U^{\lin}_{j} + \frac{1}{2} (b_{2}^{\mathrm{L}})^2 \xi_{\lin}^2 + 2 i b_{1}^{\mathrm{L}} b_{2}^{\mathrm{L}} \xi_{\lin} k_{i} U^{\lin}_{i} \right. \nonumber\\ 
    &\,\, \left. - b_{\G_{2}}^{\mathrm{L}} k_{i} k_{j} \Upsilon_{\G_{2}, ij} + 2 i b_{\G_{2}}^{\mathrm{L}} k_{i} V^{10}_{\G_{2}, i} + 2 i b_{1}^{\mathrm{L}} b_{\G_{2}}^{\mathrm{L}} k_{i} V^{12}_{\G_{2}, i} + b_{2}^{\mathrm{L}} b_{\G_{2}}^{\mathrm{L}} \chi_{\G_{2}} + (b_{\G_{2}}^{\mathrm{L}})^2 \zeta_{\G_{2}} + 2 i b_{\Gamma_{3}}^{\mathrm{L}} k_{i} U_{\Gamma_{3}, i} + 2 b_{1}^{\mathrm{L}} b_{\Gamma_{3}}^{\mathrm{L}} \theta_{\Gamma_{3}} \right\}.
    \label{eq:pkmu_lpt_1loop}
\end{align}
Similarly to the EPT, $\{ b_{3}^{\mathrm{L}}, b_{\G_{3}}^{\mathrm{L}}, b_{\delta \G_{2}}^{\mathrm{L}} \}$ do not appear in the final one-loop power spectrum due to the renormalization (see also Appendix~\ref{sec:lpt_formulas}), and we have dropped the $b_{\nabla^{2} \delta}^{\mathrm{L}}$ terms because their behavior is similar to the counterterms \cite{Vlah:2016bcl}.

The redshift-space distortions are introduced by shifting the displacement vectors along the line-of-sight $\hat{\bn}$ \cite{Matsubara:2007wj}.
The apparent displacement in redshift space is
\begin{align}
    \bPsi^{\mathrm{s}} = \bPsi + \frac{\hat{\bn} \cdot \dot{\bPsi}}{H} \hat{\bn}.
\end{align}
Under the approximation of the time-independent LPT kernels, the time dependence of the $n$th-order displacement is given by $\Psi^{(n)} \propto D^{n}$, where $D$ is the linear growth factor.
Hence, its time derivative scales as $\dot{\bPsi}^{(n)} = n H f \bPsi^{(n)}$, and the $n$th-order displacement in redshift space becomes \cite{Matsubara:2007wj}
\begin{align}
    \Psi^{\mathrm{s}(n)}_{i} = (\delta_{ij} + n f \hat{n}_i \hat{n}_j) \Psi^{(n)}_{j} \equiv R^{(n)}_{ij} \Psi^{(n)}_{j},
    \label{eq:displacement_rsd}
\end{align}
where we defined the matrices $R^{(n)}_{ij}$ that maps real-space displacements to redshift-space displacements at each perturbative order.
The redshift-space LPT power spectrum is obtained by replacing the displacement in Eq.~(\ref{eq:ps_1loop_lpt}) with the redshift-space displacement (\ref{eq:displacement_rsd}).

\subsection{Counterterms and stochasticity}
\label{sec:ctr_and_stoch}

For the EFT counterterms, we adopt the same default parametrization for the EPT and LPT implementations, while also providing an alternative LPT implementation that follows the standard LPT convention (see Appendix~\ref{sec:lpt_ctr}).
The leading-order counterterm contribution to the redshift-space one-loop power spectrum can be written as
\begin{align}
    P_{\mathrm{ctr}}(\bk) &= 2 Z_{1}(\bk) \left( \tilde{c}_{0} + \tilde{c}_{2} \mu^2 + \tilde{c}_{4} \mu^4 \right) k^{2} P_{\lin}(k).
    \label{eq:pkmu_ctr_lo}
\end{align}
These terms account for the UV sensitivity associated with short-scale physics and with the renormalization of the composite operators appearing in the perturbative expansion of the redshift-space mapping in Eq.~(\ref{eq:rsd_to_3rd_order_fourier}) \cite{Senatore:2014vja}.
In addition, following \cite{Ivanov:2019pdj,Chudaykin:2020aoj,Mergulhao:2023zso,Rubira:2025scu}, we include the next-to-leading-order (NLO) counterterm to capture sensitivity to small-scale velocity dispersion or the Fingers-of-God effect \cite{Jackson:1971sky}.
To absorb the UV sensitivity related to the composite operators that arises from the expansion (\ref{eq:rsd_to_3rd_order_fourier}) extended to the next order, \textit{e.g.}, $k_{\parallel}^{4} [u_{\parallel}]^{4}(\bk)$, one can include counterterms $\delta_{\mathrm{ctr}}^{\mathrm{NLO}}(\bk) \supset k^4 \mu^{4} \delta^{(1)}(\bk), k^4 \mu^6 \delta^{(1)}(\bk)$.
We therefore include the NLO contribution
\begin{align}
    P_{\mathrm{ctr, NLO}}(\bk) &= 2 Z_{1}(\bk) \left( \tilde{c}_{44} + \tilde{c}_{46} \mu^{2} \right)  (k \mu)^{4} P_{\lin}(k),
    \label{eq:pkmu_ctr_nlo}
\end{align}
to the power spectrum.
When IR resummation is employed, the counterterms are resummed consistently with the other one-loop contributions, and $P_{\lin}(k)$ in the above is replaced with the IR-resummed linear power spectrum, $P_{\lin}(k) \rightarrow P_{\mathrm{nw}}(k) + e^{-k^{2} \Sigma_{\mathrm{s}}^{2}(\mu)} P_{\mathrm{w}}(k)$.

For both the EPT and LPT power spectra, we model the stochastic contribution as \cite{Perko:2016puo}
\begin{align}
    P_{\mathrm{stoch}}(\bk) &= \epsilon_{00} + (\epsilon_{20} + \epsilon_{22} \mu^2) k^{2},
    \label{eq:pkmu_stoch}
\end{align}
and the code interface does not require the normalizing factors such as the tracer number density or the characteristic scale.
The users can set such factors as in Eq.~\eqref{eq:pkmu_stoch_auto} according to their convention.

\subsection{Cross power spectrum}

Our code also supports the cross power spectrum between two different tracers, denoted by $\mathrm{A}$ and $\mathrm{B}$.
The two tracers can have independent sets of bias parameters.
For each bias operator $O$, the corresponding combinations appearing in the auto-power spectrum are generalized to the cross spectrum according to
\begin{align}
    b_{O} \mapsto \frac{1}{2} (b_{O}^{\mathrm{A}} + b_{O}^{\mathrm{B}}), \quad b_{O}^{2} \mapsto b_{O}^{\mathrm{A}} b_{O}^{\mathrm{B}}, \quad b_{O} b_{O'} \mapsto \frac{1}{2} (b_{O}^{\mathrm{A}} b_{O'}^{\mathrm{B}} + b_{O}^{\mathrm{B}} b_{O'}^{\mathrm{A}}).
\end{align}
If the EFT coefficients are assigned independently to the two tracers, the cross-spectrum counterterms take the form
\begin{align}
    P_{\mathrm{ctr}}^{\mathrm{AB}}(\bk) &= \left[ Z_{1}^{\mathrm{A}}(\bk) \left( \tilde{c}_{0}^{\mathrm{B}} + \tilde{c}_{2}^{\mathrm{B}} \mu^{2} + \tilde{c}_{4}^{\mathrm{B}} \mu^{4} \right) + (\mathrm{A} \leftrightarrow \mathrm{B}) \right] k^{2} P_{\lin}(k), \\
    P_{\mathrm{ctr, NLO}}^{\mathrm{AB}}(\bk) &= \left[ Z_{1}^{\mathrm{A}}(\bk) \left( \tilde{c}_{44}^{\mathrm{B}} + \tilde{c}_{46}^{\mathrm{B}} \mu^{2} \right) + (\mathrm{A} \leftrightarrow \mathrm{B}) \right] (k \mu)^{4} P_{\lin}(k),
\end{align}
which reduce to Eqs.~(\ref{eq:pkmu_ctr_lo}) and (\ref{eq:pkmu_ctr_nlo}) in the case of auto power spectrum.
For the public interface of the code, however, we adopt a more flexible reparametrization of these contributions, similar to those used, \textit{e.g.}, in \cite{Chudaykin:2020aoj,Linde:2024uzr},
\begin{align}
    P_{\mathrm{ctr}}(\bk) &= - 2 \left( c_{0} + c_{2} f \mu^2 + c_{4} f^{2} \mu^4 + c_{6} f^{3} \mu^{6} \right) k^{2} P_{\lin}(k), \label{eq:pkmu_ept_ctr} \\
    P_{\mathrm{ctr, NLO}}(\bk) &= -2 \left( c_{44} + c_{46}  f \mu^{2} + c_{48} f^{2} \mu^{4} \right) (k \mu f)^{4}  P_{\lin}(k). \label{eq:pkmu_ept_ctr_nlo}
\end{align}
These parametrizations can be used for both auto- and cross-power spectra, with the coefficients $c_i$ interpreted as effective counterterm parameters for the spectrum under consideration.
This keeps the code interface independent of a particular convention for assigning EFT coefficients to individual tracers and allows users to map the $c_i$ onto their preferred counterterm basis.

We follow the same philosophy for the stochastic contribution in Eq.~(\ref{eq:pkmu_stoch}), whose coefficients are treated as effective stochastic parameters of the auto- or cross-spectrum under consideration.

\subsection{Alcock--Paczy\'{n}ski effect}
\label{sec:ap_effect}

In galaxy spectroscopic surveys, we directly observe the redshift and the angular position (RA and Dec) of each target galaxy, and transform them into the comoving coordinates assuming some cosmological model as the reference (or fiducial) cosmology.
As a result, the power spectra measured from actual survey catalogs are distorted due to the coordinate change arising from the difference between the unknown true cosmology and the reference cosmology, which is called the Alcock--Paczy\'{n}ski (AP) effect \cite{Alcock:1979mp,Matsubara:1996nf}.
To take into account the AP effect, we distort the wavevector in Fourier space as
\begin{align}
    k^{\mathrm{ref}}_{\perp} = \frac{ D_{\mathrm{A}}(z) }{ D^{\mathrm{ref}}_{\mathrm{A}}(z) } k_{\perp} \equiv \alpha_{\perp} k_{\perp}, \text{ } k^{\mathrm{ref}}_{\parallel} = \frac{ H^{\mathrm{ref}}(z) }{ H(z) } k_{\parallel} \equiv \alpha_{\parallel} k_{\parallel},
\end{align}
where $D_{\mathrm{A}}$ and $H(z)$ are the angular diameter distance and the Hubble expansion rate (respectively) at the redshift $z$ of the target galaxies.
According to the above coordinate change, the AP-distorted redshift-space power spectrum in terms of $(k^{\mathrm{ref}}, \mu^{\mathrm{ref}})$ is written as
\begin{align}
    P^{\mathrm{ref}}(k^{\mathrm{ref}}, \mu^{\mathrm{ref}}) &= \frac{1}{ \alpha_{\perp}^{2} \alpha_{\parallel} } P( k(k^{\mathrm{ref}}, \mu^{\mathrm{ref}} ), \mu(\mu^{\mathrm{ref}}) ),
\end{align}
and the multipole moments that are usually directly measured from the survey catalogs are obtained by integrating with the Legendre polynomial with respect to $\mu^{\mathrm{ref}}$, the cosine of the angle between the wavevector and the line-of-sight in the reference coordinates,
\begin{align}
    P_{\ell}^{\mathrm{ref}}(k^{\mathrm{ref}}) &= (2 \ell + 1) \int_{-1}^{1} \frac{\mathrm{d} \mu_{\mathrm{ref}}}{2} P^{\mathrm{ref}}( k^{\mathrm{ref}}, \mu^{\mathrm{ref}} ) \L_{\ell}(\mu^{\mathrm{ref}}).
\end{align}
In the above, $k$ and $\mu$ in the true coordinates are written as a function of $k^{\mathrm{ref}}$ or $\mu^{\mathrm{ref}}$:
\begin{align}
    k(k^{\mathrm{ref}}, \mu^{\mathrm{ref}}) = \frac{k^{\mathrm{ref}}}{\alpha_{\perp}} \left[ 1 + (\mu^{\mathrm{ref}})^{2} \left( \frac{ \alpha_{\perp}^{2} }{ \alpha_{\parallel}^{2} } - 1 \right) \right]^{\frac{1}{2}}, \text{ }
    \mu(\mu^{\mathrm{ref}}) = \mu^{\mathrm{ref}} \frac{ \alpha_{\perp} }{ \alpha_{\parallel} } \left[ 1 + (\mu^{\mathrm{ref}})^{2} \left( \frac{ \alpha_{\perp}^{2} }{ \alpha_{\parallel}^{2} } - 1 \right) \right]^{-\frac{1}{2}},
\end{align}
and the $\mu_{\mathrm{ref}}$ integral is numerically done by Gauss--Legendre quadrature.

\section{Implementation}
\label{sec:implementation}

In this section, we describe the implementation details of our code.
For implementation of one-loop integrals, we use two different methods developed in \cite{Schmittfull:2016jsw} and \cite{Simonovic:2017mhp}, which we will refer to as the \textsc{Hankel} method (Sec.~\ref{sec:hankel_method}) and the \textsc{Matrix} method (Sec.~\ref{sec:matrix_method}), respectively.
In terms of computational complexity, the \textsc{Hankel} method is more efficient than the \textsc{Matrix} method: when evaluating the loop terms at $N_{\mathrm{FFT}}$ points of $k$, for which the FFTs are performed, computational complexity of the former scales as $\O(N_{\mathrm{FFT}} \log N_{\mathrm{FFT}})$ for both the 22- and 13-terms, whereas that of the latter scales as $\O(N_{\mathrm{FFT}}^{3})$ for the 22-terms and $\O(N_{\mathrm{FFT}}^{2})$ for the 13-terms (however, very recently \cite{Harder:2026uei} has devised a way to reduce the complexity).
Hence, we employ the \textsc{Hankel} method for EPT and LPT, and the \textsc{Matrix} method only for EPT as an option.
We clarify the mathematical relation between the two methods in Appendix~\ref{sec:relation_btw_hankel_matrix}.
We describe the implementation of the infrared resummation for the EPT power spectrum (Sec.~\ref{sec:ir_resum_impl}) and the angular integrals that involve the exponential factor for the LPT power spectrum in redshift space (Sec.~\ref{sec:angular_int_lpt}).

\subsection{\textsc{Hankel} method: one-loop integrals using Hankel transforms}
\label{sec:hankel_method}

As the primary scheme for implementing one-loop integrals, we employ the \textsc{Hankel} method \cite{Schmittfull:2016jsw}, which is based on what its authors call the ``generalized'' correlation functions defined as
\begin{align}
    \xi^{\ell}_{n}(r) &\equiv \int_{0}^{\infty} \frac{k^2 \mathrm{d}k}{2 \pi^2} j_{\ell}(k r) k^{n} P_{\lin}(k),
    \label{eq:xi_ln}
\end{align}
where $j_{\ell}(\cdot)$ is the spherical Bessel function of the first kind.
The case of $(\ell, n) = (0, 0)$ corresponds to the ordinary linear correlation function, \textit{i.e.}, $\xi^{0}_{0}(r) = \xi_{\lin}(r)$. 
Equation~(\ref{eq:xi_ln}) is computed with Hankel transforms of the linear power spectrum using 1D FFTs~\cite{Hamilton:1999uv}.
In the default setting, we perform FFTs on the linear power spectrum sampled on logarithmically equidistant $N_{\mathrm{FFT}} = 512$ bins over the range $[k_{\mathrm{min}}, k_{\mathrm{max}} ] = [10^{-5}, 10^{2}] \, h \, \mathrm{Mpc}^{-1}$.

One-loop terms of the EPT power spectrum consist of the Hankel transforms of $\xi^{\ell}_{n}$'s with different combinations of $(\ell, n)$.
In short, each of the 22-terms and of the 13-terms is expressed by a linear combination of
\begin{align}
    P_{22}(k) &\supset \int_{\bp} p^{n_{1}} |\bk - \bp|^{n_{2}} \L_{\ell}(\hat{\bp} \cdot \widehat{\bk - \bp}) P_{\lin}(p) P_{\lin}(|\bk - \bp|) 
    = (-1)^{\ell} \int_{0}^{\infty} 4 \pi r^2 \mathrm{d}r j_{0}(k r) \xi^{\ell}_{n_{1}}(r) \xi^{\ell}_{n_{2}}(r), 
    \label{eq:ept_P22}
    \\
    P_{13}(k) &\supset k^{m} P_{\lin}(k) \int_{\bp} \frac{p^{n} \L_{\ell}(\hat{\bk} \cdot \hat{\bp})}{|\bk - \bp|^2} P_{\lin}(p) 
    = k^{m} P_{\lin}(k) \int_{0}^{\infty} r \mathrm{d}r j_{\ell}(k r) \xi^{\ell}_{n}(r),
    \label{eq:ept_P13}
\end{align}
respectively.
Aside from the EPT power spectrum, we also employ this method to implement the LPT power spectrum entirely, including the $Q_{n}$ and $R_{n}$ integrals (see Appendix~\ref{sec:lpt_formulas} for the details) as described already in \cite{Schmittfull:2016jsw} and also employed in the implementation of the existing code \texttt{velocileptors} \cite{Chen:2020zjt}.

For the $\Lambda$CDM linear power spectrum, the generalized correlation functions (\ref{eq:xi_ln}) with $n>0$ are increasingly sensitive to the UV part of the spectrum.\footnote{Since $P_{\lin}$ scales roughly $\propto k^{-3}$ at high-$k$, the integral in Eq.~(\ref{eq:xi_ln}) scales like $\xi^{\ell}_{n}(r) \sim \int_{0}^{\infty} \mathrm{d}k j_{\ell}(kr) k^{n-1}$, which diverges for $n \ge 2$ at UV).}
Consequently, an artificial high-$k$ damping of the input linear power spectrum is often required to obtain numerically stable Hankel transforms.
In this work, we have found that this high-$k$ damping can be completely removed as follows.
The decomposition of the 22-terms is actually restricted to the cases of $n_{1} = -n_{2} (= n)$, and hence we can rewrite Eq.~\eqref{eq:ept_P22} as
\begin{align}
    P_{22}(k) &\supset \int_{\bp_{1},\bp_{2}} \dD(\bk - \bp_{12}) \frac{1}{2} \left[ \left( \frac{p_{1}}{p_{2}} \right)^{n} + \left( \frac{p_{2}}{p_{1}} \right)^{n} \right] \L_{\ell}(\mu_{12}) P_{\lin}(p_{1}) P_{\lin}(p_{2}),
\end{align}
where we have defined $\mu_{12} \equiv \hat{\bp}_{1} \cdot \hat{\bp}_{2}$.
Using the cosine rule $k^{2} = p_{1}^{2} + p_{2}^{2} + 2 p_{1} p_{2} \mu_{12}$, positive powers of $p_{1}$ and $p_{2}$ can be rewritten in terms of inverse powers of $p_{1}$ and $p_{2}$ together with polynomials of $\mu_{12}$. For example, for $n = 1$ and $2$,
\begin{align}
    \frac{p_{1}}{p_{2}} + \frac{p_{2}}{p_{1}} = \frac{k^{2}}{p_{1} p_{2}} - 2 \mu_{12}, \text{  } \left( \frac{p_{1}}{p_{2}} \right)^{2} + \left( \frac{p_{2}}{p_{1}} \right)^{2} = \frac{k^{4}}{p_{1}^{2} p_{2}^{2}} - 4 \frac{k^{2}}{p_{1} p_{2}} \mu_{12} + 4 \mu_{12}^{2} - 2.
\end{align}
Combined with Bonnet's recursion formula $(2 \ell + 1) \mu \L_{\ell}(\mu) = (\ell + 1) \L_{\ell+1}(\mu) + \ell \L_{\ell-1}(\mu)$, all 22-type terms can be recast into a linear combination of
\begin{align}
    P_{22}^{(\ell, n, m)}(k) &\equiv k^{m} \int_{\bp} p^{n} |\bk-\bp|^{n} \L_{\ell}(\hat{\bp} \cdot \widehat{\bk-\bp}) P_{\lin}(p) P_{\lin}(|\bk-\bp|) = (-1)^{\ell} k^{m} \int_{0}^{\infty} 4 \pi r^{2} \mathrm{d}r j_{0}(kr) \left[ \xi^{\ell}_{n}(r) \right]^{2},
    \label{eq:pk_22_lnm}
\end{align}
with $n \le 0$.
This eliminates the need for generalized correlation functions with positive $n$, thereby avoiding the strong high-$k$ damping that would otherwise appear in the corresponding Hankel transforms.

We also implement the 13-type contribution using generalized correlation functions $\xi^{\ell}_{n}$ with $n \le 0$ only.
As described in \cite{Schmittfull:2016jsw}, the decomposition of the 13-type kernel is not unique because of the cosine rule $p^{2} = |\bk - \bp|^{2} - k^{2} + 2 k p (\hat{\bk} \cdot \hat{\bp})$.
Specifically, let us denote the integral appearing in Eq.~(\ref{eq:ept_P13}) as $I_{13}^{(\ell, n)}(k) \equiv \int_{\bp} \frac{p^{n} \L_{\ell}(\hat{\bk} \cdot \hat{\bp})}{|\bk - \bp|^{2}} P_{\lin}(p)$.
Combining the above cosine rule and Bonnet's recursion formula, we have
\begin{align}
    I_{13}^{(\ell, n)} &= \left[ \int_{\bp} p^{n-2} P_{\lin}(p) \right] \delta_{\ell,0} - k^{2} I_{13}^{(\ell, n-2)} + 2 k \left[ \frac{\ell + 1}{2 \ell + 1} I_{13}^{(\ell+1, n-1)} + \frac{\ell}{2 \ell + 1} I_{13}^{(\ell-1, n-1)} \right].
\end{align}
which recursively lowers the index $n$.
The first term on the right-hand side corresponds to a zero-lag contribution, since it is proportional to the moment of the linear power spectrum and can be identified with $\xi_{n-2}^{0}(r=0)$.
Such terms are generated only as intermediate artifacts of the recursive reduction and are not present in the original 13-type kernel.
Hence, after combining all contributions, these zero-lag terms cancel identically, leaving an expression that depends only on generalized correlation functions with $n \le 0$.

\subsection{\textsc{Matrix} method: one-loop integrals using power-law decomposition of linear spectrum}
\label{sec:matrix_method}

As for the EPT power spectrum, we also implement another FFT-based method \cite{Simonovic:2017mhp,Chudaykin:2020aoj} (see also \cite{Lee:2024nqu}) to approximate the linear power spectrum by a linear combination of the power-law functions of $k$ with complex exponents:
\begin{align}
    P_{\lin}(k) = \sum_{m=-N_{\mathrm{FFT}} / 2}^{N_{\mathrm{FFT}} / 2-1} c_{m} k^{\nu + i \eta_{m}},
    \label{eq:decomp_pk_lin}
\end{align}
where the coefficients $c_m$ and exponents $\eta_m$ are given by
\begin{align}
    c_{m} = k_{\mathrm{min}}^{- i \eta_{m}} \frac{1}{N_{\mathrm{FFT}}} \sum_{j=0}^{N_{\mathrm{FFT}}-1} \frac{P_{\lin}(k_{j})}{k_{j}^{\nu}} e^{- \frac{2 \pi i m j}{ N_{\mathrm{FFT}} } }, \qquad\eta_{m} = \frac{N_{\mathrm{FFT}}-1}{N_{\mathrm{FFT}}} \frac{2\pi m}{\ln (k_\mathrm{max} / k_\mathrm{min})} = \frac{2\pi m}{N_{\mathrm{FFT}}\Delta_{\mathrm{FFT}}}.
    \label{eq:coef_power}
\end{align}
On the other hand, one can decompose one-loop PT kernels into a linear combination of the rational functions of $\{p, |\bk-\bp|, \hat{\bn} \cdot \bp \}$ of the form $\frac{ (\hat{\bn} \cdot \bp)^{n_{z}} }{ p^{2 n_{1}} |\bk - \bp|^{2 n_{2}} }$, and insert the analytic solution constructed by
\begin{equation}
    \int_{\bp} \frac{(\hat{\bn} \cdot \bp)^{n_z}}{ p^{2\nu_1} |\bk - \bp|^{2\nu_2}} = k^{3-2(\nu_1+\nu_2)+n_z} I^{(n_z)}(\mu,\nu_1,\nu_2),
\end{equation}
where $\mu = \hat{\bk} \cdot \hat{\bn}$, and $I^{(n_z)}(\mu,\nu_1,\nu_2)$ is an $n_z$th-order polynomial of $\mu$.
Consequently, one can express each one-loop integral as a linear combination of these analytic solutions. 
The explicit formulas for $I^{(n_z)}(\mu,\nu_1,\nu_2)$ can be found in \textit{e.g.}, Appendix~B of \cite{Linde:2024uzr} (and originally in \cite{Chudaykin:2020aoj}).

The implementation results depend on the choice of the FFT bias parameter $\nu$ that appears in Eq.~(\ref{eq:decomp_pk_lin}).
Following \cite{Chudaykin:2020aoj,Linde:2024uzr}, we adopt $\nu = -1.6$ for the 13-type terms that depend on $b_{\G_{2}}$ or $b_{\Gamma_{3}}$ and $\nu = -0.3$ for the terms that depend on none of $b_{2}$, $b_{\G_{2}}$ and $b_{\Gamma_{3}}$. For the 22-type terms that depend on $b_{2}$ or $b_{\G_{2}}$, we use $\nu = -1$ instead (Sec.~\ref{sec:discretization}).
Furthermore, under our setting of $\nu$, the 13-term also misses the UV part of the term implemented with $\nu = -0.3$, as described in \cite{Simonovic:2017mhp}.
We thus add by hand the UV part missed in the FFTLog-based calculation,
\begin{align}
    P_{13}^{\mathrm{UV}}(k, \mu) &= (b_{1} + f \mu^2) \left\{ - \frac{61}{315} b_{1} - \left[ \left( \frac{3}{5} - \frac{2}{105} b_{1} \right) + \left( \frac{16}{35} + \frac{1}{3} b_{1} \right) f \right] f \mu^{2} - \left[ \frac{46}{105} + \frac{1}{3} f \right] f^2 \mu^{4} \right\} \nonumber\\
    &\,\, \times k^2 P_{\lin}(k) \int_{0}^{\infty} \frac{\mathrm{d} p}{2 \pi^{2}} P_{\lin}(p).
    \label{eq:P_13_UV}
\end{align}
Note that Eq.~(\ref{eq:P_13_UV}) is not equal to the UV limit of the whole 13-term in Eq.~(\ref{eq:pkmu_ept_1loop}), but rather the UV limit after the parts that depend on $b_{\G_{2}}$ or $b_{\Gamma_{3}}$ (the 13-term does not depend on $b_{2}$ after the bias renormalization \cite{McDonald:2006mx,Assassi:2014fva}) are subtracted.

\subsection{Discretization and zero-wavenumber terms}
\label{sec:discretization}
The \textsc{Hankel} method evaluates every transform in the same way as $\xi^{\ell}_{n}$ in Eq.~\eqref{eq:xi_ln_in_g_l}.
The sampled input is decomposed into the complex power laws of Eqs.~\eqref{eq:decomp_pk_lin} and \eqref{eq:coef_power}, with a bias $\nu$ chosen for each transform, and each power law is transformed analytically with Eq.~\eqref{eq:g_l}.
Unlike in the \textsc{Matrix} method, the input is first extended by $N_{\mathrm{FFT}}/2$ nodes of the same spacing $\Delta_{\mathrm{FFT}}$ on each side, which we call the padded band, so that Eqs.~\eqref{eq:decomp_pk_lin} and \eqref{eq:coef_power} hold with $N_{\mathrm{FFT}} \to 2 N_{\mathrm{FFT}}$ and with $k_{\mathrm{min}}$ replaced by the lowest padded node.
The sum (Eq.~\eqref{eq:decomp_pk_lin}) is periodic in $\ln k$ with the period of the grid and passes through every sample, so that the result is the exact transform of this interpolating function rather than an approximation to the transform of the input itself.
The accuracy is therefore set by how well the interpolating function represents the input, which depends on which samples enter Eq.~\eqref{eq:coef_power} and on how they are weighted.

Since $k_{\mathrm{min}}^{-i \eta_{m}} e^{-2 \pi i m j / N_{\mathrm{FFT}}} = k_{j}^{-i \eta_{m}}$, the coefficient $N_{\mathrm{FFT}} \Delta_{\mathrm{FFT}} \, c_{m}$ in Eq.~\eqref{eq:coef_power} equals $\Delta_{\mathrm{FFT}} \sum_{j} P_{\lin}(k_{j}) ~ k_{j}^{-\nu - i \eta_{m}}$, \textit{i.e.}, the rectangle-rule approximation, with the same weight on every node, to $\int \mathrm{d} \ln k ~P_{\lin}(k) ~ k^{-\nu - i \eta_{m}}$ over one period.
Suppose that the input is smooth on the nodes $j_1 \le j \le j_2$ and vanishes at all other nodes, as the zero-padded input spectrum does with $k_{j_1} = k_{\mathrm{min}}$ and $k_{j_2} = k_{\mathrm{max}}$.
Giving the two end nodes the weight $w$ and the interior nodes the weight 1, the Euler--Maclaurin formula for a smooth function $h$ of $\ln k$ reads
\begin{align}
    & \Delta_{\mathrm{FFT}} \left[ w ~ h(k_{j_{1}}) + \sum_{j=j_1+1}^{j_2-1} h(k_j) + w  ~  h(k_{j_2}) \right] \nonumber \\
    & \quad = \int_{k_{j_1}}^{k_{j_2}} h ~ \mathrm{d} \ln k + \left( w - \frac12 \right) \Delta_{\mathrm{FFT}} \left[ h(k_{j_1}) + h(k_{j_2}) \right] + \frac{\Delta_{\mathrm{FFT}}^2}{12} \left[ h'(k_{j_2}) - h'(k_{j_1}) \right] + \O(\Delta_{\mathrm{FFT}}^4),
    \label{eq:euler_maclaurin}
\end{align}
where the prime denotes $\mathrm{d}/\mathrm{d} \ln k$.
The FFT uses $w = 1$, for which the second term on the right-hand side is of order $\Delta_{\mathrm{FFT}}$. 
The sum counts half a cell beyond each end node, as if the input continued to $k_{\mathrm{max}} e^{\Delta_{\mathrm{FFT}}/2}$ (1.6\% beyond $k_{\mathrm{max}}$ in the default setting).
Multiplying the two end samples by $\frac{1}{2}$ before the FFT sets $w = \frac{1}{2}$ and can remove this term.
In this case, the sum becomes the trapezoidal rule, whose error is the $\O(\Delta_{\mathrm{FFT}}^{2})$ third term.
For $m \neq 0$, the derivative of $h = P_{\lin} k^{-\nu - i \eta_{m}}$ contains $-i \eta_{m} h$, so that this holds for the coefficients with $|\eta_{m}| \Delta_{\mathrm{FFT}} \ll 1$, which determine the integrals of the input and the slowly varying part of its transform.

We therefore give the weight $\frac12$ to the end nodes of every FFT input whose support ends at a node inside the padded grid.
These are the end nodes of the zero-padded input spectrum at $k_{\mathrm{min}}$ and $k_{\mathrm{max}}$, of the LPT sources $Q_n$ and $R_n$ at the ends of their supports $[k_{\mathrm{min}}, 2 k_{\mathrm{max}}]$ and $[k_{\mathrm{min}}, k_{\mathrm{max}}]$, and of the $q$-space integrands at the largest sampled separation $q_{\mathrm{max}}$.
Note that the weight is applied once to the complete input of each transform.

The 22-type contribution to the EPT power spectrum has a non-vanishing $k \to 0$ limit,
\begin{equation}
    P_{22}(0) = \frac{b_2^2}{2} \int_0^\infty \frac{p^2 \mathrm{d}p}{2 \pi^2} \left[ P_{\lin}(p) \right]^2,
    \label{eq:pk_22_k0_limit}
\end{equation}
Following \cite{McDonald:2006mx}, this contribution is usually absorbed into the intrinsic shot-noise parameter.
When the shot noise is low, as for the Lyman-alpha forest or 21~cm intensity mapping, this contribution can dominate and must be kept.
We retain it by default and provide an option to subtract it.

In the \textsc{Hankel} method, Eq.~\eqref{eq:pk_22_lnm} with $m = 0$ approaches the constant
\begin{equation}
    B_{\ell n} \equiv (-1)^\ell \int_0^\infty 4 \pi r^2 \mathrm{d}r ~ \left[ \xi^\ell_n(r) \right]^{2}
\end{equation}
as $k \to 0$, and these constants, multiplied by the coefficients of the blocks, add up to Eq.~(\ref{eq:pk_22_k0_limit}).
We evaluate each $m = 0$ block as the residual
\begin{equation}
    P_{22}^{(\ell, n, 0)}(k) - B_{\ell n} = (-1)^\ell \int_0^\infty 4 \pi r^2 \mathrm{d}r \left[ j_0(kr) - 1 \right] \left[ \xi^\ell_n(r) \right]^2,
    \label{eq:m0_residual}
\end{equation}
computing the first term with FFTLog and $B_{\ell n}$ as the weighted sum of the same samples $(-1)^\ell 4 \pi r_j^3 [\xi^\ell_n(r_j)]^2$ that enter the transform, end weight included, as discussed above.
Both terms then approximate the same sampled function, and the residual is as accurate as the transform itself.
In the \textsc{Matrix} method, the corresponding subtraction follows from choosing $\nu > -3/2$ in Eq.~\eqref{eq:decomp_pk_lin} for the 22-type terms that depend on $b_2$ or $b_{\G_2}$ \cite{Simonovic:2017mhp}.
The decomposition then crosses the pole of $I(\nu_1, \nu_2)$ in Eq.~\eqref{eq:I_nu1_nu2} at $\nu_1 + \nu_{2} = 3/2$, and the matrix product evaluates the 22-type integral minus its $k \to 0$ limit by analytic continuation.
We use $\nu = -1$ for these terms.
In both methods, the constant (\ref{eq:pk_22_k0_limit}) is the weighted sum of $\frac{b_2^2}{2} k_j^3 [P_{\lin}(k_j)]^2 / (2 \pi^2)$, with the end weight applied to $P_{\lin}^2$.
It is added once, after the no-wiggle and wiggle parts are combined in Eq.~\eqref{eq:ir_resummed_pkmu}, because it is a $k$-independent contact term without BAO feature.

The LPT power spectrum contains terms of the same kind.
First, the term $\frac12 (b_2^{\mathrm{L}})^2 \xi_{\lin}^2$ in Eq.~(\ref{eq:pkmu_lpt_1loop}) is multiplied by the exponential factor $e^{-\frac12 k_i k_j A^{<}_{ij}}$.
We split it into $\frac12 (b_2^{\mathrm{L}})^2 \xi_{\lin}^2$ and $\frac12 (b_2^{\mathrm{L}})^2 \xi_{\lin}^2 (e^{-\frac12 k_i k_j A^{<}_{ij}} - 1)$, and include only the latter in the final radial integral (Sec.~\ref{sec:angular_int_lpt}).
The former depends neither on $\mu$ nor on $f$.
We transform it once, as a residual of the form (Eq.~\eqref{eq:m0_residual}) plus its $k \to 0$ limit $\frac12 (b_2^{\mathrm{L}})^2 \int \mathrm{d}^3q ~ \xi_{\lin}^2$, which by Parseval's theorem equals Eq.~\eqref{eq:pk_22_k0_limit} with $b_2 \to b_2^{\mathrm{L}}$ and is evaluated by the same weighted sum.
Second, the $\ell = 0$ integrand of the final radial integral approaches a non-zero constant at large $q$, where the cumulants approach constants or vanish.
The transform of this constant is proportional to $\delta_{\mathrm{D}}(\bk)$, and we remove it by subtracting the value of the integrand at $q_{\mathrm{max}}$ before the transform.
Third, $\chi_{\G_2}$ and $\zeta_{\G_2}$ are total derivatives (Eq.~\eqref{eq:chi_zeta_total_derivative}) so that their integrals over $0 < q < \infty$ vanish and so do their transforms at $k \to 0$.
The final radial integral, however, covers only $q_{\mathrm{min}} \le q \le q_{\mathrm{max}}$.
As these functions are negligible at $q_{\mathrm{max}}$, the missing part is the inner ball $q < q_{\mathrm{min}}$, whose transform does not depend on $k$ for $k q_{\mathrm{min}} \ll 1$ and equals $-\int_{q_{\mathrm{min}}}^{q_{\mathrm{max}}} 4 \pi q^2 \chi_{\G_2} \, \mathrm{d}q$.
We add this constant as the weighted sum of $-4 \pi q_j^3 \chi_{\G_2}(q_j)$ over the sampled interval, and likewise for $\zeta_{\G_2}$.
Fourth, $\theta_{\Gamma_3}$ approaches a non-zero constant as $q \to 0$, and the periodic image of this plateau affects its samples near $q_{\mathrm{max}}$, so that the weighted sum over the sampled interval is unreliable.
We instead add the inner-ball contribution directly, using the total derivative (Eq.~\eqref{eq:theta_total_derivative}):
\begin{equation}
    \int_0^{q_{\mathrm{min}}} 4 \pi q^2 \mathrm{d}q ~ \theta_{\Gamma_3}(q) = \frac{96 \pi}{5} q_{\mathrm{min}}^2 U^{(3)}(q_{\mathrm{min}}).
\end{equation}

\subsection{Infrared resummation}
\label{sec:ir_resum_impl}

Our implementation of the IR resummation described in Sec.~\ref{sec:ir_resum} is based on splitting the linear power spectrum into wiggle and no-wiggle components \cite{Baldauf:2015xfa}.
This decomposition is not unique, and various splitting schemes have been adopted or investigated in the literature, including Gaussian filtering (\textit{e.g.}, \cite{Vlah:2015zda,MoradinezhadDizgah:2020whw,Linde:2024uzr}), Savitzky--Golay (SG) filtering (\textit{e.g.}, \cite{Chen:2020fxs}), B-spline regression (\textit{e.g.}, \cite{MoradinezhadDizgah:2020whw}), and the discrete sine transform (DST; \textit{e.g.}, \cite{Hamann:2010pw,Chudaykin:2020aoj,MoradinezhadDizgah:2020whw,Linde:2024uzr}).
In our code, we adopt the DST-based wiggle--no-wiggle splitting as the default and provide the SG filter as an alternative.
Implementation details of the splitting procedures are given in Appendix~\ref{sec:wiggle_splitting}.

\begin{figure}[htbp]
\centering
\begin{minipage}{0.48\linewidth}
\centering
\includegraphics[width=\linewidth]{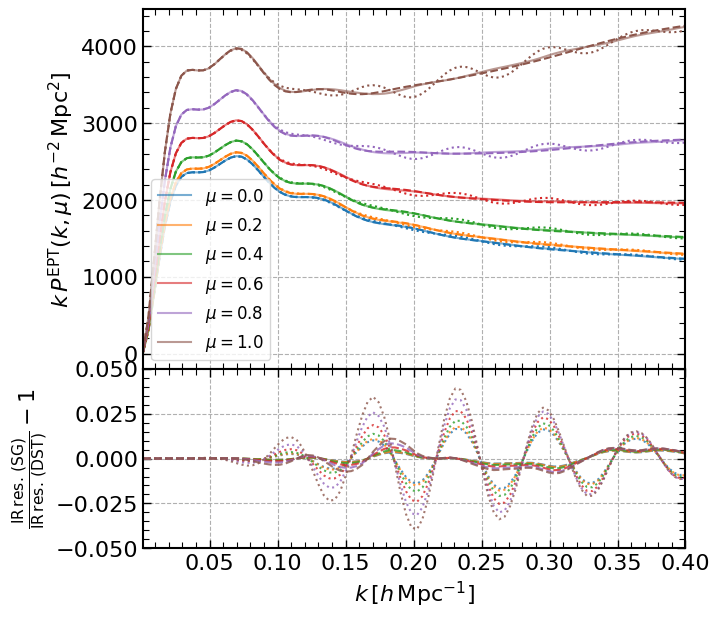}
\end{minipage}
\begin{minipage}{0.48\linewidth}
\centering
\includegraphics[width=\linewidth]{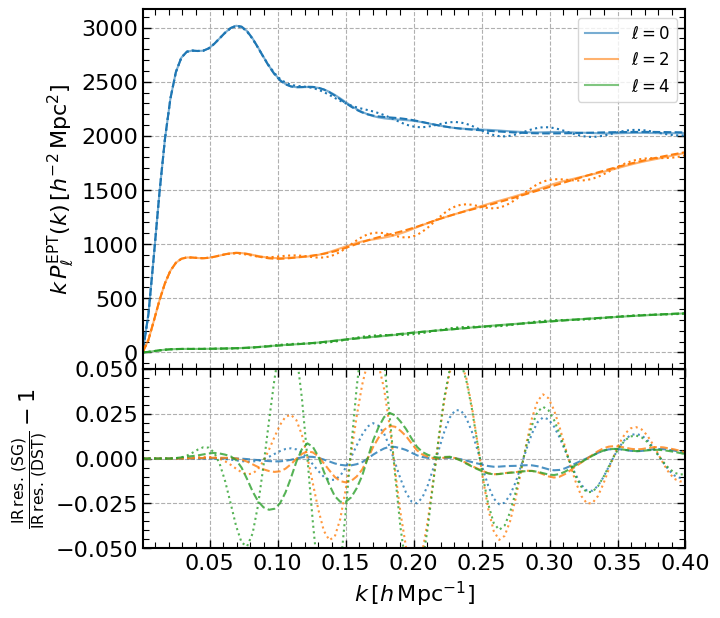}
\end{minipage}
\caption{
Comparison of different wiggle--no-wiggle splitting schemes for the IR-resummed EPT power spectrum.
The left panels show $P(k,\mu)$ for several values of $\mu$, while the right panels show the multipoles $\ell=0,2,4$.
In the upper panels, solid and dashed lines correspond to IR resummation using the DST-based method and the SG filter, respectively, while dotted lines show the result without IR resummation.
The lower panels show the fractional differences of the SG-filter (dashed) and non-resummed (dotted) results relative to the DST result.
}
\label{fig:irres_split_methods}
\end{figure}

Figure~\ref{fig:irres_split_methods} illustrates the impact of the wiggle--no-wiggle splitting scheme on the final IR-resummed power spectrum.
The DST- and SG-based results agree over the scales of interest, but with differences of order $1\%$ around $k \sim \O(0.1)\,h\,\mathrm{Mpc}^{-1}$.
This would be the largest uncertainty in the EPT calculation.
For comparison, the non-resummed result exhibits substantially larger oscillatory deviations, illustrating the effect of IR resummation itself.

\subsection{Angular integrals with exponential factor in LPT}
\label{sec:angular_int_lpt}

The implementation of the LPT power spectrum requires dealing with the angular integrals involving an exponential factor in Eq.~(\ref{eq:ps_1loop_lpt}).
For example, the simplest case is the real-space matter power spectrum in the Zel'dovich approximation,
\begin{align}
    P_{\mathrm{mm}}^{\mathrm{ZA}}(k) &= \int_{\bq} e^{i \bk \cdot \bq} e^{-\frac{1}{2} k_{i} k_{j} A^{\lin}_{ij}(\bq)},
    \label{eq:pk_mm_zel}
\end{align}
where $A^{\lin}_{ij}$ is the leading-order part of $A_{ij} \equiv \avrgc{\Delta_{i} \Delta_{j}}$.
By symmetry, $A^{\lin}_{ij}$ is expressed as $A_{ij}^{\lin}(\bq) = X^{\lin}(q) \delta_{ij} + Y^{\lin}(q) \hat{q}_{i} \hat{q}_{j}$, and hence Eq.~(\ref{eq:pk_mm_zel}) reduces to
\begin{align}
    P_{\mathrm{mm}}^{\mathrm{ZA}}(k) &= \int_{0}^{\infty} 4 \pi q^{2} \mathrm{d}q e^{ -\frac{1}{2} k^{2} X^{\lin}(q) } \int_{-1}^{1} \frac{\mathrm{d}\mu_{\bq}}{2} e^{i k q \mu_{\bq} - \frac{1}{2} k^{2} Y^{\lin}(q) \mu_{\bq}^{2}},
\end{align}
where we have defined $\mu_{\bq} = \hat{\bk} \cdot \hat{\bq}$, \textit{i.e.}, the cosine of the angle between $\bk$ and $\bq$.
Importantly, the angular integral in the above equation is computed analytically using the identity \cite{1995MNRAS.273..475S, Vlah:2014nta}
\begin{align}
    \int_{-1}^{1} \frac{\mathrm{d} \mu_{\bq}}{2} e^{i A \mu_{\bq} - \frac{1}{2} B \mu_{\bq}^{2}} = e^{- \frac{1}{2} B} \sum_{\ell=0}^{\infty} \left( \frac{B}{A} \right)^{\ell} j_{\ell}(A),
    \label{eq:angular_int_real_space_lpt}
\end{align}
and only the radial integral is left,
\begin{align}
    P_{\mathrm{mm}}^{\mathrm{ZA}}(k) = \sum_{\ell=0}^{\infty} \int_{0}^{\infty} 4\pi q^2 \mathrm{d}q ~ e^{-\frac{1}{2} k^2 [X^{\lin}(q) + Y^{\lin}(q)]} \left[ \frac{k Y^{\lin}(q)}{q} \right]^{\ell} j_{\ell}(kq).
\end{align}
Since it has the spherical Bessel function $j_{\ell}(kq)$, this radial integral can be efficiently computed using Hankel transforms with different order $\ell$.

In redshift space, the angular integrals become harder to compute efficiently due to the additional dependence of the integrand on the line-of-sight direction.
We follow the approach called Method II in \cite{Vlah:2018ygt,Chen:2019lpf,Chen:2020zjt}, where we define the alternative wavevector $K_{i} \equiv R^{(1)}_{ij} k_{j}$ by applying the transformation to promote the real-space displacement to the redshift-space one at linear order, $R^{(1)}_{ij} = \delta_{ij} + f \hat{n}_{i} \hat{n}_{j}$ (see Eq.~(\ref{eq:displacement_rsd})).
Accordingly, we redefine $\mu_{\bq}$ as $\mu_{\bq} \equiv \hat{\bK} \cdot \hat{\bq}$, \textit{i.e.}, the cosine of the angle between the new wavevector $\bK$ and $\bq$, and $\mu_{\bq,\bn} \equiv \hat{\bq} \cdot \hat{\bn}$, with $\hat{\bn}$ being the line-of-sight.
In terms of these quantities, the angular integrals in the redshift-space LPT power spectrum are expressed in the following form:
\begin{align}
    I^{(a,b)}(A, B, C) &\equiv \int_{-1}^{1} \frac{\mathrm{d}\mu_{\bq}}{2} \int_{0}^{2\pi} \frac{\mathrm{d}\phi}{2\pi} e^{i A \mu_{\bq} + B \mu_{\bq}^2 -i C \sqrt{1-\mu_{\bq}^2} \cos \phi} \mu_{\bq}^{a} \mu_{\bq,\bn}^{b}, \label{eq:Iab_definition}
\end{align}
where $\phi$ is the azimuthal angle of $\bq$ around $\bK$.
Since $\mu_{\bq,\bn}$ is expressed by \cite{Chen:2019lpf,Chen:2020zjt}
\begin{align}
    \mu_{\bq,\bn} = \frac{\mu (1 + f)}{\sqrt{ 1 + f (2 + f) \mu^2 }} \mu_{\bq} + \sqrt{\frac{ 1 - \mu^2 }{ 1 + f (2 + f) \mu^2 }} \sqrt{1 - \mu_{\bq}^2} \cos \phi,
\end{align}
where $\mu = \hat{\bk} \cdot \hat{\bn}$, the angular integral (\ref{eq:Iab_definition}) becomes a linear combination of the integrals of the form
\begin{align}
    I_{m,n}(A, B, C) &= i^{m-n} \int_{-1}^{1} \frac{\mathrm{d}\mu_{\bq}}{2} \int_{0}^{2\pi} \frac{\mathrm{d}\phi}{2\pi} e^{i A \mu_{\bq} + B \mu_{\bq}^2 -i C \sqrt{1-\mu_{\bq}^2} \cos \phi} \mu_{\bq}^{m} \left( \sqrt{ 1 - \mu_{\bq}^2 } \cos \phi \right)^{n} \\
    &= e^{B} \sum_{\ell = 0}^{\infty} \left(\frac{-2}{\rho}\right)^{\ell} G^{(m,n)}_{\ell}(A, B, C) j_{\ell}(\rho), \label{eq:Imn_formula}
\end{align}
where $\rho \equiv \sqrt{A^2 + C^2}$ and $G_{\ell}^{(m,n)}$ are some analytic formulas that depend on $\ell$ and a pair of non-negative integers $(m, n)$.
$G_{\ell}^{(m,n)}$ is constructed from $G_{\ell}^{(0,0)}$ by applying the recursion relations obtained by taking derivatives of $I_{m,n}$ with respect to $A$ or $C$ \cite{Chen:2019lpf,Chen:2020zjt},
\begin{align}
    G_{\ell}^{(m+1,n)} = \frac{\partial G_{\ell}^{(m,n)}}{\partial A} + \frac{A}{2} G_{\ell-1}^{(m,n)}, \text{  } G_{\ell}^{(m,n+1)} = \frac{\partial G_{\ell}^{(m,n)}}{\partial C} + \frac{C}{2} G_{\ell-1}^{(m,n)}.
    \label{eq:Gmn_recursion}
\end{align}
The combinations of $(m,n)$ needed to compute the one-loop LPT power spectrum are
\begin{align}
    (m,n) &= (0,0), (0,1), (0,2), (1,0), (1,1), (2,0), (2,1), (3,0), (4,0).
\end{align}
For $(m,n) = (0,0)$, Ref.~\cite{Vlah:2018ygt} has shown the analytic expression as an infinite sum of the Gauss hypergeometric functions ${_2F_1}$,
\begin{align}
    G_{\ell}^{(0,0)}(A, B, C) &= \sum_{n = \ell}^{\infty} \frac{\Gamma(\ell + n + \frac{1}{2})}{\Gamma(\ell + 1) \Gamma(n + \frac{1}{2}) \Gamma(1-\ell+n)} \left( \frac{B A^2}{\rho^2} \right)^{n} {_2F_1}\left(\frac{1}{2}-n, -n; \frac{1}{2} - \ell - n; \frac{\rho^2}{A^2} \right), \label{eq:G00_vlah_white}
\end{align}
where $\Gamma(\cdot)$ denotes the Gamma function.
This formula arises already in the Zel'dovich power spectrum in redshift space \cite{Vlah:2018ygt}.
In this work, we have newly found its closed-form expression (see Appendix~\ref{sec:G00_closed_form} for the derivation),
\begin{align}
    G_{\ell}^{(0,0)}(A, B, C) 
    &= \exp \left( \frac{- B C^2}{\rho^2} \right) \sum_{k=0}^{\ell} \sum_{i=0}^{\ell} 
    c_{(\ell,k,i)} (-B)^{\ell+i} \left( \frac{C^2}{\rho^2} \right)^{\ell+i-k},
    \label{eq:G00_closed_form}
\end{align}
where the coefficients $c_{(\ell, k, i)}$ are rational numbers given by
\begin{align}
    c_{(\ell, k, i)} &= \frac{ (-1)^{i-k} }{ (\ell-k)! i! } \sum_{n=0}^{i} (-1)^{n} \binom{i}{n} \binom{\ell+n}{k} \left( \ell + n + \frac{1}{2} \right)_{\ell - k},
    \label{eq:G00_coeff}
\end{align}
with $(x)_{n} \equiv \Gamma(x + n) / \Gamma(x) = x (x + 1) \cdots (x + n - 1)$ being the Pochhammer symbol.
The coefficient (\ref{eq:G00_coeff}) does not depend on cosmology, redshift, or EFT parameters, and thus can be precomputed and tabulated in the code.
As for $G_{\ell}^{(m, n)}$ with $(m, n) \ne (0, 0)$, we can also find the closed-form expressions by applying the recursion relations (\ref{eq:Gmn_recursion}) on Eq.~(\ref{eq:G00_closed_form}).
\footnote{In a JAX implementation, we can also employ automatic differentiation to compute $G_{\ell}^{(m,n)}$ using Eqs.~(\ref{eq:Gmn_recursion}) and (\ref{eq:G00_closed_form}). However, we have found that it takes longer time for the JIT compilation.}

The remaining radial integrals along $q$ are evaluated using Hankel transforms of order $\ell$ thanks to the spherical Bessel function $j_{\ell}$ in Eq.~\eqref{eq:Imn_formula}.
In practice, we truncate the summation over $\ell$ in Eq.~\eqref{eq:Imn_formula} at a finite multipole, $\ell_{\mathrm{max}}$, and set $\ell_{\mathrm{max}} = 5$ by default.
Since each term in Eqs.~\eqref{eq:pkmu_lpt_tree} and \eqref{eq:pkmu_lpt_1loop} is linear in $I_{m,n}$, we first sum up the coefficients of each $I_{m,n}$ over all the terms, and thus only a single radial integral is needed for each $\ell$ at each $(k, \mu)$.
Because its integrand itself depends on $k$, this integral is needed only at that $k$.
We thus precompute the rows of the discrete Hankel transform at the $N_{\mathrm{FFT}}$ nodes of the $k$ grid, and evaluate the integral as the inner product of the sampled integrand and these rows interpolated in $\ln k$.
This requires only $\O(\ell_{\mathrm{max}} N_{\mathrm{FFT}})$ operations per $(k, \mu)$, instead of an FFT for each $(k, \mu)$.

For numerical stability, we combine $e^{B}$ in Eq.~\eqref{eq:Imn_formula} and $e^{-B C^{2}/\rho^{2}}$ in Eq.~\eqref{eq:G00_closed_form} with $e^{-\frac{1}{2} K^{2} X^{<}}$, where $A^{<}_{ij} = X^{<} \delta_{ij} + Y^{<} \hat{q}_{i} \hat{q}_{j}$, into a single exponential, which never exceeds unity, and evaluate the polynomial in $-B$ with Horner's rule.
Furthermore, the part of the integrand that is linear in $A^{\lin}_{ij}$, $U^{\lin}_{i}$, and $\xi_{\lin}$ remains almost constant up to the largest separation $q_{\mathrm{max}}$, and the truncation at $q_{\mathrm{max}}$ thus leads to an error of $\O((k q_{\mathrm{max}})^{-2})$ at low $k$.
We remove this error by subtracting this part multiplied by $D \equiv e^{-\frac{1}{2} K^{2} [X^{<}(q_{\mathrm{max}}) + Y^{<}(q_{\mathrm{max}})]}$ from the integrand before the transform, and by adding back its exact transform, $D (1 + b_{1}^{\mathrm{L}} + f \mu^{2})^{2} P_{\lin}(k)$.

\section{Code performance}
\label{sec:code_performance}

In this section, we assess the numerical accuracy and computational performance of our JAX implementation of the one-loop EPT and LPT galaxy power spectra.
We first validate the accuracy of the FFTLog-based implementations by comparing them with brute-force numerical integration and with public codes.
We then present benchmark timings on CPUs and a GPU.

\subsection{Accuracy}

\subsubsection{Eulerian perturbation theory}
\label{sec:code_accuracy_ept}

\begin{figure}[htbp]
\centering
\begin{minipage}{0.48\linewidth}
\centering
\includegraphics[width=\linewidth]{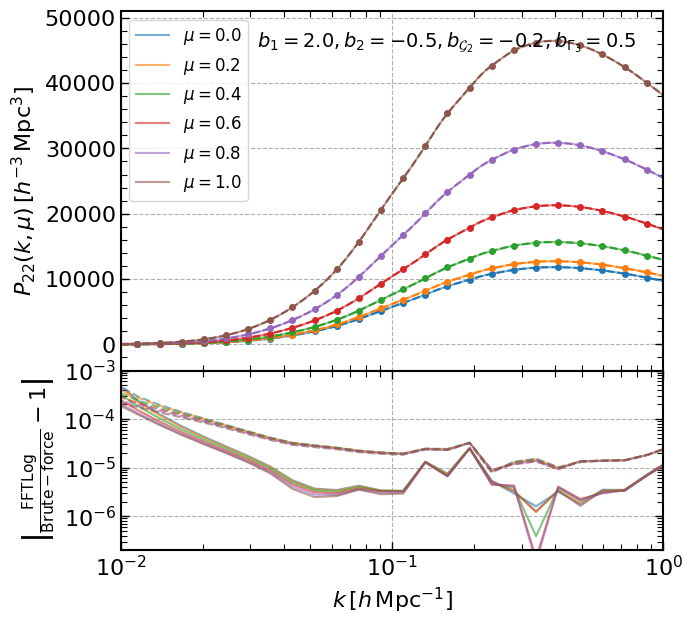}
\end{minipage}
\begin{minipage}{0.48\linewidth}
\centering
\includegraphics[width=\linewidth]{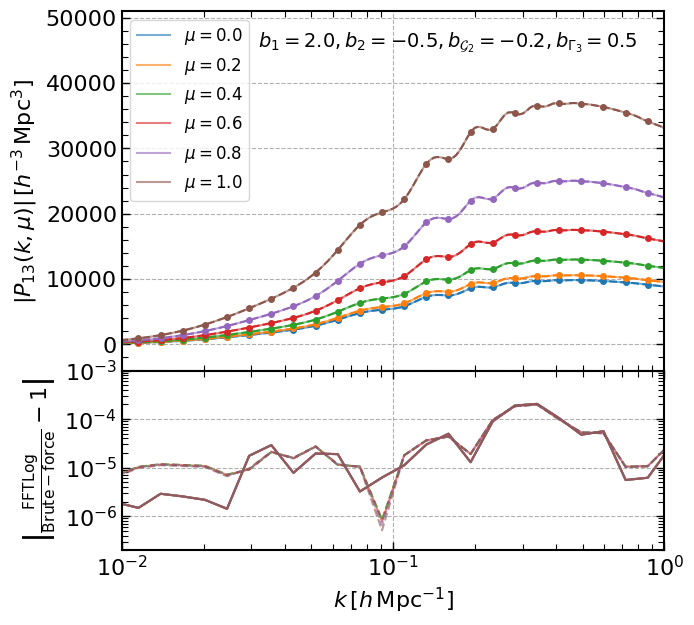}
\end{minipage}
\caption{
Comparison of the one-loop contributions to the redshift-space EPT power spectrum computed with our two FFTLog-based implementations and with brute-force numerical integration.
The left and right panels show the $P_{22}$ and $P_{13}$ contributions, respectively, with the absolute value $|P_{13}|$ shown in the latter.
Solid and dashed lines denote the results obtained with the \textsc{Hankel} and \textsc{Matrix} methods, respectively, while the symbols show the brute-force results.
Different colors correspond to different values of $\mu$, as indicated in the legends.
The lower panels show the absolute fractional differences of the two FFTLog-based calculations relative to the brute-force results.
The bias parameters are fixed to the values indicated in the upper panels.
}
\label{fig:comparison_to_bf_ept}
\end{figure}

Figure~\ref{fig:comparison_to_bf_ept} compares the one-loop EPT terms implemented using FFTLog-based methods to those computed with brute-force numerical integration of one-loop integrals. 
We show the 22-term (the left panel) and 13-term (the right panel) of the redshift-space galaxy power spectrum, assuming the values of the Eulerian bias parameters $b_{1} = 2, b_{2} = -0.5, b_{\G_{2}} = -0.2, b_{\Gamma_{3}} = 0.5$.
Different colors indicate the different values of $\mu = \hat{\bk} \cdot \hat{\bn}$.
The brute-force integration is based on a two-dimensinal Gauss--Legendre quadrature and a one-dimensional trapezoidal rule.
Both the \textsc{Hankel} and \textsc{Matrix} methods agree with the brute-force integration in $\sim 0.01 \%$ over the wavenumber range where the one-loop contributions become important, \textit{i.e.}, $k \sim \O(0.01\text{--}0.1)\, h\,\mathrm{Mpc}^{-1}$.

\begin{figure}[htbp]
\centering
\begin{minipage}{0.48\linewidth}
\centering
\includegraphics[width=\linewidth]{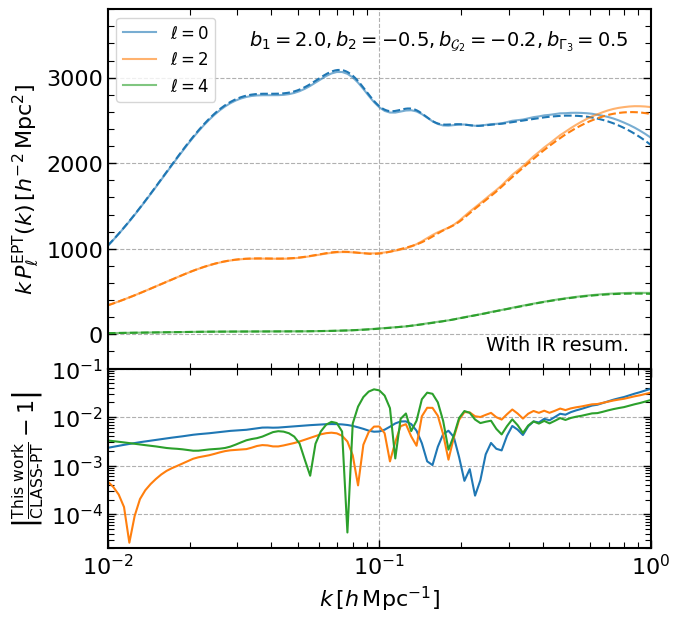}
\end{minipage}
\begin{minipage}{0.48\linewidth}
\centering
\includegraphics[width=\linewidth]{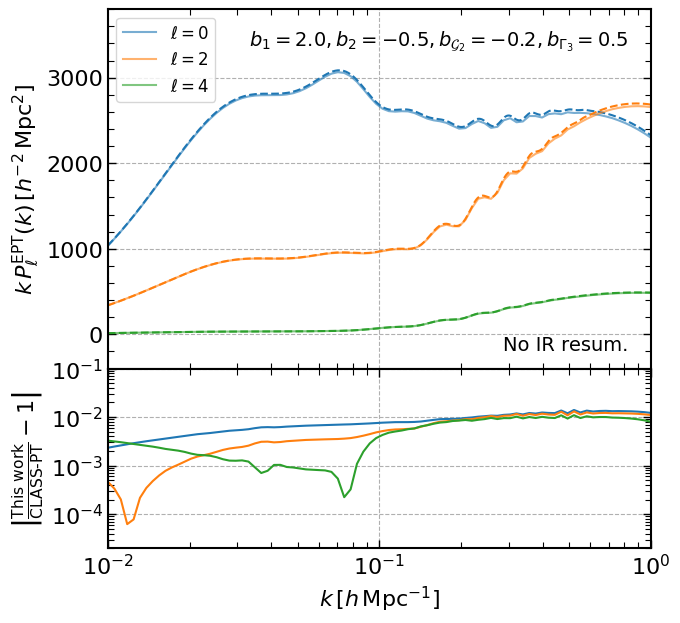}
\end{minipage}
\caption{
Comparison of the redshift-space EPT power-spectrum multipoles computed with our JAX implementation (solid lines) and \texttt{CLASS-PT} (dashed lines).
The monopole, quadrupole, and hexadecapole ($\ell=0,2,4$) are shown in different colors, as indicated in the legends.
The lower panels show the absolute fractional differences.
The left panels show the results with IR resummation, for which our implementation uses the DST-based wiggle--no-wiggle decomposition, while the right panels show the results without IR resummation.
The one-loop integrals in our code are evaluated using the \textsc{Hankel} method described in Sec.~\ref{sec:hankel_method}.
The bias parameters are fixed to the values indicated in the upper panels, while all EFT counterterms and stochastic contributions are set to zero.
}
\label{fig:comparison_to_class-pt}
\end{figure}

We compare our JAX code of EPT power spectrum multipoles to the public \texttt{CLASS-PT} \cite{Chudaykin:2020aoj} code in Fig.~\ref{fig:comparison_to_class-pt}.
The left and right panels show the case with and without the IR resummation, respectively.
In both cases, we set the counterterms and stochasticity terms to be zero.
Since \texttt{CLASS-PT} primarily outputs the power spectrum multipoles of $\ell = \{0,2,4\}$, we compare these multipoles to our implementation.
Note that we use the default FFT settings for both our code and \texttt{CLASS-PT}, resulting in different FFT grids between the two codes.
Our implementation agrees with \texttt{CLASS-PT} to within roughly 1\% for all three multipoles across most of the wavenumber range shown in this figure.
However, there are percent-level differences in the IR-resummed power spectrum.
A similar amount of discrepancy is also observed in the comparison between \texttt{CLASS-PT} and the more recent \texttt{CLASS-OneLoop} code (see Fig.~3 of \cite{Linde:2024uzr}).
As we have shown in Sec.~\ref{sec:ir_resum_impl}, the implementation details of the IR resummation could lead to this level of discrepancy.

\subsubsection{Lagrangian perturbation theory}
\label{sec:code_accuracy_lpt}

\begin{figure}[htbp]
\centering
\begin{minipage}{0.48\linewidth}
\centering
\includegraphics[width=\linewidth]{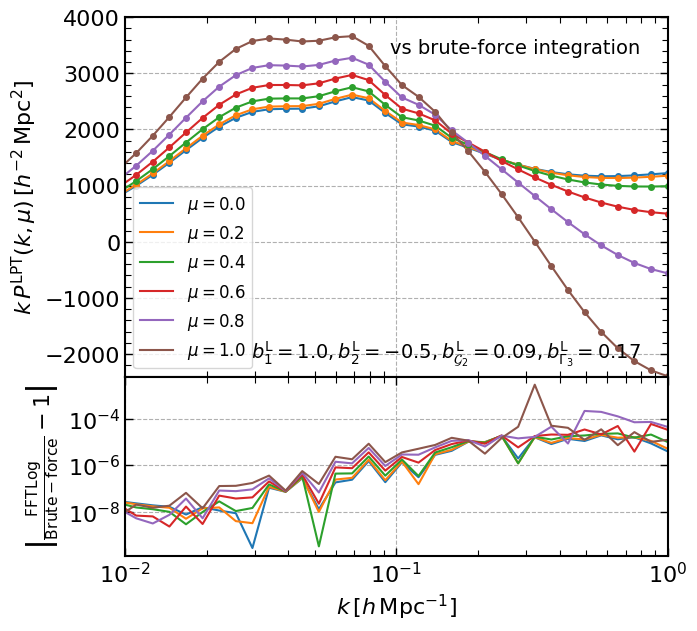}
\end{minipage}
\begin{minipage}{0.48\linewidth}
\centering
\includegraphics[width=\linewidth]{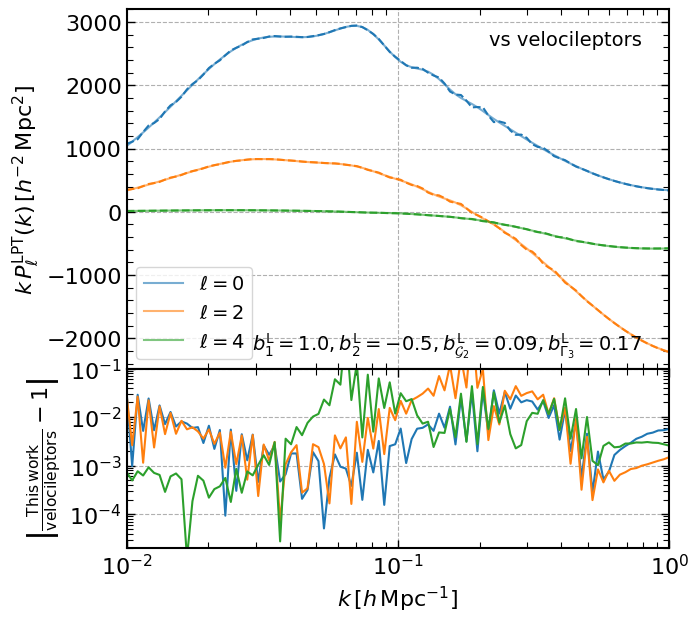}
\end{minipage}
\caption{
Validation of our LPT implementation against brute-force integration (left) and \texttt{velocileptors} (right).
The left panels show the redshift-space power spectrum $P(k,\mu)$ for several values of $\mu$, with solid lines and symbols denoting our JAX implementation and brute-force integration, respectively.
The right panels show the power spectrum multipoles $\ell=0,2,4$, with solid and dashed lines denoting our JAX implementation and \texttt{velocileptors}, respectively.
The lower panels show the absolute fractional differences relative to the corresponding reference calculations.
For these comparisons, we adopt the Lagrangian bias parameters indicated in the upper panels and set the EFT counterterms and stochastic contributions to zero.
}
\label{fig:comparison_to_bf_lpt}
\end{figure}

The left panel of Fig.~\ref{fig:comparison_to_bf_lpt} compares our implementation of the LPT power spectrum with a brute-force calculation in which all Hankel transforms are evaluated using composite Gauss--Legendre quadrature instead of FFTs.
We choose the Lagrangian bias parameters
$b_{1}^{\mathrm{L}} = 1.0$,
$b_{2}^{\mathrm{L}} = -0.5$,
$b_{\G_{2}}^{\mathrm{L}} = 0.09$, and
$b_{\Gamma_{3}}^{\mathrm{L}} = 0.17$
such that they correspond to the Eulerian bias parameters adopted in the previous figures, using the Eulerian--Lagrangian bias relations (see Sec.~\ref{sec:inference} and Appendix~\ref{sec:relation_btw_eulerian_and_largangian_biases} for details).
Here we show only the total power spectrum, with the counterterms and stochastic contributions set to zero; comparisons of the individual components are presented in Appendix~\ref{sec:comarison_to_bf_lpt}.
The upper panel shows the two-dimensional redshift-space power spectrum $P(k,\mu)$ for several values of $\mu$, while the lower panel shows the absolute fractional difference of our JAX implementation relative to the brute-force result.
The two calculations agree to within $\sim 0.01\%$ over the scales shown.

We further validate our implementation against the public \texttt{velocileptors} code \cite{Chen:2020zjt}, as shown in the right panel of Fig.~\ref{fig:comparison_to_bf_lpt}.
Unlike the brute-force comparison, here we compare the power spectrum multipoles $\ell=\{0,2,4\}$, which also tests the angular integration used to construct the multipoles.
The \texttt{velocileptors} code adopts the Lagrangian bias basis of Ref.~\cite{McDonald:2009dh}, which is linearly related to our default basis.
In particular, Ref.~\cite{McDonald:2009dh} uses
$s^{2}(\bq)=s_{ij}(\bq)s_{ij}(\bq)$, with
$s_{ij}(\bq)=\left(\frac{\nabla_i\nabla_j}{\nabla^2}-\frac{1}{3}\delta_{ij}\right)\delta(\bq)$,
as the second-order shear operator instead of $\G_{2}(\bq)$.
We therefore translate the above bias parameters to the \texttt{velocileptors} convention: the quadratic bias $b_{2}^{\mathrm{L}}$, second-order shear bias $b_{s}^{\mathrm{L}}$, and third-order bias $b_{3}^{\mathrm{L}}$ in \texttt{velocileptors} are related to those in our default basis, Eq.~\eqref{eq:lagrangian_bias_expansion}, as
$b_{2}^{\mathrm{L}}-\frac{4}{3}b_{\G_{2}}^{\mathrm{L}}$,
$b_{\G_{2}}^{\mathrm{L}}$, and
$-2b_{\Gamma_{3}}^{\mathrm{L}}$, respectively.
We use $N_{\mathrm{FFT}}=512$ and the FFT range
$[k_{\mathrm{min}},k_{\mathrm{max}}]=[10^{-5},10^{2}]\,h\,\mathrm{Mpc}^{-1}$
for both codes.\footnote{
We also multiply the input linear power spectrum by a high-$k$ cutoff factor
$e^{-(k/k_{\mathrm{cut}})^2}$ with
$k_{\mathrm{cut}}=10\,h\,\mathrm{Mpc}^{-1}$
in our code, consistently with the setting adopted in \texttt{velocileptors}.
}
The resulting multipoles agree with \texttt{velocileptors} at approximately the percent level, despite subtle differences in the FFT implementations.

\subsection{Computation time} 

\begin{table}
\begin{center}
\begin{tabular}{c|ccc} \hline \hline
    Processor & AMD EPYC 7713 CPU & Apple M3 CPU & NVIDIA A100 GPU \\ \hline

    \multicolumn{4}{c}{Power spectrum multipoles} \\
    \hline
    EPT & $6.34 \pm 0.07$ & $1.80 \pm 0.24$ & $2.34 \pm 0.22$ \\
    LPT & $11.9 \pm 0.10$ & $12.0 \pm 0.55$ & $2.98 \pm 0.06$ \\

    \hline
    \multicolumn{4}{c}{Log likelihood and parameter gradient} \\
    \hline
    EPT & $14.6 \pm 0.31$ & $4.51 \pm 0.42$ & $6.56 \pm 0.21$ \\
    LPT & $201 \pm 1.68$ & $91.9 \pm 17.0$ & $9.22 \pm 0.34$ \\
    \hline
\end{tabular}
\caption{
Execution times in milliseconds for the EPT and LPT implementations on three different processors: a server-class CPU, a laptop-class CPU, and a GPU.
The upper block reports the time required to evaluate the power spectrum multipoles with $\ell=\{0,2,4\}$, while the lower block reports the time required to evaluate the Gaussian log likelihood together with its gradient with respect to the model parameters.
We use the default FFT array size, $N_{\mathrm{FFT}}=512$, and evaluate the power spectrum multipoles over 20 $k$-bins.
The quoted values are the mean $\pm$ standard deviation of the execution time for a single evaluation after JIT compilation (see the main text for details).
}
\label{tab:computation_time}
\end{center}
\end{table}

By implementing our code in JAX, we achieve efficient computation of the one-loop galaxy power spectrum.
The code is fully compatible with JAX's just-in-time (JIT) compilation and vectorized execution on GPUs.
To quantify its computational performance, we benchmark both the evaluation of the one-loop EPT and LPT power spectra and the evaluation of a Gaussian log likelihood together with its parameter gradient, as summarized in Table~\ref{tab:computation_time}.

For the CPU benchmarks, we use an AMD EPYC 7713 as a representative server-class CPU and an Apple M3 as a representative laptop-class CPU.
For the GPU benchmark, we use an NVIDIA A100 GPU with 40 GB of memory.
We adopt the default setting $N_{\mathrm{FFT}}=512$ for the FFTs and compute the power-spectrum multipoles $\ell=\{0,2,4\}$ at 20 linearly spaced $k$ bins over the range $k\in[0.01,0.2]\,h\,\mathrm{Mpc}^{-1}$.
The runtime of the LPT calculation depends on the number of $k$ bins because the final radial integral is evaluated separately at each $(k,\mu)$, as the inner product of an integrand sampled on the $N_{\mathrm{FFT}}$ nodes of the $q$ grid with the precomputed transform rows (Sec.~\ref{sec:angular_int_lpt}), although these evaluations are vectorized using \texttt{jax.vmap}.
In contrast, the EPT loop integrals are evaluated on fixed FFT grids and subsequently interpolated to the requested $k$ bins, so the EPT runtime depends only weakly on the number of output $k$ bins.

Since JAX uses asynchronous dispatch by default, which can otherwise lead to misleading timing measurements, we explicitly synchronize the computation using \texttt{jax.block\_until\_ready}.
The reported runtimes are the mean $\pm$ standard deviation per evaluation, measured with \texttt{IPython}'s \texttt{\%timeit} using 10 runs and 100 loops per run.

In addition to the power spectrum evaluation, we benchmark the Gaussian log likelihood introduced in Eq.~\eqref{eq:pkl_likelihood} of the next section and its gradient with respect to the model parameters.
The likelihood uses the three multipoles evaluated at the same 20 $k$ bins.
For this benchmark, the linear power spectrum is computed using the JAX-based differentiable emulator \texttt{CosmoPower-JAX} \cite{Piras:2023aub}.
We use JIT-compiled \texttt{jax.value\_and\_grad} to simultaneously evaluate the log likelihood and its derivatives with respect to the five cosmological parameters
$\{\omega_{\mathrm{b}},\omega_{\mathrm{cdm}},h,\ln(10^{10}A_{\mathrm{s}}),n_{\mathrm{s}}\}$
and all EFT parameters.

As shown in Table~\ref{tab:computation_time}, the NVIDIA A100 provides particularly strong performance for both implementations, with the power-spectrum evaluation taking only a few milliseconds for both EPT and LPT.
This is up to approximately two orders of magnitude faster than existing non-JAX, non-emulator public implementations of the one-loop EPT and LPT power spectra, such as \texttt{CLASS-PT}, \texttt{PyBird}, and \texttt{velocileptors}, which typically require $\mathcal{O}(100)\,\mathrm{ms}$ on CPUs.
The evaluation of the log likelihood together with its full parameter gradient also remains below $10\,\mathrm{ms}$ on the A100 for both PT models (however, when vectorizing multiple chains, the computation speed seems to be affected by the memory bandwidth as well, see Sec.~\ref{sec:inference_result}).

On CPUs, the EPT power-spectrum calculation still achieves a runtime of a few milliseconds, whereas the LPT calculation takes $\mathcal{O}(10)\,\mathrm{ms}$.
The difference between the CPU and GPU performance is considerably more pronounced for the LPT implementation, particularly when parameter derivatives are included.
This behavior reflects the extensive use of vectorized operations in the LPT calculation, which can be executed particularly efficiently on GPUs.

\section{Parameter inference using Hamiltonian Monte Carlo}
\label{sec:inference}

In this section, we demonstrate our JAX code in parameter inference using the Hamiltonian Monte Carlo (HMC) algorithm \cite{Neal:2011mrf,2017arXiv170102434B}.
HMC is a variant of the MCMC algorithm that numerically solves the Hamiltonian dynamics of particles moving in the potential, which is specified by the target probability density.
Specifically, each MCMC particle residing in the parameter space that we are interested in is assigned the auxiliary momentum $\bp$ and the Hamiltonian given by the sum of the kinetic energy $K(\bp)$ and the potential energy $U(\bx)$,
\begin{align}
    K(\bp) = \frac{1}{2} \bp^{T} M^{-1} \bp, \text{ } U(\bx) = - \log p(\bx),
\end{align}
where the matrix $M$ is called the mass matrix, which affects how far the particle can travel in one step, and $p(\bx)$ is the target probability density, \textit{i.e.}, the posterior density of the parameters $\bx$ we want to infer.

HMC regards $e^{-H(\bx, \bp)} = e^{-K(\bp)} p(\bx)$ as the (unnormalized) target density over the phase space $(\bx,\bp)$ to apply the Metropolis acceptance/rejection procedure, and provides MCMC steps by Hamiltonian dynamics realized by the leapfrog integrator.
Since Hamiltonian dynamics preserves the total energy $H(\bx,\bp) = K(\bp) + U(\bx)$ (in the absence of numerical errors), it leads to a high acceptance probability. 
Hence, it scales to inference in high-dimensional spaces, which is the main advantage of the HMC over the ordinary Metropolis--Hastings algorithm.

\subsection{Data and covariance}
\label{sec:data_and_covariance}

As the data vector to be analyzed, we use the power spectrum multipoles of dark matter halos generated from $N$-body simulations, for a cosmology from \textit{Planck} 2018 $\Lambda$CDM TT,TE,EE+lowE+lensing+BAO results \cite{Planck:2018vyg}, \textit{i.e.}, $\omega_{\mathrm{b}} = 0.02242, \omega_{\mathrm{cdm}} = 0.11933, h = 0.6766, \ln(10^{10}A_{\mathrm{s}}) = 3.047$, and $n_s = 0.9665$.
We use 10 realizations of $N$-body simulation of volume $1  \, (h^{-1} \, \mathrm{Gpc})^{3}$ with $1536^3$ particles, initialized at $z=39$ using the \texttt{2LPTIC}~\cite{Crocce:2006ve}, and evolved with \texttt{Gadget-4} code~\cite{Springel:2020plp}.
The halos are identified using the public halo finder \texttt{Rockstar} \cite{Behroozi:2011ju} and only the parent halos are kept in the catalogs.
We further rank the \texttt{Rockstar} parent halos by the virial mass $M_{\mathrm{vir}}$ and keep only the most massive ones such that the mean number density $\bar{n}$ equals $10^{-4} \, (h^{-1} \, \mathrm{Mpc})^{-3}$.
We construct the redshift-space positions of the halos by displacing the real-space positions with their core velocities along the $z$-axis.
We measure the monopole, quadrupole, and hexadecapole moments of the power spectrum of the halos using the estimator:
\begin{align}
    \hat{P}_{\ell}(k_{i}) &= \frac{2 \ell + 1}{N_{i}} \sum_{\bk \in \bin~i} \L_{\ell}(\hat{\bk} \cdot \hat{\bn}) \left[ \delta(\bk) \delta(-\bk) - \frac{1}{\bar{n}} \right],
    \label{eq:pkl_estimator}
\end{align}
where $N_{i}$ is the number of wavevector modes in the $i$th $k$-bin, $\delta(\bk)$ is the Fourier transform of the halo density contrast, and $\bar{n}$ is the mean number density of the halos.
$\hat{\bn}$ is the unit vector along the line-of-sight, which is taken to be the $z$-axis.
To obtain $\delta(\bk)$, we assign the halo positions in redshift space onto $1024^3$ cubic grids using the cloud-in-cells (CIC) interpolation kernel and perform a three-dimensional FFT.
In doing this, we also apply the interlacing technique to reduce the aliasing effect, as well as perform deconvolution of the CIC interpolation window in Fourier space \cite{1988csup.book.....H,Sefusatti:2015aex}.
This grid setting means the Nyquist wavenumber $k_{\mathrm{Ny}} \approx 3.2 \, h \, \mathrm{Mpc}^{-1}$.
We measure the power spectrum signals in the linear equidistant $k$-bins in the range $k \in [0.0, 0.2) \, h \, \mathrm{Mpc}^{-1}$ with width $\Delta k = 0.01 \, h \, \mathrm{Mpc}^{-1}$, and use the mean signals of 10 realizations of $1  \, (h^{-1} \, \mathrm{Gpc})^{3}$ simulations, which amounts to the $10 \, (h^{-1} \, \mathrm{Gpc})^{3}$ volume.

As for the covariance, we use the Gaussian covariance measured from the same halo catalogs using the estimator:
\begin{align}
    C^{\mathrm{G}}_{\ell_{1}, \ell_{2}}(k_{i}, k_{j}) &= \delta_{ij}^{\mathrm{K}} \frac{(2 \ell_{1} + 1) (2 \ell_{2} + 1)}{N_{i}^{2}} \sum_{\bk \in \bin~i} \L_{\ell_{1}}(\hat{\bk} \cdot \hat{\bn}) \L_{\ell_{2}}(\hat{\bk} \cdot \hat{\bn}) \left[ \delta(\bk) \delta(-\bk) \right]^{2},
\end{align}
where $\delta_{ij}^{\mathrm{K}}$ is the Kronecker delta.
Note that the above formula includes the shot-noise contribution associated with the shot-noise-subtracted estimator Eq.~(\ref{eq:pkl_estimator}).
We obtain the mean of the estimated covariance of the 10 realizations, and rescale it to match the volume to be $10 \, (h^{-1} \, \mathrm{Gpc})^{3}$.

\subsection{Likelihood and priors}
\label{sec:likelihood_and_priors}

We sample the posterior distribution of the model parameters $\btheta$ constructed from the likelihood and the prior distribution,
\begin{align}
    p(\btheta | \bd) &\propto p(\bd | \btheta) p(\btheta), 
\end{align}
where $\bd$ denotes the data vector of the power spectrum multipoles, \textit{i.e.}, $\bd = \{ \hat{P}_{\ell}(k_{i}) \}$.
We assume the Gaussianity of the likelihood $p(\bd | \btheta)$, \textit{i.e.}
\begin{align}
    p(\bd | \btheta) &\propto \exp \left[ - \frac{1}{2} \sum_{\ell_{1}, \ell_{2}} \sum_{i,j} \left[ \hat{P}_{\ell_{1}}(k_{i}) - P_{\ell_{1}}(k_{i} | \btheta) \right] C^{-1}_{\ell_{1}, \ell_{2}}(k_{i}, k_{j}) \left[ \hat{P}_{\ell_{2}}(k_{j}) - P_{\ell_{2}}(k_{j} | \btheta) \right]  \right],
    \label{eq:pkl_likelihood}
\end{align}
where $P_{\ell}(k_{i} | \btheta)$ denotes the theory prediction of the $\ell$th multipole moment of the power spectrum at the $i$th $k$-bin for the input model parameter $\btheta$.
In computing the theory predictions during the parameter inference, we use the differentiable emulator \texttt{CosmoPower-JAX} \cite{Piras:2023aub} to compute the input linear matter power spectrum.
To take into account the AP effect in theory prediction, we assume the reference flat $\Lambda$CDM cosmology with $\Omega_{\mathrm{m}} = 0.30964$, \textit{i.e.}, the same as the input cosmology used for the simulation halo catalogs.
For computing the EPT power spectrum, we use \textsc{Hankel} method (Sec.~\ref{sec:hankel_method}) for one-loop integrals and the DST-based wiggle splitting for the IR resummation (Sec.~\ref{sec:ir_resum_impl}).

We sample the 16 model parameters (five cosmological, four bias, four counterterm, and three stochasticity parameters)
for both the EPT and LPT cases.
The prior distributions on these parameters, $p(\btheta)$, are set as follows.
We impose the uniform priors on three cosmological parameters $\{ \omega_{\mathrm{cdm}}, h, \ln (10^{10} A_{\mathrm{s}}) \}$ over the ranges covered by \texttt{CosmoPower-JAX} linear matter power spectrum emulator, \textit{i.e.}
\begin{align}
    \omega_{\mathrm{cdm}} \sim \U(0.05, 0.255), h \sim \U(0.64, 0.82), \ln (10^{10} A_{\mathrm{s}}) \sim \U(1.61, 3.91),
\end{align}
and the Gaussian priors on $\omega_{\mathrm{b}}$ and $n_{\mathrm{s}}$ based on \textit{Planck} 2018 $\Lambda$CDM TT,TE,EE+lowE+lensing+BAO results \cite{Planck:2018vyg}, \textit{i.e.}
\begin{align}
    \omega_{\mathrm{b}} \sim \N(0.02242, 0.00014^2), n_{\mathrm{s}} \sim \N(0.9665, 0.0038^2),
\end{align}
whose mean values are the same as the input cosmology for the simulation data described in Sec.~\ref{sec:data_and_covariance}. 
The other cosmological parameters, such as $\Omega_{\mathrm{m}}$ and $\sigma_{8}$, are obtained as the derived parameters.
For the Eulerian bias parameters, we impose a uniform prior on the linear bias as well as Gaussian priors on the nonlinear biases as
\begin{align}
    b_{1} \sim \U(1, 4), \{ b_{2}, b_{\G_{2}}, b_{\Gamma_{3}} \} \sim \N(0,5^2).
    \label{eq:priors_on_eulerian_biases}
\end{align}
On the other hand, we set the priors on the Lagrangian bias parameters to match those on the Eulerian bias parameters to make the priors for EPT and LPT as close to each other.
Namely, we can derive the mapping between the Eulerian and Lagrangian biases (see Appendix~\ref{sec:relation_btw_eulerian_and_largangian_biases} for the derivation):
\begin{align}
    b_{1} = 1 + b_{1}^{\mathrm{L}}, b_{2} = b_{2}^{\mathrm{L}}, b_{\G_{2}} = b_{\G_{2}}^{\mathrm{L}} - \frac{2}{7} b_{1}^{\mathrm{L}}, b_{\Gamma_{3}} = b_{\Gamma_{3}}^{\mathrm{L}} + \frac{23}{42} b_{1}^{\mathrm{L}} - \frac{5}{2} b_{\G_{2}}^{\mathrm{L}},
    \label{eq:eulerian_biases_in_terms_of_lagrangian_biases}
\end{align}
and inversely write the Lagrangian biases in terms of the Eulerian biases,
\begin{align}
    b_{1}^{\mathrm{L}} = b_{1} - 1, b_{2}^{\mathrm{L}} = b_{2}, b_{\G_{2}}^{\mathrm{L}} = b_{\G_{2}} + \frac{2}{7} (b_{1} - 1), b_{\Gamma_{3}}^{\mathrm{L}} = b_{\Gamma_{3}} + \frac{1}{6} (b_{1} - 1) + \frac{5}{2} b_{\G_{2}}.
    \label{eq:lagrangian_biases_in_terms_of_eulerian_biases}
\end{align}
We combine the priors on the Eulerian biases (\ref{eq:priors_on_eulerian_biases}) and Eq.~(\ref{eq:lagrangian_biases_in_terms_of_eulerian_biases}) to derive the priors on the Lagrangian bias parameters.
Note that we subtract the $k \rightarrow 0$ limit of the $b_{2}^{2}$ term \eqref{eq:pk_22_k0_limit} during in the inference.
As described in Sec.~\ref{sec:ctr_and_stoch}, we use the same counterterms for both EPT and LPT, \textit{i.e.}, Eqs.~(\ref{eq:pkmu_ept_ctr}) and (\ref{eq:pkmu_ept_ctr_nlo}) with $P_{\lin}(k)$ being the IR-resummed linear power spectrum.
We impose the Gaussian priors on the counterterm parameters as,
\footnote{Note that we find that varying $c_{44}$ does not actually affect the cosmological and bias parameters for the data and analysis setting we use in this paper.}
\begin{align}
    \{c_{0}, c_{2}, c_{4}\} / k_{\mathrm{NL}}^{2} \sim \N(0, 100^2), \quad c_{44} / k_{\mathrm{NL}}^{4} \sim \N(0, 100^2),
    \label{eq:priors_on_ctr_params}
\end{align}
where we set $k_{\mathrm{NL}} = 0.406 \, h \, \mathrm{Mpc}^{-1}$ and fix $c_{6} = c_{46} = c_{48} = 0$ since we focus on the three multipoles of $\ell = \{0, 2, 4\}$ and expect that these are strongly degenerate with other parameters.
Finally, we reparameterize the stochasticity term (\ref{eq:pkmu_stoch}) as
\begin{align}
    P_{\mathrm{stoch}}(\bk) &= \frac{1}{\bar{n}} \left[ P_{\mathrm{shot}} + (a_{0} + a_{2} \mu^2) \left( \frac{k}{k_{\mathrm{NL}}} \right)^2 \right],
    \label{eq:pkmu_stoch_auto}
\end{align}
where $\bar{n} = 10^{-4} \, (h^{-1} \, \mathrm{Mpc})^{-3}$ is the number density of the halo samples, and we impose the Gaussian priors on the parameters $P_{\mathrm{shot}}, a_{0},$ and $a_{2}$ as
\begin{align}
    \{ P_{\mathrm{shot}}, a_{0}, a_{2} \} \sim \N(0, 5^2).
\end{align}

\subsection{Inference}
\label{sec:inference_result}

Among various variants of HMC, we exclusively use the No-U-Turn Sampler (NUTS; \cite{JMLR:v15:hoffman14a}) in this work.
NUTS adaptively determines the trajectory length, \textit{i.e.}, the number of leapfrog steps used to generate each MCMC sample, terminating the trajectory before it begins to make a U-turn and retrace its path.
At each MCMC iteration, it generates a binary tree of candidate phase-space states by integrating the Hamiltonian dynamics either forward or backward in time.
The tree building is terminated once the U-turn condition is met or until it reaches the pre-determined maximum tree depth, and the next state is selected from the candidate states according to the NUTS transition rule.
For a maximum tree depth of $j$, the trajectory therefore contains at most $2^{j}-1$ leapfrog steps, requiring the same number of new evaluations of the posterior density and its gradient, up to the evaluation at the initial state.

We perform parameter sampling using the NUTS algorithm implemented in \texttt{NumPyro} \cite{phan2019composable,bingham2019pyro}, a publicly available probabilistic programming package built on JAX.
To determine the initial points of the chains, we first obtain a maximum-a-posteriori (MAP) estimate using stochastic variational inference with the \textsc{AutoDelta} guide in \texttt{NumPyro}.
We subsequently run a warm-up phase of 1000 steps, during which the inverse mass matrix and the step size are adapted.
We run 10 independent chains vectorized on a single GPU.
After the warm-up phase, with the inverse mass matrix and step size fixed, we continue the sampling until both the Gelman--Rubin statistic $\hat{R}<1.01$ and the effective sample size $\mathrm{ESS}>1000$, computed from the 10 chains, are satisfied for all sampled parameters.
Throughout the inference, we set the maximum binary-tree depth to $\mathtt{max\_tree\_depth}=10$, such that generating a single sample requires at most $2^{10}-1=1023$ leapfrog steps.

\begin{figure}[htbp]
\centering
\includegraphics[width=1\linewidth]{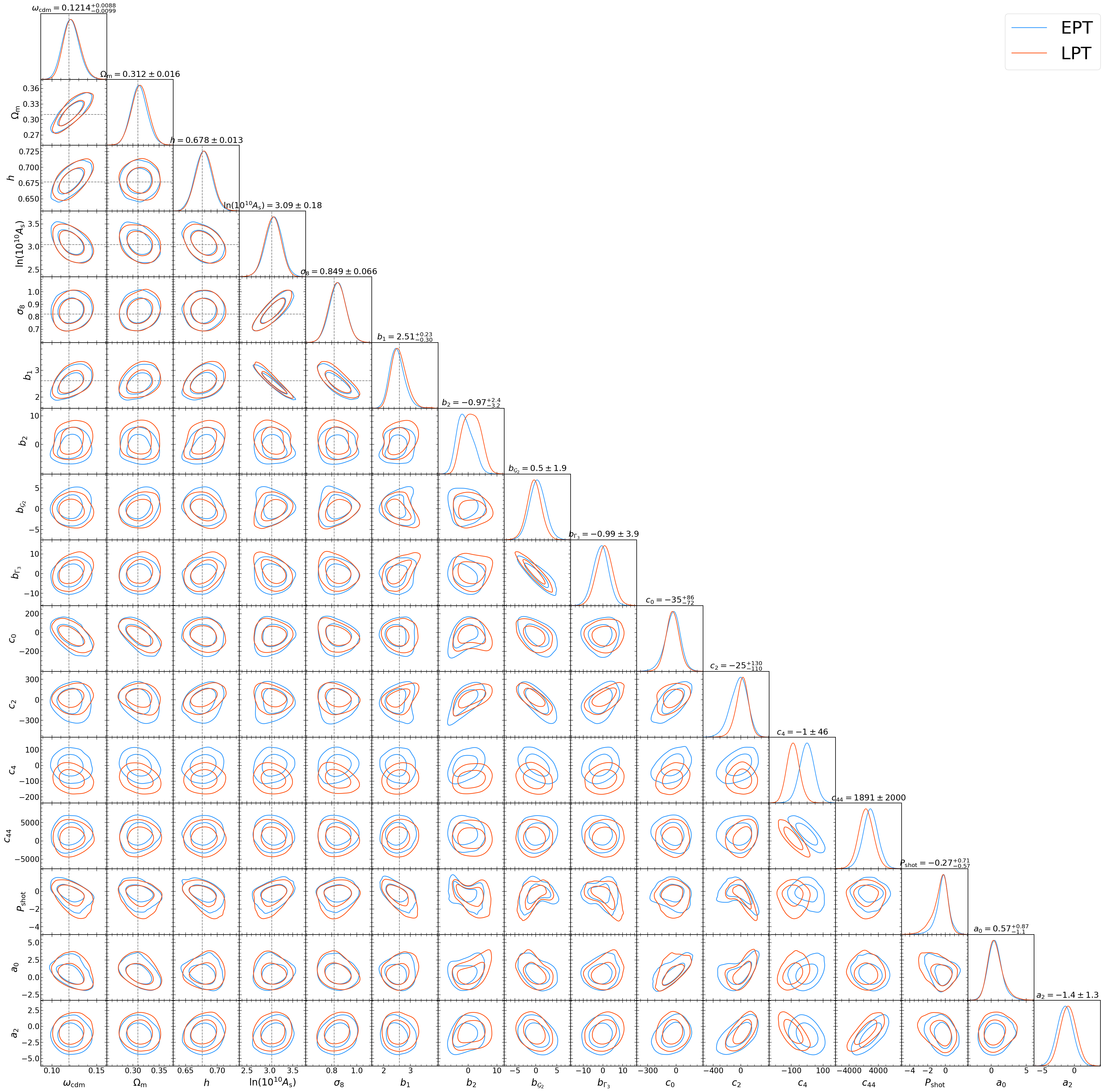}
\caption{
The two-dimensional marginal posterior distributions of the cosmological and (Eulerian) EFT parameters inferred from the halo power spectrum multipoles using the NUTS sampler.
The blue and orange contours show the results obtained using the EPT and LPT power spectra as the theoretical model, respectively.
For the LPT result, the Eulerian bias parameters are obtained from the corresponding Lagrangian bias parameters through the mapping in Eq. (\ref{eq:eulerian_biases_in_terms_of_lagrangian_biases}).
Both cases assume a maximum wavenumber of $k_{\mathrm{max}} = 0.2 \, h \, \mathrm{Mpc}^{-1}$ and a survey volume of $V = 10 \, (h^{-1} \, \mathrm{Gpc})^{3}$.
}
\label{fig:nuts_ept_vs_lpt_halo}
\end{figure}

Figure~\ref{fig:nuts_ept_vs_lpt_halo} shows the two-dimensional marginalized posterior distributions of the cosmological and EFT parameters inferred from the halo power-spectrum multipoles up to $k_{\mathrm{max}}=0.2\,h\,\mathrm{Mpc}^{-1}$, as described in Sec.~\ref{sec:data_and_covariance}.
The blue and orange contours show the posteriors obtained using the EPT and LPT models, respectively, while the horizontal and vertical gray dashed lines indicate the input parameter values.
The input values are recovered within the $1\sigma$ uncertainties in both cases.
For the bias parameters, we show the four Eulerian parameters ${b_{1},b_{2},b_{\G_{2}},b_{\Gamma_{3}}}$ for both the EPT and LPT results.
For the LPT result, the sampled Lagrangian bias parameters are mapped to the corresponding Eulerian parameters using Eq.~(\ref{eq:eulerian_biases_in_terms_of_lagrangian_biases}).
The EPT and LPT results exhibit broadly similar posterior distributions for both the cosmological and bias parameters, with the main exception of $b_{2}$, which is only weakly constrained by the power spectrum alone.
The counterterm parameters exhibit more noticeable differences between the EPT and LPT cases.
Even if the bias parameters are matched through the relation (\ref{eq:lagrangian_biases_in_terms_of_eulerian_biases}), the EPT and LPT power spectra appear different from each other, because of the large-scale displacements exponentiated in the LPT.
Varying the counterterms absorbs this residual difference between the two theoretical models.
The agreement of the inferred cosmological and bias parameters between the two cases therefore provides an important validation check of our implementations of the EPT and LPT models.
Apart from the comparison between the two PT models, another potentially important source of numerical systematics is the wiggle--no-wiggle splitting used for IR resummation, which leads to percent-level differences in the power spectrum as we have shown in Sec.~\ref{sec:ir_resum_impl}.
In Appendix~\ref{sec:wiggle_splitting}, we therefore compare parameter inference with the EPT model when the IR resummation is performed using the DST-based method and the SG filter (see Fig.~\ref{fig:nuts_ept_DST_vs_SG}).

On a single NVIDIA A100 40 GB GPU, these parameter inference runs using \texttt{NumPyro} NUTS and our JAX-based power spectrum code converge within several hours for the EPT model and take more than ten hours for the LPT model.
On average, a single MCMC step requires approximately $800$--$900$ leapfrog steps in both the EPT and LPT cases.
This indicates that the NUTS trajectory frequently reaches the maximum binary-tree depth.
The average computation time per leapfrog step is $\sim 7$ msec for EPT and $\sim 20$ msec for LPT.
Although the roughly threefold increase in the LPT case may appear inconsistent with the similar likelihood-and-gradient evaluation times reported in Table~\ref{tab:computation_time}, we find that it does not arise primarily from a difference in the computational cost of a single likelihood-and-gradient evaluation.
Instead, it is largely associated with the vectorization of the 10 parallel chains.
The LPT implementation requires substantially more GPU memory than the EPT implementation, because the integrand of the final radial integral is held on the $q$ grid for each order $\ell$ at every $(k,\mu)$, and the reverse-mode differentiation stores these arrays for the gradient.
When 10 chains are vectorized, the LPT computation appears to become limited by the memory bandwidth of the GPU.

\section{Summary and conclusion}
\label{sec:conclusion}

In this paper, we have described a differentiable implementation of the one-loop EFT power spectrum in redshift space based on both Eulerian and Lagrangian perturbation theory.
Implemented using JAX, the code is differentiable with respect to the model parameters such as cosmology (when combined with a differentiable implementation of the input linear spectrum), galaxy bias, or other EFT parameters.
It is also fully compatible with JAX's JIT compilation that enables fast execution of the code, as well as accelerated computation on a GPU.
On a GPU, our code computes both the one-loop EPT and LPT galaxy power spectrum in $\O(1)$ milliseconds after the JIT compilation, which is roughly two orders of magnitude faster than most of the currently available Python codes run on CPUs, \textit{e.g.}, \cite{DAmico:2020kxu,Chudaykin:2020aoj,Chen:2020fxs}.
Note, however, that the \texttt{COBRA} \cite{Bakx:2024zgu} method has been recently devised for the EPT loop computation and has shown quite high computation speed and accuracy, as well as applicability to higher-order loops \cite{Bakx:2025jwa}.
Also, there are already several fast differentiable emulators of the one-loop EPT power spectrum~\cite{Bonici:2025ltp,Reeves:2025bxn} based on machine learning using neural networks.
Hence, in this research direction, the primary contribution of our work is presenting a fast differentiable implementation of the LPT one-loop power spectrum, which involves complicated integrals with an exponential factor, as described in Sec.~\ref{sec:angular_int_lpt}.
In addition, we have demonstrated parameter inference using an HMC sampler on $N$-body simulation halo power spectra and confirmed that parameter inference on a GPU is two orders of magnitude faster than the \texttt{Cobaya} standard MCMC sampling on CPUs, which is widely used in the community (see Sec.~\ref{sec:inference}).

Finally, we mention several potential future directions of this work.
First, the code currently implements the standard scenario of structure formation, without primordial non-Gaussianity~\cite{Cabass:2022wjy,Cabass:2022ymb,DAmico:2022gki,Chudaykin:2025vdh}, massive neutrino dynamics~\cite{Senatore:2017hyk}, or primordial features~\cite{Beutler:2019ojk,Chen:2020ckc,Calderon:2025xod}.
Incorporating these extensions into our differentiable framework would broaden its applicability to a wider range of beyond-$\Lambda$CDM cosmologies targeted by ongoing (\textit{e.g.}, DESI, \textit{Euclid}, SPHEREx, Roman, etc.) and upcoming (\textit{e.g.}, DESI-II and MegaMapper) surveys.
Second, it could be interesting to extend our JAX EFT code beyond the ordinary galaxy power spectrum to statistics such as the bispectrum \cite{Philcox:2022frc,DAmico:2022osl,DAmico:2022ukl,Chen:2024pyp} (but see \cite{Bakx:2025pop} that uses the \texttt{COBRA}~\cite{Bakx:2024zgu} scheme), skew spectra~\cite{Schmittfull:2020hoi,Chen:2024bdg}, Lyman-$\alpha$ forest~\cite{Chen:2021rnb,Ivanov:2023yla}, peculiar velocity statistics~\cite{Chen:2025aqr}, intrinsic alignment statistics~\cite{Bakx:2023mld,Chen:2023yyb,Akitsu:2026wzu}, or correlators of shifted operators in the field-level forward model \cite{Schmittfull:2018yuk,Schmittfull:2020trd,Kokron:2022iok,Akitsu:2025boy}.
Third, it would be valuable to explore applications to high-dimensional parameter inference involving many redshift bins and multi-tracer samples, combining with different gradient-based sampling methods, \textit{e.g.}, \cite{pmlr-v130-hoffman21a,Robnik:2022bzs,Robnik:2023pgt}.
Differentiability is also useful for Fisher forecasts and score-based compression~\cite{Alsing:2017var}.
We could also investigate other inference techniques such as variational inference~\cite{2016arXiv160100670B,JMLR:v18:16-107} or other variants of HMC (\textit{e.g.},~\cite{pmlr-v130-hoffman21a}) that take advantage of differentiability.

\acknowledgments
We are grateful to 
Giovanni Cabass, 
Mikhail Ivanov, 
Azadeh Moradinezhad Dizgah, 
and
Oliver Philcox 
for kind advice on the implementation of the one-loop EPT power spectrum.
We also thank 
Xiao Fang
and
Elisabeth Krause
for useful discussions in the early phase of this work.
We thank Takahiro Nishimichi and Satoshi Tanaka, who set up their NVIDIA A100 GPU machine and were willing to allow us to use it for this work.
Numerical computation was carried out on the Cray XD2000 and the GPU cluster at the Center for Computational Astrophysics, National Astronomical Observatory of Japan.
YK acknowledges the Grant-in-Aid (No. 24KJ0211) for Japan Society for the Promotion of Science (JSPS) Fellows.
YK also acknowledges support from the SPHEREx project under a contract from the NASA/GODDARD Space Flight Center to the California Institute of Technology, as well as support from the David and Lucile Packard Foundation.

\appendix

\section{Explicit formulas for LPT power spectrum}
\label{sec:lpt_formulas}

In Sec.~\ref{sec:model_lpt}, we have briefly outlined the one-loop LPT model of the galaxy power spectrum that we implement.
Each term in Eq.~(\ref{eq:ps_1loop_lpt}) is defined in the previous literature, \textit{e.g.} \cite{Matsubara:2007wj,Matsubara:2008wx,Carlson:2012bu,Vlah:2016bcl,Chen:2020fxs,Chen:2020zjt}.
In this appendix, for completeness, we put the explicit formulas that arise in the calculation of these LPT terms.

\subsection{$Q_{n}$ and $R_{n}$ integrals}

Many of the one-loop LPT terms are calculated from the integrals that are conventionally denoted as $Q_{n}$ and $R_{n}$.
The integrals used in our code are listed below.
\begin{align}
    Q_{1}(k) &\equiv \int_{\bp} \left[ \frac{k^2 p^2 - (\bk \cdot \bp)^2}{p^2 |\bk - \bp|^2} \right]^2 P_{\lin}(p) P_{\lin}(|\bk-\bp|),  \label{eq:Q1}\\
    Q_{2}(k) &\equiv \int_{\bp} \frac{(\bk \cdot \bp) (k^2 - \bk \cdot \bp) [k^2 p^2 - (\bk \cdot \bp)^2]}{p^4 |\bk-\bp|^4} P_{\lin}(p) P_{\lin}(|\bk-\bp|), \label{eq:Q2}\\
    Q_{5}(k) &\equiv \int_{\bp} \frac{(\bk \cdot \bp) [k^2 p^2 - (\bk \cdot \bp)^2]}{p^4 |\bk - \bp|^2} P_{\lin}(p) P_{\lin}(|\bk - \bp|), \label{eq:Q5}\\
    Q_{8}(k) &\equiv \int_{\bp} \frac{k^2 p^2 - (\bk \cdot \bp)^2}{p^2 |\bk - \bp|^2} P_{\lin}(p) P_{\lin}(|\bk-\bp|), \label{eq:Q8}\\
    Q_{\G_{2}}(k) &\equiv \int_{\bp} \frac{[p^2 - (\bk \cdot \bp)] [k^2 p^2 - (\bk \cdot \bp)^2]}{p^4 |\bk - \bp|^4} P_{\lin}(p) P_{\lin}(|\bk-\bp|), \\
    R_{1}(k) &\equiv P_{\lin}(k) \int_{\bp} \left[ \frac{k^2 p^2 - (\bk \cdot \bp)^2}{k p^2 |\bk - \bp|} \right]^2 P_{\lin}(p), \label{eq:R1}\\
    R_{2}(k) &\equiv P_{\lin}(k) \int_{\bp} \frac{(\bk \cdot \bp) (k^2 - \bk \cdot \bp) [k^2 p^2 - (\bk \cdot \bp)^2]}{k^2 p^4 |\bk-\bp|^2} P_{\lin}(p).
\end{align}
where we follow the original numbering notation in \cite{Matsubara:2007wj,Matsubara:2008wx}.
Note that one can find $Q_{5}(k) = \frac{1}{2} [Q_{1}(k) + Q_{2}(k)]$.
Since $Q_{n}$ and $R_{n}$ integrals have the same integral structures as the EPT 22- and 13-terms, respectively, they can be implemented using the FFTLog-based methods.
In our code, we implement them using the \textsc{Hankel} method \cite{Schmittfull:2016jsw} described in Sec.~\ref{sec:hankel_method}, similarly to $\texttt{velocileptors}$.

\subsection{Cumulants}

Here, we explicitly write each term that appears in the LPT power spectrum, Eqs.~(\ref{eq:pkmu_lpt_tree}) and (\ref{eq:pkmu_lpt_1loop}).
In the matter power spectrum, the cumulant expansion of Eq.~(\ref{eq:lpt_generating_function}) generates the cumulants of the difference of the displacements between the two Lagrangian positions, $\bDelta = \bPsi(\bq_{2}) - \bPsi(\bq_{1})$.
At one-loop order of the power spectrum, we include the second- and third-order cumulants, \textit{i.e.}, $A_{ij} \equiv \avrgc{ \Delta_{i} \Delta_{j} }$ and $W_{ijk} \equiv \avrgc{ \Delta_{i} \Delta_{j} \Delta_{k} }$. 
As mentioned in Sec.~\ref{sec:model_lpt}, $A_{ij}$ has the linear and one-loop order parts.
The linear part amounts to the correlation between two linear-order parts of $\bDelta$, \textit{i.e.},
\begin{align}
    A_{ij}^{\lin}(\bq) &\equiv \avrgc{ \Delta^{(1)}_{i} \Delta^{(1)}_{j} } = 2 \int_{\bk} [1 - \cos(\bk \cdot \bq)] \frac{k_{i} k_{j}}{k^{4}} P_{\lin}(k).
\end{align}
By rotational symmetry, it takes the form of $A_{ij}^{\lin}(\bq) = X^{\lin}(q) \delta_{ij} + Y^{\lin}(q) \hat{q}_{i} \hat{q}_{j}$, where $\hat{q}_{i}$ denotes the $i$th component of the unit vector along $\bq$, and the funcitons $X^{\lin}$ and $Y^{\lin}$ depend on $\bq$ only through its norm $q \equiv |\bq|$.
By taking two different contractions of $A_{ij}^{\lin}$ with $\delta^{ij}$ or $\hat{q}^{i} \hat{q}^{j}$ and solving the system of those two equations, one can obtain the explicit formulas of $X^{\lin}$ and $Y^{\lin}$ as
\begin{align}
    X^{\lin}(q) &= \frac{2}{3} \int_{0}^{\infty} \frac{\mathrm{d}k}{2 \pi^{2}}  [1 - j_{0}(kq) - j_{2}(kq)] P_{\lin}(k), \\
    Y^{\lin}(q) &= 2 \int_{0}^{\infty} \frac{\mathrm{d}k}{2\pi^2} j_{2}(kq) P_{\lin}(k).
\end{align}
The one-loop part consists of the correlation of two second-order displacements and those of a linear- and a third-order displacements, \textit{i.e.}, $A_{ij}^{\text{1-loop}} = A_{ij}^{(22)} + A_{ij}^{(13)} + A_{ij}^{(31)}$, where $A_{ij}^{(mn)} \equiv \avrgc{ \Delta^{(m)}_{i} \Delta^{(n)}_{j} }$.
Similarly, we can express them as $A_{ij}^{(22)}(\bq) = X_{22}(q) \delta_{ij} + Y_{22}(q) \hat{q}_{i} \hat{q}_{j}$ and $A_{ij}^{(13)}(\bq) = A_{ij}^{(31)}(\bq) = X_{13}(q) \delta_{ij} + Y_{13}(q) \hat{q}_{i} \hat{q}_{j}$ with
\begin{align}
    X_{22}(q) &= \frac{2}{3} \int_{0}^{\infty} \frac{\mathrm{d}k}{2 \pi^2} \left[ 1 - j_{0}(kq) - j_{2}(kq) \right] \frac{9}{98} Q_{1}(k), \\
    Y_{22}(q) &= 2 \int_{0}^{\infty} \frac{\mathrm{d}k}{2 \pi^2} j_{2}(kq) \frac{9}{98} Q_{1}(k), \\
    X_{13}(q) &= \frac{2}{3} \int_{0}^{\infty} \frac{\mathrm{d}k}{2 \pi^2} \left[ 1 - j_{0}(kq) - j_{2}(kq) \right] \frac{5}{21} R_{1}(k), \\
    Y_{13}(q) &= 2 \int_{0}^{\infty} \frac{\mathrm{d}k}{2 \pi^2} j_{2}(kq) \frac{5}{21} R_{1}(k),
\end{align}
where $Q_{1}(k)$ and $R_{1}(k)$ are defined in the previous subsection.
The third-order cumulant, $W_{ijk}(\bq) \equiv \avrgc{ \Delta_{i} \Delta_{j} \Delta_{k} }$, consists of two linear-order and one second-order displacements for the one-loop power spectrum.
That is, $W_{ijk}(\bq) = W_{ijk}^{(112)}(\bq) + W_{ijk}^{(121)}(\bq) + W_{ijk}^{(211)}(\bq)$, where each piece is defined as $W_{ijk}^{(lmn)} \equiv \avrgc{ \Delta^{(l)}_{i} \Delta^{(m)}_{j} \Delta^{(n)}_{k} }$.
By symmetry, the $W_{ijk}^{(112)}$ is expressed as
\begin{align}
    W_{ijk}^{(112)}(\bq) &= V_{1}(q) \hat{q}_{i} \delta_{jk} + V_{2}(q) \hat{q}_{j} \delta_{ki} + V_{3}(q) \hat{q}_{k} \delta_{ij} + T(q) \hat{q}_{i} \hat{q}_{j} \hat{q}_{k},
\end{align}
where $V_{1}(q), V_{2}(q), V_{3}(q)$ and $T(q)$ are obtained by taking contractions of $W_{ijk}^{(112)}$ with four different combinations of $\hat{q}^{i}$ and $\delta^{ij}$.
The explicit expressions for them are
\begin{align}
    T(q) &= - \frac{3}{7} \int_{0}^{\infty} \frac{\mathrm{d}k}{2 \pi^2} \frac{1}{k} [Q_{1}(k) + 2 Q_{2}(k) + 2 R_{1}(k) + 4 R_{2}(k)] j_{3}(kq), 
    \label{eq:T}
    \\
    V_{1}(q) = V_{2}(q) &= \frac{3}{35} \int_{0}^{\infty} \frac{\mathrm{d}k}{2 \pi^2} \frac{1}{k} [Q_{1}(k) + 2 Q_{2}(k) - 3 R_{1}(k) + 4 R_{2}(k)] j_{1}(kq) - \frac{1}{5} T(q), \\
    V_{3}(q) &= - \frac{3}{35} \int_{0}^{\infty} \frac{\mathrm{d}k}{2 \pi^2} \frac{1}{k} [4 Q_{1}(k) - 2 Q_{2}(k) - 2 R_{1}(k) - 4 R_{2}(k)] j_{1}(kq) - \frac{1}{5} T(q).
\end{align}

In the case of biased tracers, the cumulant expansion generates additional cumulants according to the Lagrangian bias expansion such as Eq.~(\ref{eq:lagrangian_bias_expansion}).
For the local-in-matter-density (LIMD) bias, we first have $U^{10}_{i} \equiv \avrgc{ \Delta_{i} \delta }$ that has the linear and one-loop order parts, $U^{10}_{i}(\bq) = U^{\lin}_{i}(\bq) + U^{(3)}_{i}(\bq)$, where
\begin{align}
    U^{\lin}_{i}(\bq) &\equiv \avrgc{ \Delta_{i}^{(1)} \delta } = - \hat{q}_{i} \int_{0}^{\infty} \frac{k \mathrm{d}k}{2 \pi^2} j_{1}(kq) P_{\lin}(k), \\
    U^{(3)}_{i}(\bq) &\equiv \avrgc{ \Delta_{i}^{(3)} \delta } = - \frac{5}{21} \hat{q}_{i} \int_{0}^{\infty} \frac{k \mathrm{d}k}{2 \pi^2} j_{1}(kq) R_{1}(k).
\end{align}
All the other bias terms arise at one-loop order.
Up to the second-order LIMD bias, we have
\begin{align}
    U^{11}_{i}(\bq) &\equiv \avrgc{ \Delta_{i} \delta(\bq_{1}) \delta(\bq_{2}) } = - \frac{6}{7} \hat{q}_{i} \int_{0}^{\infty} \frac{k \mathrm{d}k}{2 \pi^2} j_{1}(kq) [R_{1}(k) + R_{2}(k)], \\
    U^{20}_{i}(\bq) &\equiv \avrgc{ \Delta_{i} \delta^{2} } = - \frac{3}{7} \hat{q}_{i} \int_{0}^{\infty} \frac{k \mathrm{d}k}{2 \pi^2} j_{1}(kq) Q_{8}(k).
\end{align}
We also include a cumulant involving two displacement differences,
\begin{align}
    A^{10}_{ij}(\bq) &\equiv \avrgc{ \Delta_{i} \Delta_{j} \delta } = X_{10}(q) \delta_{ij} + Y_{10}(q) \hat{q}_{i} \hat{q}_{j},
\end{align}
where the isotropic and anisotropic parts are written as
\begin{align}
    X_{10}(q) &= \frac{2}{7} \int_{0}^{\infty} \frac{\mathrm{d}k}{2 \pi^2} R_{1}(k) - \frac{2}{7} \int_{0}^{\infty} \frac{\mathrm{d}k}{2 \pi^2} [Q_{5}(k) + 2 R_{2}(k)] j_{0}(kq) - \frac{1}{7} \int_{0}^{\infty} \frac{\mathrm{d}k}{2 \pi^2} [2 Q_{5}(k) + 3 R_{1}(k) + 4 R_{2}(k)] j_{2}(kq), \label{eq:X10} \\
    Y_{10}(q) &= \frac{3}{7} \int_{0}^{\infty} \frac{\mathrm{d}k}{2 \pi^2} [2 Q_{5}(k) + 3 R_{1}(k) + 4 R_{2}(k)] j_{2}(kq). \label{eq:Y10}
\end{align}
Note that we have dropped $R_{2}(k)$ in the integrand of the first term on the right-hand side in Eq.~(\ref{eq:X10}), unlike in the literature (\textit{e.g.} \cite{Carlson:2012bu}), since $\int_{0}^{\infty} \frac{\mathrm{d}k}{2 \pi^2} R_{2}(k) = 0$, which follows from the angular symmetry of the integrand.

For the second-order shear bias $\G_{2}$, we have the correlations between one displacement difference and the bias operators,
\begin{align}
    V^{10}_{\G_{2}, i}(\bq) &\equiv \avrgc{ \Delta_{i} \G_{2} } = \frac{3}{7} \hat{q}_{i} \int_{0}^{\infty} \frac{k \mathrm{d}k}{2 \pi^2} j_{1}(kq) Q_{1}(k), \\
    V^{12}_{\G_{2}, i}(\bq) &\equiv \avrgc{ \Delta_{i} \delta(\bq_{1}) \G_{2}(\bq_{2}) } = 2 \hat{q}_{i} \int_{0}^{\infty} \frac{k \mathrm{d}k}{2 \pi^2} j_{1}(kq) Q_{5}(k),
\end{align}
We also have the correlation between two displacement differences and $\G_{2}$ at one point,
\begin{align}
    \Upsilon_{\G_{2}, ij}(\bq) \equiv \avrgc{ \Delta_{i} \Delta_{j} \G_{2} }.
\end{align}
Note that, at the order relevant for the one-loop LPT power spectrum, we have $\Upsilon_{\G_{2}, ij}(\bq) = \avrgc{ \Psi_{i}(\bq_{1}) \Psi_{j}(\bq_{1}) \G_{2}(\bq_{2}) }$.
By symmetry, it can be decomposed as
\begin{align}
    \Upsilon_{\G_{2}, ij}(\bq) = X_{\G_{2}}(q) \delta_{ij} + Y_{\G_{2}}(q) \hat{q}_{i} \hat{q}_{j},
\end{align}
where
\footnote{
In \cite{White:2014gfa,Vlah:2016bcl}, the corresponding term is defined using $s^{2}(\bq) = s_{ij}(\bq) s_{ij}(\bq)$ where $s_{ij}(\bq) = \left( \frac{\nabla_{i} \nabla_{j}}{\nabla^{2}} - \frac{1}{3} \delta_{ij} \right) \delta(\bq)$.
In this case, $\Upsilon_{s^{2}, ij}(\bq) \equiv \avrgc{ \Delta_{i} \Delta_{j} s^{2} } = \avrgc{ \Psi_{i}(\bq_{1}) \Psi_{j}(\bq_{1}) s^{2}(\bq_{2}) }$ is also written as a contraction of the two cumulants,
\begin{align}
    \Upsilon_{s^{2}, ij}(\bq) = 2 \avrgc{ \Psi_{i}(\bq_{1}) s_{mn}(\bq_{2}) } \avrgc{ \Psi_{j}(\bq_{1}) s_{mn}(\bq_{2}) }.
\end{align}
The cumulant above can be further decomposed as \cite{White:2014gfa}
\begin{align}
    \avrgc{ \Psi_{i}(\bq_{1}) s_{mn}(\bq_{2}) } = \J_{2}(q) \hat{q}_{i} \delta_{mn} + \J_{3}(q) ( \delta_{im} \hat{q}_{n} + \delta_{in} \hat{q}_{m} ) + \J_{4}(q) \hat{q}_{i} \hat{q}_{m} \hat{q}_{n},
\end{align}
where each of the $\J_{i}$ functions is a linear combination of $\xi^{\ell}_{n}(q)$'s.
Thus, upon decomposing $\Upsilon_{s^{2},ij}$ as $X_{s^{2}}(q) \delta_{ij} + Y_{s^{2}}(q) \hat{q}_{i} \hat{q}_{j}$, both $X_{s^{2}}$ and $Y_{s^{2}}$ can be expressed as linear combinations of products of two $\J_{i}$'s, and hence require no additional Hankel transforms.
This factorized form follows from the representation $s^{2} = s_{mn} s_{mn}$, and such a factorization is not manifest when $\G_{2}$ is used to define $\Upsilon_{ij}$.
However, since $\G_{2} = s^{2} - \frac{2}{3} \delta^{2}$, we have $\Upsilon_{\G_{2},ij}(\bq) = \Upsilon_{s^{2},ij}(\bq) - \frac{4}{3} U^{\lin}_{i}(\bq) U^{\lin}_{j}(\bq)$.
Therefore, $X_{\G_{2}}$ and $Y_{\G_{2}}$ in Eqs.~(\ref{eq:X_G2}) and (\ref{eq:Y_G2}) can alternatively be computed without additional Hankel transforms as $X_{\G_{2}}(q) = X_{s^{2}}(q)$ and $Y_{\G_{2}}(q) = Y_{s^{2}}(q) - \frac{4}{3} [U^{\lin}(q)]^{2}$.
}
\begin{align}
    X_{\G_{2}}(q) &= \frac{1}{3} \int_{0}^{\infty} \frac{\mathrm{d}k}{2 \pi^2} \left\{ j_{0}(kq) [ Q_{1}(k) - Q_{2}(k) - 3 k^2 Q_{\G_{2}}(k)] + j_{2}(kq) [ Q_{1}(k) + 2 Q_{2}(k) ] \right\}, \label{eq:X_G2} \\
    Y_{\G_{2}}(q) &= - \int_{0}^{\infty} \frac{\mathrm{d}k}{2 \pi^2} \left\{  j_{0}(kq) [ Q_{1}(k) - Q_{2}(k) - k^2 Q_{\G_{2}}(k) ] + j_{2}(k q) [ Q_{1}(k) + 2 Q_{2}(k) ] \right\}. \label{eq:Y_G2}
\end{align}
Aside from the cumulants that involve the displacements, the correlations between the second-order bias terms,
\begin{align}
    \chi_{\G_{2}}(q) &\equiv \avrgc{ \delta^{2}(\bq_{1}) \G_{2}(\bq_{2}) } = \frac{4}{3} \left\{ [\xi^{2}_{0}(q)]^{2} - [\xi^{0}_{0}(q)]^{2} \right\},
    \label{eq:chi_G2}\\
    \zeta_{\G_{2}}(q) &\equiv \avrgc{ \G_{2}(\bq_{1}) \G_{2}(\bq_{2}) } =  \frac{16}{15} [\xi^{0}_{0}(q)]^{2} - \frac{32}{21} [\xi^{2}_{0}(q)]^{2} + \frac{16}{35} [\xi^{4}_{0}(q)]^{2}.
    \label{eq:zeta_G2}
\end{align}
are included as well.
In the above, $\xi^{\ell}_{n}(q)$ is the generalized correlation function defined as Eq.~(\ref{eq:xi_ln}). 

Finally, the third-order bias terms consist of two pieces, \textit{i.e.},
\begin{align}
    U_{\Gamma_{3}, i}(\bq) &\equiv \avrgc{ \Delta_{i} \Gamma_{3} } = - \frac{2}{5} \hat{q}_{i} \int_{0}^{\infty} \frac{k \mathrm{d}k}{2 \pi^2} j_{1}(kq) \F_{\G_{2}}(k), \\
    \theta_{\Gamma_{3}}(q) &\equiv \avrgc{ \delta(\bq_{1}) \Gamma_{3}(\bq_{2}) } = \frac{2}{5} \int_{0}^{\infty} \frac{k^2 \mathrm{d}k}{2 \pi^2} j_{0}(kq) \F_{\G_{2}}(k).
\end{align}
where we have used the notation in \cite{Assassi:2014fva} for $\F_{\G_{2}}$, \textit{i.e.},
\begin{align}
    \F_{\G_{2}}(k) &\equiv 4 P_{\lin}(k) \int_{\bp} K_{\G_{2}}(\bk-\bp,\bp) F_{2}(\bk,-\bp) P_{\lin}(p),
\end{align}
with the kernel $K_{\G_{2}}$ being defined in Eq.~(\ref{eq:KG2}).
This function reduces to
\begin{align}
    \F_{\G_{2}}(k) &= - \frac{32}{21} k^2 P_{\lin}(k) \int_{0}^{\infty} q \mathrm{d}q  \left[ j_{0}(k q) \xi^{0}_{0}(q) - \frac{10}{7} j_{2}(k q) \xi^{2}_{0}(q) + \frac{3}{7} j_{4}(k q) \xi^{4}_{0}(q) \right],
    \label{eq:FG2}
\end{align}
and is implemented using the \textsc{Hankel} method.
The cumulants involving the third-order bias operators other than $\Gamma_{3}$ do not appear in Eq.~(\ref{eq:pkmu_lpt_1loop}) due to the renormalization.
For example, $U_{\delta \G_{2}, i}(\bq) \equiv \avrgc{ \Delta_{i} \delta \G_{2} }$ also arises in the cumulant expansion of Eq.~(\ref{eq:lpt_generating_function}), which equals $- \frac{4}{3} \sigma^{2} U^{\lin}_{i}(q)$ at one-loop order (in \cite{Vlah:2016bcl,Vlah:2018ygt}, the term corresponding to this is denoted by $V^{11}_{i} \equiv \avrgc{ \Delta_{i} \delta s^{2} }$, which equals zero).
We have another correlation of $\delta \G_{2}$, \textit{i.e.}, $\theta_{\delta \G_{2}}(q) \equiv \avrgc{ \delta(\bq_{1}) (\delta \G_{2})(\bq_{2}) } = - \frac{4}{3} \sigma^{2} \xi_{\lin}(q)$.
However, both terms cancel with the contributions from the $- \frac{4}{3} \sigma^{2} \delta$ part subtracted in the bias expansion (\ref{eq:lagrangian_bias_expansion}).
These cancellations reflect the renormalization of bias operators, ensuring that no explicit dependence on $\sigma^2$ remains in the final power spectrum.

\subsection{Useful identities}

In this subsection we introduce several compact notations. Vector cumulants are represented by their scalar coefficients, \textit{e.g.}, $U^{\rm lin}_i = \hat{q}_i U^{\rm lin}$, a prime denotes ${\mathrm d}/{\mathrm d}q$, and the Hankel transform of a function $S(k)$ is written as
\begin{equation}
    {\cal H}_\ell^n[S](q) \equiv \int_0^\infty \frac{k^2 {\mathrm d}k}{2\pi^2} k^n S(k) j_\ell(kq),
\end{equation}
so that $\xi_n^\ell = {\cal H}_\ell^n[P_{\rm lin}]$.
With this notation, the LPT cumulants of the previous subsection that are relevant below read
\begin{align}
    U^{\rm lin} &= -\xi^1_{-1},  \qquad Y^{\rm lin} = 2\xi^2_{-2},  \qquad U^{(3)} = - \frac{5}{21}{\cal H}_1^{-1}[R_1],  \qquad Y_{22} = \frac{9}{49}{\cal H}_2^{-2}[Q_1],
    \nonumber\\
    U^{20} & = -\frac37 {\cal H}^{-1}_1[Q_8], \qquad V^{10}_{{\cal G}_2} = \frac37{\cal H}^{-1}_1[Q_1],  \qquad  V^{12}_{{\cal G}_2} = 2{\cal H}^{-1}_1[Q_5],
    \nonumber\\
    U_{\Gamma_3} &= -\frac25{\cal H}_1^{-1}[{\cal F}_{{\cal G}_2}], \qquad \theta_{\Gamma_3} = \frac25 {\cal H}_0^0[{\cal F}_{{\cal G}_2}].
    \label{eq:cumulants_H}
\end{align}
The spherical-Bessel relations $\frac{\mathrm{d}}{\mathrm{d}q}[q^{-\ell} j_{\ell}(kq)] = -k q^{-\ell} j_{\ell+1}(kq)$, ~$\frac{\mathrm{d}}{\mathrm{d}q}[q^{\ell+1} j_{\ell}(kq)] = k q^{\ell+1} j_{\ell-1}(kq)$ and $j_{\ell-1}(x) + j_{\ell+1}(x) = (2\ell+1) j_{\ell}(x)/x$ give, 
\begin{equation}
    \mathcal{H}^{n+1}_{\ell+1} = -\left( \mathcal{H}^n_{\ell} \right)' + \frac{\ell}{q} \mathcal{H}^n_{\ell}, \qquad
    \mathcal{H}^{n+1}_{\ell-1} = \left( \mathcal{H}^n_{\ell} \right)' + \frac{\ell+1}{q} \mathcal{H}^n_{\ell}, \qquad
    \mathcal{H}^{n-1}_{\ell} = \frac{q}{2\ell+1} \left( \mathcal{H}^{n}_{\ell-1} + \mathcal{H}^n_{\ell+1} \right).
    \label{eq:H_recurrences}
\end{equation}
The first two relations indicate
\begin{align}
    \xi^1_{-1} &= (\xi^2_{-2})' + \frac{3}{q} \xi^2_{-2}, \qquad
    \xi^3_{-1} = -(\xi^2_{-2})' + \frac{2}{q} \xi^2_{-2}, \nonumber \\
    \xi^0_0 &= (\xi^1_{-1})' + \frac{2}{q} \xi^1_{-1}, \qquad
    \xi^2_0 = -(\xi^1_{-1})' + \frac{1}{q} \xi^1_{-1}, \qquad
    \xi^4_0 = -(\xi^3_{-1})' + \frac{3}{q} \xi^3_{-1}.
    \label{eq:xi_ladder}
\end{align}
In each of the pairs $(\xi^1_{-1}, \xi^3_{-1})$ and $(\xi^0_0, \xi^2_0)$, the sum contains no derivative, $\xi^1_{-1} + \xi^3_{-1} = 5 \xi^2_{-2}/q$ and $\xi^0_0 + \xi^2_0 = 3 \xi^1_{-1}/q$.
Then, the differences of the squares can be expressed by total derivatives,
\begin{align}
    (\xi^1_{-1})^2 - (\xi^3_{-1})^2 &= \frac{5}{q} \xi^2_{-2} \left[ 2 (\xi^2_{-2})' + \frac{1}{q} \xi^2_{-2} \right] = \frac{5}{q^2} \left[ q (\xi^2_{-2})^2 \right]', \label{eq:xi_squares_1} \\
    (\xi^0_0)^2 - (\xi^2_0)^2 &= \frac{3}{q} \xi^1_{-1} \left[ 2 (\xi^1_{-1})' + \frac{1}{q} \xi^1_{-1} \right] = \frac{3}{q^2} \left[ q (\xi^1_{-1})^2 \right]'. \label{eq:xi_squares_0}
\end{align}
Substituting the expressions for $\xi^1_{-1}$ and $\xi^3_{-1}$ in Eq.~\eqref{eq:xi_ladder} into those for $\xi^0_0$, $\xi^2_0$ and $\xi^4_0$ expresses the latter via $\xi^2_{-2}$,
\begin{equation}
    \xi^0_0 = (\xi^2_{-2})'' + \frac{5 (\xi^2_{-2})'}{q} + \frac{3 \xi^2_{-2}}{q^2}, \qquad
    \xi^2_0 = -(\xi^2_{-2})'' - \frac{2 (\xi^2_{-2})'}{q} + \frac{6 \xi^2_{-2}}{q^2}, \qquad
    \xi^4_0 = (\xi^2_{-2})'' - \frac{5 (\xi^2_{-2})'}{q} + \frac{8 \xi^2_{-2}}{q^2}.
    \label{eq:xi_l0_from_xi2}
\end{equation}
Combining these relations with $\frac{8}{15} (\xi^0_0)^{2} - \frac{16}{21} (\xi^2_0)^2 + \frac{8}{35} (\xi^4_0)^{2}$, the terms $[(\xi^2_{-2})'']^{2}$ and $(\xi^2_{-2})'' (\xi^2_{-2})'$ cancel, and the remaining terms form a total derivative,
\begin{equation}
    \frac{8}{15} (\xi^0_0)^2 - \frac{16}{21} (\xi^2_0)^2 + \frac{8}{35} (\xi^4_0)^2 = \frac{8}{q^2} \left\{ \frac{1}{q} \left[ q (\xi^2_{-2})^{2} \right]' \right\}'.
    \label{eq:xi_squares_2}
\end{equation}
Rewriting Eqs.~\eqref{eq:chi_G2} and \eqref{eq:zeta_G2} using Eqs.~\eqref{eq:xi_squares_0},~\eqref{eq:xi_squares_1}, and \eqref{eq:xi_squares_2} yields
\begin{equation}
    q^2 \chi_{\G_2} = -4 \left[ q (\xi^1_{-1})^2 \right]', \qquad
    q^2 \zeta_{\G_2} = \frac{16}{5} \left\{ q \left[ (\xi^1_{-1})^2 - (\xi^3_{-1})^2 \right] \right\}'.
    \label{eq:chi_zeta_total_derivative}
\end{equation}
These identites mean that the integrals of $q^2 \chi_{\G_2}$ and $q^2 \zeta_{\G_2}$ over $0<q<\infty$ vanish, as both $\xi^1_{-1}$ and $\xi^3_{-1}$ vanish at $q=0$ and $q\to \infty$.
As the $Q_n(k)$ integrals are $22$-type, one can use Eqs.~\eqref{eq:ept_P22} and \eqref{eq:pk_22_lnm} to find
\begin{align}
    Q_1(k) &= 4 \pi \int_0^\infty r^2 \mathrm{d}r ~ j_0(kr)~ \frac12 \zeta_{\G_2}(r), \qquad
    Q_8(k) = -4 \pi \int_0^\infty r^2 \mathrm{d}r ~ j_0(kr)~\frac12 \chi_{\G_2}(r), \nonumber \\
    Q_1(k) - Q_2(k) &= 4 \pi k^2 \int_0^\infty r^2 \mathrm{d}r~j_0(kr)~\frac25 \left[ (\xi^1_{-1})^2 - (\xi^3_{-1})^2 \right].
    \label{eq:Q_as_ft}
\end{align}

We also use identities for the isotropic source function $S(k)$.
For a source of the form $S(k) = 4 \pi \int_{0}^{\infty} r^{2} \mathrm{d}r ~j_{0}(kr) F(r)$, we have
\begin{align}
    \mathcal{H}^{-1}_1[S](q) &= \frac{1}{q^{2}} \int_0^q r^2 F(r) ~\mathrm{d}r, \qquad
    \mathcal{H}^{-1}_1[k^{2} S](q) = -F'(q), \nonumber \\
    \mathcal{H}^{-2}_2[S](q) &= \frac{1}{2 q^3} \int_0^q r^2 (q^2 - r^2) F(r) ~\mathrm{d}r,
    \label{eq:finite_range}
\end{align}
which follow from $\int_0^q r^2 j_{0}(kr) \, \mathrm{d}r = q^2 j_{1}(kq)/k$ and $\int_0^q r^2 (q^2 - r^2) j_0(kr) \, \mathrm{d}r = 2 q^3 j_2(kq)/k^2$, and from $k^2 S$ being the transform of $-r^{-2} (r^2 F')'$.
Hence, these transforms depend on $F(r)$ only at $r \le q$.

Inserting Eq.~\eqref{eq:Q_as_ft} into Eqs.~\eqref{eq:cumulants_H} and \eqref{eq:finite_range} and integrating Eq.~\eqref{eq:chi_zeta_total_derivative} gives
\begin{equation}
    U^{20} = \frac{3}{14 q^2} \int_0^q r^{2} \chi_{\G_2} ~ \mathrm{d}r = -\frac67 \frac{(U^{\lin})^2}{q}, \qquad
    V^{10}_{\G_2} = \frac{3}{14 q^2} \int_0^q r^2 \zeta_{\G_2} ~ \mathrm{d}r = \frac{24}{35 q} \left[ (\xi^1_{-1})^2 - (\xi^3_{-1})^2 \right].
    \label{eq:U20_V10}
\end{equation}
Since $Q_5 = \frac12 (Q_1 + Q_2)$, we have $2 Q_5 = 2 Q_1 - (Q_1 - Q_2)$, and the relation $\mathcal{H}^{-1}_1[k^2 S] = -F'$ of Eq.~(\ref{eq:finite_range}) gives
\begin{equation}
    V^{12}_{\G_2} = \frac{14}{3} V^{10}_{\G_2} + \frac25 \frac{\mathrm{d}}{\mathrm{d}q} \left[ (\xi^1_{-1})^2 - (\xi^3_{-1})^2 \right].
    \label{eq:V12}
\end{equation}
For $Y_{22}$, Eq.~(\ref{eq:xi_squares_2}) gives $r^2 \zeta_{\G_2} = 16 \{ r^{-1} [r (\xi^2_{-2})^2]' \}'$.
Inserting this into the kernel of $\mathcal{H}^{-2}_2$ and integrating by parts,
\begin{equation}
    \mathcal{H}^{-2}_2[Q_1](q) = \frac{4}{q^3} \int_0^q (q^2 - r^2) \left\{ \frac{1}{r} \left[ r (\xi^2_{-2})^2 \right]' \right\}' \mathrm{d}r
    = \frac{8}{q^3} \int_0^q \left[ r (\xi^2_{-2})^2 \right]' \mathrm{d}r = \frac{8 (\xi^2_{-2})^2}{q^2},
\end{equation}
where the boundary terms vanish because $q^2 - r^2 = 0$ at $r = q$ and $\xi^2_{-2} = \mathcal{O}(r^2)$ as $r \to 0$. Hence
\begin{equation}
    Y_{22} = \frac{18}{49} \frac{(Y^{\lin})^2}{q^2}.
    \label{eq:Y22}
\end{equation}
The shear correlator $\Upsilon_{\G_2, ij}$ needs no transform either.
The coefficients of the decomposition of $\langle \Psi_{i}(\bq_1) s_{mn}(\bq_2) \rangle_{\mathrm{c}}$ in the previous subsection are
\begin{equation}
    \J_2 = -\frac{2}{15} \xi^1_{-1} + \frac{1}{5} \xi^3_{-1}, \qquad
    \J_3 = \frac{1}{5} \left( \xi^1_{-1} + \xi^3_{-1} \right), \qquad
    \J_4 = -\xi^{3}_{-1},
\end{equation}
which satisfy $3 \J_2 + 2 \J_3 + \J_4 = 0$ because $s_{mn}$ is traceless.
Contracting the two cumulants gives $X_{s^2} = 4 \J_3^2$ and $Y_{s^2} = 12 \J_2^2 - 4 \J_3^2$, so that
\begin{equation}
    X_{\G_2} = 4 \J_3^2 = \frac{(Y^{\lin})^{2}}{q^2}, \qquad
    Y_{\G_2} = 12 \J_2^2 - 4 \J_3^2 - \frac43 (U^{\lin})^2,
    \label{eq:Upsilon}
\end{equation}
where the second equality for $X_{\G_{2}}$ uses $\xi^{1}_{-1} + \xi^{3}_{-1} = 5 \xi^{2}_{-2}/q$.

With $x = \hat{\bk} \cdot \hat{\bp}$ and $r = |\bk - \bp|$, the kernel of $\F_{\G_2}$ (Eq.~\eqref{eq:FG2}) can be recasted as 
\begin{equation}
    4 K_{\G_2}(\bk - \bp, \bp) F_2(\bk, -\bp) = -\frac{20}{7} \frac{k^2 (1 - x^2)^2}{r^2} + \frac{2k}{p} x (1 - x^2).
\end{equation}
The second term is odd in $x$ and hence vanishes on angular integration, while the first gives rise to $-\frac{20}{7}$ times the kernel of $R_{1}$. 
Hence $\F_{\G_2} = -\frac{20}{7} R_1$.
Eq.~\eqref{eq:cumulants_H} then gives $U_{\Gamma_3} = -\frac{24}{5} U^{(3)}$ and $\theta_{\Gamma_{3}} = -\frac87 \mathcal{H}^0_0[R_1]$.
Since $[q^2 j_1(kq)]^\prime = k q^2 j_0(kq)$ implies $(q^2 \mathcal{H}^{-1}_1)^\prime = q^2 \mathcal{H}^0_0$, we get
\begin{equation}
    q^2 \theta_{\Gamma_{3}} = \frac{24}{5} \left[ q^2 U^{(3)} \right]'.
    \label{eq:theta_total_derivative}
\end{equation}

The $q$-independent constant terms of $X_{22}$, $X_{13}$ and $X_{10}$ are 
\begin{equation}
    M_{22} = \int \frac{\mathrm{d}k}{2 \pi^{2}} \frac{9}{98} Q_{1}(k),\qquad 
    M_{13} = \int \frac{\mathrm{d}k}{2 \pi^{2}} \frac{5}{21} R_{1}(k),\qquad
    M_{10} = \int \frac{\mathrm{d}k}{2 \pi^{2}} \frac{2}{7} R_1(k).
\end{equation}
Given Eqs.~\eqref{eq:Q1} and \eqref{eq:R1}, it is straightforward to show that $\int \mathrm{d} k~ Q_1(k) = \int \mathrm{d}k~ R_1(k)$.
Thus, we have
\begin{equation}
    M_{13} = \frac{70}{27}M_{22},\qquad M_{10} = \frac{28}{9}M_{22}.
    \label{eq:zero_lag_q}
\end{equation}
We evaluate only $M_{22}$, as the weighted sum of the samples of $\frac{9}{98} Q_1$ (Sec.~\ref{sec:discretization}), and obtain $M_{13}$ and $M_{10}$ from Eq.~\eqref{eq:zero_lag_q}.

The last relation of Eq.~\eqref{eq:H_recurrences} leads to
\begin{equation}
    Y_{\lin} = \frac{2q}{5} \left( \xi^1_{-1} + \xi^3_{-1} \right), \qquad
    \mathcal{H}^{-3}_3 = \frac{q}{7} \left( \mathcal{H}^{-2}_2 + \mathcal{H}^{-2}_4 \right), \qquad
    \mathcal{H}^{-3}_1 = \frac{q}{3} \left( \mathcal{H}^{-2}_0 + \mathcal{H}^{-2}_2 \right).
    \label{eq:lower_n}
\end{equation}
We use these for $Y_{\lin}$, for $T$, which is an $\mathcal{H}^{-3}_3$ transform of a combination of $Q_n$ and $R_n$ (Eq.~\eqref{eq:T}), and for the $\mathcal{H}^{-3}_1$ parts of $V_1$ and $V3$, because the input $k^{n+3}S(k)$ of a transform with lower $n$ weights the low-$k$ end of the source more strongly, where the source is cut, and the resulting error dominates at small $q$.

The identities for $V^{10}_{\G_{2}}$, $Y_{22}$ and the zero-lag constants assume that $Q_n$ and $R_n$ are the full integrals of the input spectrum.
This holds for the zero-padded input that we use for LPT, whose $Q_n$ and $R_n$ vanish beyond $2 k_{\mathrm{max}}$ and $k_{\mathrm{max}}$.

\section{Relation between EPT and LPT at one-loop order}
\label{sec:relation_btw_ept_lpt}

In this appendix, we show explicitly that the one-loop LPT expression reduces to the standard Eulerian result when the exponentiated terms are expanded perturbatively.
For simplicity, we focus on the redshift-space matter power spectrum in LPT up to one-loop order (see also, \textit{e.g.}, \cite{Vlah:2014nta,Vlah:2015sea,Vlah:2015zda} for similar calculations on the real-space power spectrum),
\begin{align}
    \dD(\bk) + P(\bk)
    = \int_{\bq} e^{i \bk \cdot \bq} \avrg{ e^{i \bk \cdot \bDelta} }
    = \int_{\bq} e^{i \bk \cdot \bq}
    \exp \left[ - \frac{1}{2} k_{i} k_{j} A_{ij}(\bq) - \frac{i}{6} k_{i} k_{j} k_{k} W_{ijk}(\bq) \right] + \O(P_{\lin}^{3}).
    \label{eq:lpt_matter_power}
\end{align}
where all contributions relevant up to one-loop order are kept in the exponential.
Expanding the exponential consistently up to one-loop order, Eq.~(\ref{eq:lpt_matter_power}) becomes
\begin{align}
    &\dD(\bk) + P(\bk) \nonumber\\
    &\approx \int_{\bq} e^{i \bk \cdot \bq} \left[ 1
    - \frac{1}{2} k_{i} k_{j} A_{ij}^{\lin}(\bq)
    - \frac{1}{2} k_{i} k_{j} A_{ij}^{\text{1-loop}}(\bq)
    - \frac{i}{6} k_{i} k_{j} k_{k} W_{ijk}(\bq)
    + \frac{1}{8} k_{i} k_{j} k_{k} k_{l} A_{ij}^{\lin}(\bq) A_{kl}^{\lin}(\bq) \right] + \O(P_{\lin}^{3}).
    \label{eq:lpt_matter_power_all_expanded}
\end{align}
This expression reproduces the one-loop Eulerian result.
The individual contributions proportional to $A_{ij}^{\lin}, A_{ij}^{\text{1-loop}}$, and $W_{ijk}$ can be written in terms of the quantities defined in Appendix~\ref{sec:lpt_formulas} as
\begin{align}
    - \frac{1}{2} k_{i} k_{j} \int_{\bq} e^{i \bk \cdot \bq} A_{ij}^{\lin}(\bq)
    &= (1 + f \mu^2)^2 P_{\lin}(k), \\
    - \frac{1}{2} k_{i} k_{j} \int_{\bq} e^{i \bk \cdot \bq} A_{ij}^{\text{1-loop}}(\bq)
    &= \frac{9}{98} (1 + 2 f \mu^2)^2 Q_{1}(k) + \frac{10}{21} (1 + f \mu^2) (1 + 3 f \mu^2) R_{1}(k) + \alpha_{1} (1 + f \mu^{2})^{2} k^{2} P_{\lin}(k), \label{eq:Aij_1loop_fourier} \\
    - \frac{i}{6} k_{i} k_{j} k_{k} \int_{\bq} e^{i \bk \cdot \bq} W_{ijk}(\bq)
    &= \frac{3}{7} (1 + f \mu^2)^2 (1 + 2 f \mu^2) [Q_{2}(k) + 2 R_{2} (k)] \nonumber\\
    &\,\, + \frac{3}{14} f^2 \mu^2 (\mu^2 - 1) [ (1 + 2 f \mu^2) Q_{1}(k) + 4 (1 + f \mu^2) R_{1}(k) ] \nonumber\\
    &\,\, + \left\{ \alpha_{2} [1 + f (2 + f) \mu^{2}] + 2 \alpha_{3} (1 + f \mu^{2}) \right\} (1 + f \mu^{2})^{2} k^{2} P_{\lin}(k),
    \label{eq:Wijk_fourier}
\end{align}
Note that Eqs.~(\ref{eq:Aij_1loop_fourier}) and (\ref{eq:Wijk_fourier}) contain the contributions arising from the EFT counterterms.
The contribution involving $A_{ij}^{\lin} A_{kl}^{\lin}$ in Eq.~(\ref{eq:lpt_matter_power_all_expanded}) becomes
\begin{align}
    \frac{1}{8} k_{i} k_{j} k_{k} k_{l} \int_{\bq} e^{i \bk \cdot \bq} A_{ij}^{\lin}(\bq) A_{kl}^{\lin}(\bq)
    &= \frac12 Q_{3}^{\mathrm{s}}(\bk) - k^{2} (1 + f \mu^2)^2 [1 + f (2 + f) \mu^2] \sigma_{\mathrm{v}}^{2} P_{\lin}(k), \label{eq:AijAkl_fourier}
\end{align}
where we have defined $\sigma_{\mathrm{v}}^{2} \equiv \frac{1}{3} \int \frac{\mathrm{d} p}{2 \pi^{2}} P_{\lin}(p)$ and
\begin{align}
    Q_{3}^{\mathrm{s}}(\bk) &\equiv \int_{\bp}
    \frac{(\bK \cdot \bp)^{2}}{p^{4}}
    \frac{(\bK \cdot (\bk - \bp))^{2}}{|\bk - \bp|^{4}}
    P_{\lin}(p) P_{\lin}(|\bk - \bp|).
    \label{eq:Q3_rsd}
\end{align}
Although this integral depends on $f$ and $\mu$ through $K_{i} = R^{(1)}_{ij} k_{j}$, this dependence can be made explicit by decomposing the tensor structure.
To this end, we write this integral as $Q_{3}^{\mathrm{s}}(\bk) = K_{i} K_{j} K_{k} K_{l} \Q_{3}^{ijkl}(\bk)$ where
\begin{align}
    \Q_{3}^{ijkl}(\bk) &= \int_{\bp_{1},\bp_{2}} \dD(\bk-\bp_{12}) \frac{p_{1i} p_{1j}}{p_{1}^{4}} \frac{p_{2k} p_{2l}}{p_{2}^{4}} P_{\lin}(p_{1}) P_{\lin}(p_{2}).
    \label{eq:Q3_ijkl}
\end{align}
By rotational symmetry as well as the symmetries under the index exchanges $i \leftrightarrow j$, $k \leftrightarrow l$, and $(i,j) \leftrightarrow (k,l)$, Eq.~(\ref{eq:Q3_ijkl}) can be decomposed as
\begin{align}
    \Q_{3}^{ijkl}(\bk) &= \I_{1}(k) \delta_{ij} \delta_{kl} + \I_{2}(k) (\delta_{ik} \delta_{jl} + \delta_{il} \delta_{jk}) + \I_{3}(k) (\delta_{ij} \hat{k}_{k} \hat{k}_{l} + \delta_{kl} \hat{k}_{i} \hat{k}_{j}) \nonumber\\
    &\,\, + \I_{4}(k) (\delta_{ik} \hat{k}_{j} \hat{k}_{l} + \delta_{il} \hat{k}_{j} \hat{k}_{k} + \delta_{jk} \hat{k}_{i} \hat{k}_{l} + \delta_{jl} \hat{k}_{i} \hat{k}_{k}) + \I_{5}(k) \hat{k}_{i} \hat{k}_{j} \hat{k}_{k} \hat{k}_{l}.
\end{align}
The five functions $\I_{i}(k) (i = 1, \cdots, 5)$ are uniquely determined by contracting $\Q_{3}^{ijkl}$ with $\delta_{ij} \delta_{kl}, \delta_{ij} \hat{k}_{k} \hat{k}_{l}, \hat{k}_{i} \hat{k}_{j} \hat{k}_{k} \hat{k}_{l}, \delta_{ik} \delta_{jl},$ and $\delta_{ik} \hat{k}_{j} \hat{k}_{l}$, which yield a system of five linear equations.
In terms of $\I_{i}(k)$, Eq.~(\ref{eq:Q3_rsd}) becomes
\begin{align}
    Q_{3}^{\mathrm{s}}(\bk) &= k^{4} \left\{ [1 + f (2 + f) \mu^2]^{2} [\I_{1}(k) + 2 \I_{2}(k)] + 2 (1 + f \mu^2)^2 [1 + f (2 + f) \mu^2] [\I_{3}(k) + 2 \I_{4}(k)] + (1 + f \mu^2)^4 \I_{5}(k) \right\}.
\end{align}
Collecting these equations, one finds that Eq.~(\ref{eq:lpt_matter_power_all_expanded}) reproduces the one-loop redshift-space EPT result,
\begin{align}
    P(k, \mu)
    &= (1 + f \mu^{2})^{2} P_{\lin}(k) \nonumber\\
    &\,\, + (1 + 2 f \mu^{2}) \left[ \frac{9}{98} (1 + 2 f \mu^2) Q_{1}(k) + \frac{3}{7} (1 + f \mu^{2})^{2} Q_{2}(k) + \frac{3}{14} f^{2} \mu^{2} (\mu^{2} - 1) Q_{1}(k) \right] + \frac{1}{2} Q_{3}^{\mathrm{s}}(k,\mu) \nonumber\\
    &\,\, + (1 + f \mu^{2})  \left[ \frac{10}{21} (1 + 3 f \mu^{2}) R_{1}(k) + \frac{6}{7} (1 + f \mu^{2}) (1 + 2 f \mu^{2}) R_{2}(k) + \frac{6}{7} f^{2} \mu^{2} (\mu^{2} - 1) R_{1}(k) \right] \nonumber\\
    &\,\, - k^{2} (1 + f \mu^2)^2 [1 + f (2 + f) \mu^2] \sigma_{\mathrm{v}}^{2} P_{\lin}(k) \nonumber\\
    &\,\, + \{ \alpha_{1} + \alpha_{2} [1 + f (2 + f) \mu^{2} ] + 2 \alpha_{3} (1 + f \mu^{2}) \} (1 + f \mu^{2})^{2} k^{2} P_{\lin}(k) + \O(P_{\lin}^{3}).
\end{align}
On the right-hand side above, the first line equals the tree-level EPT, while the second line and the third and fourth lines equal the 22- and 13-terms of EPT, respectively.
The last line contains the UV counterterms.

\section{On IR resummation in redshift space}
\label{sec:ir_resum_redshift_space}

\begin{figure}[htbp]
\centering
\includegraphics[width=0.48\linewidth]{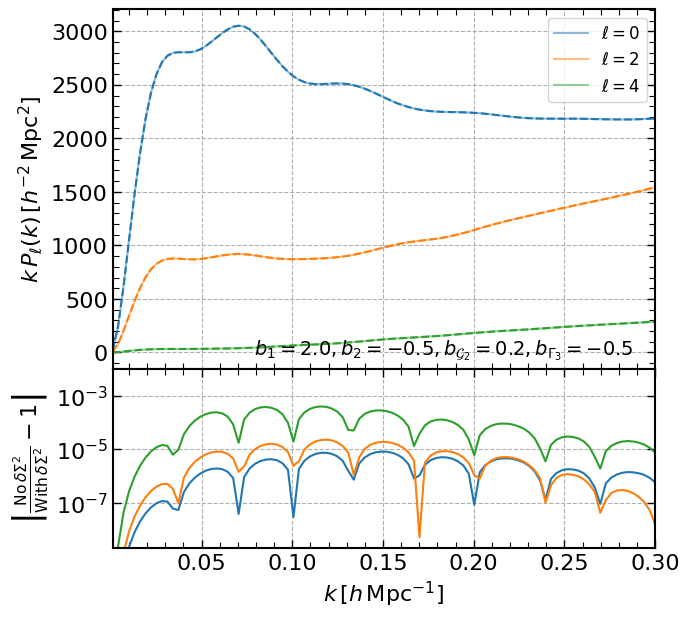}
\caption{
Comparison of the IR-resummed EPT power spectrum multipoles with and without the 
$f^{2} \mu^{2} (\mu^{2} - 1)\delta \Sigma^{2}$ contribution in the redshift-space damping factor.
The upper panel shows the multipoles of $\ell = \{0,2,4\}$ with (solid) and without (dashed) the $\delta \Sigma^{2}$ term, while the lower panel shows the fractional difference between the two cases.
The impact of the $\delta \Sigma^{2}$ term is below the sub-percent level over the scales shown.
}
\label{fig:pkl_with_and_no_delta_Sigma2}
\end{figure}

The IR resummation adopted in this work is based on \cite{Ivanov:2018gjr}. 
The redshift-space damping factor in Eq.~(\ref{eq:damping_redshift_space}) can also be understood from the argument of \cite{Baldauf:2015xfa} applied to redshift-space clustering.
The relative displacement between the two points separated by $\bx$, caused by the long-wavelength fluctuations $\delta(\bx) = \int_{\bp}^{k_{\mathrm{IR}}} \delta(\bp) \cos(\bp \cdot \bx)$, is
\begin{align}
    \Delta \bPsi^{\mathrm{IR}}(\bx) = \int_{\bp}^{k_{\mathrm{IR}}} 2 \delta(\bp) \sin\left(\frac{\bp \cdot \bx}{2}\right) \frac{\bp}{p^{2}}.
\end{align}
In redshift space, we promote the long-wavelength displacement $\Delta \bPsi^{\mathrm{IR}}$ using the mapping (\ref{eq:displacement_rsd}) at linear order,
\begin{align}
    \Delta \Psi^{\mathrm{IR,s}}_{i} = R_{ij}^{(1)} \Delta \Psi^{\mathrm{IR}}_{j} = ( \delta_{ij} + f \hat{n}_{i} \hat{n}_{j} ) \Delta \Psi^{\mathrm{IR}}_{j}.
\end{align}
Assuming Gaussian long-wavelength displacements, the cumulant expansion gives
\begin{align}
    \avrg{ e^{i \bk \cdot \Delta \bPsi^{\mathrm{IR, s}} } } &= \exp \left[ - \frac{1}{2} \avrg{ ( \bk \cdot \Delta \bPsi^{\mathrm{IR, s}} )^{2} } \right].
\end{align}
We evaluate the cumulant in the exponent using the linear theory calculation,
\begin{align}
    \avrg{ ( \bk \cdot \Delta \bPsi^{\mathrm{IR, s}} )^{2} } 
    &= 2 \int_{\bp}^{k_{\mathrm{IR}}} [1 - \cos(\bp \cdot \bx)] \frac{(\bp \cdot \bK)^{2}}{p^{4}} P_{\lin}(p).
    \label{eq:2nd_cumulant_IR}
\end{align}
This is equal to $k_{i} k_{j} A^{\lin,\mathrm{s}}_{ij}(\bx) = k_{i} k_{j} R^{(1)}_{ik} R^{(1)}_{jl} A^{\lin}_{kl}(\bx)$ where $A^{\lin}_{ij}$ is constructed from long-wavelength linear modes with $|\bp| < k_{\mathrm{IR}}$.
Hence, it is approximately equivalent to the long-wavelength part of the displacement correlator kept exponentiated in the LPT power spectrum (\ref{eq:pk_bias_lpt}) in Sec.~\ref{sec:model_lpt}.
Performing the angular integral of Eq.~(\ref{eq:2nd_cumulant_IR}) leads to
\begin{align}
    \avrg{ ( \bk \cdot \Delta \bPsi^{\mathrm{IR, s}} )^{2} } &= 2 \int_{0}^{k_{\mathrm{IR}}} \frac{\mathrm{d}p}{2 \pi^{2}} P_{\lin}(p) \left\{ \frac{1}{3} [ 1 - j_{0}(px) - j_{2}(px) ] K^{2} + j_{2}(px) (\hat{\bx} \cdot \bK)^{2} \right\}.
    \label{eq:2nd_cumulant_IR_radial}
\end{align}
Using $K^{2} = k^{2} [1 + f (2 + f) \mu^2]$ and $(\hat{\bx} \cdot \bK)^{2} \approx (\hat{\bk} \cdot \bK)^{2} = k^{2} (1 + f \mu^{2})^{2}$, and approximating the dominant contribution as arising from separations around the BAO scale, $x \approx r_{\mathrm{BAO}}$, Eq.~(\ref{eq:2nd_cumulant_IR_radial}) reduces to
\begin{align}
    \frac{1}{2} \avrg{ ( \bk \cdot \Delta \bPsi^{\mathrm{IR, s}} )^{2} } &\approx k^{2} \left\{ [1 + f (2 + f) \mu^{2}] \Sigma^{2} + f^{2} \mu^{2} (\mu^{2} - 1) \delta \Sigma^{2} \right\} = k^{2} \Sigma_{\mathrm{s}}^{2}(\mu).
\end{align}
The second term, $f^{2} \mu^{2} (\mu^{2} - 1) \delta \Sigma^{2}$, is often neglected in the literature (\textit{e.g.}, \cite{Chen:2020fxs,Chen:2025aqr}).
Indeed, this term can be safely neglected because it is roughly two orders of magnitude smaller than the first term, $[1 + f(2 + f)\mu^{2}] \Sigma^{2}$ (see Fig.~3 of \cite{Ivanov:2018gjr}).
In Fig.~\ref{fig:pkl_with_and_no_delta_Sigma2}, we compare the IR-resummed EPT power spectrum multipoles of $\ell = \{ 0,2,4 \}$ between the cases with (solid lines in the upper panel) and without (dashed lines) this second piece in the damping factor.
The lower panel shows the fractional difference between the two for each multipole moment, which stays below approximately $0.1\%$.

\section{Relation between \textsc{Hankel} and \textsc{Matrix} methods}
\label{sec:relation_btw_hankel_matrix}

In this work, we implement the one-loop integrals of the EPT power spectrum using two different FFTLog-based methods described in Sec.~\ref{sec:implementation}, \textit{i.e.}, \textsc{Hankel} \cite{Schmittfull:2016jsw} and \textsc{Matrix} \cite{Simonovic:2017mhp}.
In this appendix, we explain the mathematical relation between the two methods.

We would like to rewrite the 22- and 13-type integrals in the \textsc{Hankel} method, \textit{i.e.},
\begin{align}
    P_{22}^{(\ell, n_{1}, n_{2})}(k) &\equiv \int_{\bp} p^{n_{1}} |\bk - \bp|^{n_{2}} \L_{\ell}(\hat{\bp} \cdot \widehat{\bk - \bp}) P_{\lin}(p) P_{\lin}(|\bk - \bp|) = (-1)^{\ell} \int_{0}^{\infty} 4 \pi r^{2} \mathrm{d}r j_{0}(kr) \xi^{\ell}_{n_{1}}(r) \xi^{\ell}_{n_{2}}(r), \label{eq:P22_ln1n2} \\ 
    P_{13}^{(\ell, n, m)}(k) &\equiv k^{m} P_{\lin}(k) \int_{\bp} \frac{p^{n} \L_{\ell}(\hat{\bk} \cdot \hat{\bp}) }{ |\bk - \bp|^{2} } P_{\lin}(p) = k^{m} P_{\lin}(k) \int_{0}^{\infty} r \mathrm{d}r j_{\ell}(kr) \xi^{\ell}_{n}(r), \label{eq:P13_lnm}
\end{align}
in the form of matrix-vector multiplication as in the \textsc{Matrix} method.
Using the complex power-law decomposition of the linear power spectrum (\ref{eq:decomp_pk_lin}), the generalized correlation function (\ref{eq:xi_ln}) becomes
\begin{align}
    \xi^{\ell}_{n}(r) &= \frac{1}{2 \pi^{2}} \sum_{m=-N_{\mathrm{FFT}}/2}^{N_{\mathrm{FFT}}/2-1} c_{m} g_{\ell}(n+3-2\nu_{m}) r^{-(n+3-2\nu_{m})},
    \label{eq:xi_ln_in_g_l}
\end{align}
where we have defined the complex exponent $\nu_{m} \equiv - \frac{1}{2} (\nu + i \eta_{m})$ and the Mellin transform of spherical Bessel functions,
\begin{align}
    g_{\ell}(z) \equiv \int_{0}^{\infty} t^{z-1} j_{\ell}(t) \mathrm{d}t = \sqrt{\pi} 2^{z-2} \frac{ \Gamma(\frac{\ell + z}{2}) }{ \Gamma(\frac{3 + \ell - z}{2}) },
    \label{eq:g_l}
\end{align}
which converges for $-\ell < \Re[z] < 2$.
Using Eqs.~(\ref{eq:xi_ln_in_g_l}) and (\ref{eq:g_l}), the 22-type integral (\ref{eq:P22_ln1n2}) becomes
\begin{align}
    P_{22}^{(\ell, n_{1}, n_{2})}(k) &= k^{3 + (n_{1} + n_{2})} \sum_{m_{1},m_{2}} c_{m_{1}} k^{- 2 \nu_{m_{1}}} c_{m_{2}} k^{- 2 \nu_{m_{2}}} I_{1}^{(\ell)}\left( \nu_{m_{1}} - \frac{n_{1}}{2}, \nu_{m_{2}} - \frac{n_{2}}{2} \right),
\end{align}
where we have defined an analytic formula
\begin{align}
    I_{1}^{(\ell)}(\nu_{1}, \nu_{2}) &\equiv \frac{ (-1)^{\ell} }{8 \pi^{3/2}} \frac{ \Gamma( \frac{\ell + 3}{2} - \nu_{1} ) \Gamma( \frac{\ell+3}{2} - \nu_{2}) \Gamma( \nu_{1} + \nu_{2} - \frac{3}{2} ) }{ \Gamma( \frac{\ell}{2} + \nu_{1} ) \Gamma( \frac{\ell}{2} + \nu_{2} ) \Gamma( 3 - \nu_{1} - \nu_{2} ) }.
    \label{eq:I1_l_nu1_nu2}
\end{align}
For the 13-term, let us consider a more general integral as in \cite{Schmittfull:2016jsw},
\begin{align}
    I_{13}^{(\ell, n_{1}, n_{2})}(k) &\equiv \int_{\bp} p^{n_{1}} |\bk - \bp|^{n_{2}} \L_{\ell}(\hat{\bk} \cdot \hat{\bp}) P_{\lin}(p) = \int_{0}^{\infty} 4 \pi r^{2} \mathrm{d}r j_{\ell}(kr) \xi^{\ell}_{n_{1}}(r) \Xi_{n_{2}}(r),
    \label{eq:13_schmittfull_generalized}
\end{align}
where $\Xi_{n}(r)$ is defined as
\begin{align}
    \Xi_{n}(r) &\equiv \int_{\bk} e^{i \bk \cdot \bbr} k^{n} = \int_{0}^{\infty} \frac{k^{2} \mathrm{d}k}{ 2 \pi^{2} } k^{n} j_{0}(kr) = \frac{1}{2 \pi^{2}} g_{0}(n + 3) r^{-(n + 3)}.
\end{align}
Here, $\Xi_{n}$ differs from $\xi^{\ell}_{n}$ in that it does not contain $P_{\lin}$. 
By inserting Eq.~(\ref{eq:xi_ln_in_g_l}) to Eq.~(\ref{eq:13_schmittfull_generalized}), we obtain
\begin{align}
    I_{13}^{(\ell, n_{1}, n_{2})}(k) &= k^{3 + (n_{1} + n_{2})} \sum_{m} c_{m} k^{-2 \nu_{m}} I_{2}^{(\ell)}\left( \nu_{m} - \frac{n_{1}}{2}, - \frac{n_{2}}{2} \right),
\end{align}
where we have defined another formula,
\begin{align}
    I_{2}^{(\ell)}(\nu_{1}, \nu_{2}) &\equiv \frac{ 1 }{8 \pi^{3/2}} \frac{ \Gamma( \frac{\ell + 3}{2} - \nu_{1} ) \Gamma( \frac{3}{2} - \nu_{2}) \Gamma( \nu_{1} + \nu_{2} + \frac{\ell - 3}{2} ) }{ \Gamma( \frac{\ell}{2} + \nu_{1} ) \Gamma( \nu_{2} ) \Gamma( \frac{\ell}{2} + 3 - \nu_{1} - \nu_{2} ) }.
    \label{eq:I2_l_nu1_nu2}
\end{align}
Using the above, the 13-type integral in the \textsc{Hankel} method (\ref{eq:P13_lnm}) becomes
\begin{align}
    P_{13}^{(\ell, n, m)}(k) &= k^{m} P_{\lin}(k) I_{13}^{(\ell, n, -2)}(k) = k^{1 + (n + m)} P_{\lin}(k) \sum_{m_{1}} c_{m_{1}} k^{- 2 \nu_{m_{1}}} I_{2}^{(\ell)}\left( \nu_{m_{1}} - \frac{n}{2}, 1 \right).
\end{align}

We have seen that the 22- and 13-type integrals in the \textsc{Hankel} method reduce to the matrix operation in the \textsc{Matrix} method.
Both analytic formulas (\ref{eq:I1_l_nu1_nu2}) and (\ref{eq:I2_l_nu1_nu2}) can be viewed as a generalization of the formula arising in the \textsc{Matrix} method \cite{Simonovic:2017mhp},
\begin{align}
    I(\nu_{1}, \nu_{2}) &\equiv \frac{1}{8 \pi^{3/2}} \frac{ \Gamma( \frac{3}{2} - \nu_{1}) \Gamma( \frac{3}{2} - \nu_{2} ) \Gamma( \nu_{1} + \nu_{2} - \frac{3}{2} ) }{ \Gamma( \nu_{1} ) \Gamma( \nu_{2} ) \Gamma( 3 - \nu_{1} - \nu_{2} ) }.
    \label{eq:I_nu1_nu2}
\end{align}
Indeed, Eqs.~(\ref{eq:I1_l_nu1_nu2}) and (\ref{eq:I2_l_nu1_nu2}) reduce to Eq.~(\ref{eq:I_nu1_nu2}) for $\ell = 0$, and one finds
\begin{align}
    \int_{\bp} \frac{\L_{\ell}(\hat{\bp} \cdot \widehat{\bk - \bp})}{p^{2 \nu_{1}} |\bk-\bp|^{2 \nu_{2}}} &= k^{3 - 2 (\nu_{1} + \nu_{2})} I_{1}^{(\ell)}(\nu_{1}, \nu_{2}), \\
    \int_{\bp} \frac{\L_{\ell}(\hat{\bk} \cdot \hat{\bp})}{p^{2 \nu_{1}} |\bk-\bp|^{2 \nu_{2}}} &= k^{3 - 2 (\nu_{1} + \nu_{2})} I_{2}^{(\ell)}(\nu_{1}, \nu_{2}).
\end{align}
Since the dependence on the cosines $\hat{\bp} \cdot \widehat{\bk - \bp}$ and $\hat{\bk} \cdot \hat{\bp}$ can be converted into a linear combination of different rational functions of $\{k, p, |\bk - \bp|\}$, \textit{e.g.}, $\hat{\bk} \cdot \hat{\bp} = \frac{1}{2} \left( \frac{k}{p} + \frac{p}{k} - \frac{|\bk - \bp|^{2}}{k p} \right)$, the above Eqs.~(\ref{eq:I1_l_nu1_nu2}) and (\ref{eq:I2_l_nu1_nu2}) can be expressed as a linear combination of Eq.~(\ref{eq:I_nu1_nu2}) with different arguments $(\nu_{1}, \nu_{2})$.

\section{Wiggle--no-wiggle splitting for IR resummation}
\label{sec:wiggle_splitting}

In Sec.~\ref{sec:ir_resum_impl}, we briefly described the wiggle--no-wiggle splitting of the linear power spectrum used for IR resummation and illustrated how different splitting prescriptions affect the IR-resummed power spectrum at $z=0$.
In this appendix, we summarize the details of the splitting methods adopted in this work.

\begin{figure}[htbp]
\centering
\includegraphics[width=0.48\linewidth]{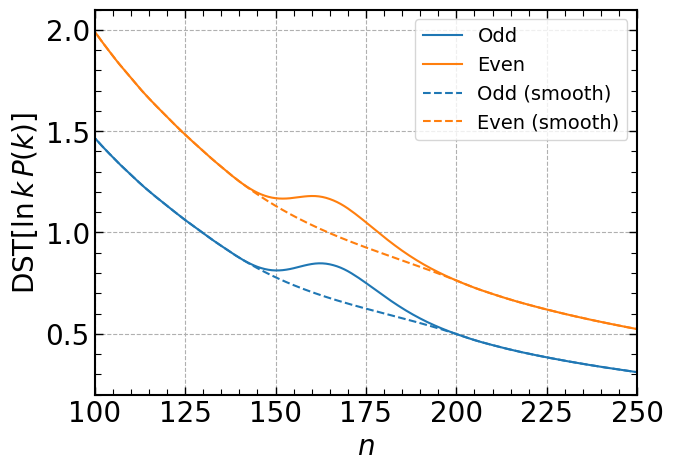}
\caption{
Smoothing of the BAO feature in the linear power spectrum using the DST-II.
The odd (blue) and even (orange) components of the DST of $\ln [k P_{\lin}(k)]$ are shown by solid lines, while the corresponding smoothed components obtained after removing the BAO bump are shown by dashed lines.
}
\label{fig:dst_odd_even_smoothing}
\end{figure}

We implement the wiggle--no-wiggle splitting based on the discrete sine transform (DST), closely following \cite{Hamann:2010pw,Chudaykin:2020aoj}.
Specifically, we sample $\ln [k P_{\lin}(k)]$ on a linearly spaced grid in $k$, measured in units of $\mathrm{Mpc}^{-1}$, using $2^{15}$ bins over the range $k \in [7\times10^{-5},\,7]\,\mathrm{Mpc}^{-1}$, and then perform the DST.
In practice, the DST is implemented by combining the \texttt{jax.scipy} implementation of the discrete cosine transform of type II (DCT-II) with the standard relation between DCT-II and DST-II.

The odd- and even-indexed coefficients of the transformed array trace two smooth envelopes, as illustrated in Fig.~\ref{fig:dst_odd_even_smoothing}, where $n$ denotes the index of the DST coefficient.
We identify the localized feature in the original DST coefficients (solid lines) over the interval $n \in [140,200]$ with the BAO bump, and remove it by cubic-spline interpolation of the odd and even envelopes after excluding the coefficients in this interval, as indicated by the dashed lines.

The BAO sound horizon is accurately approximated by \cite{BOSS:2014hhw}
\begin{align}
    r_{\mathrm{s}} \approx \frac{55.154 e^{-72.3 (\omega_{\nu} + 0.0006)^2}}{\omega_{\mathrm{b}}^{0.12807} \omega_{\mathrm{cb}}^{0.25351}} \, \mathrm{Mpc},
\end{align}
where $\omega_{\nu}$ denotes the physical density parameter of massive neutrinos.
Since $r_{\mathrm{s}}$ depends only weakly on cosmology when expressed in units of $\mathrm{Mpc}$, the location of the BAO bump in the DST coefficients is nearly cosmology independent within our setup.
We therefore adopt the fixed interval $n \in [140,200]$ for removing the BAO bump for all cosmologies considered in this work.

As a simpler alternative, we also implement the wiggle--no-wiggle splitting using a Savitzky--Golay (SG) filter \cite{1964AnaCh..36.1627S}.
For this purpose, we sample $\ln P_{\lin}(k)$ on 256 logarithmically spaced $k$ bins over the range $k \in [10^{-4},\,10]\,h\,\mathrm{Mpc}^{-1}$ and smooth it with an SG filter using a cubic polynomial and a 49-point window.
This choice is empirical, but the implementation is simple and computationally inexpensive, and it provides a satisfactory separation between the smooth and oscillatory parts of the spectrum.

\begin{figure}[htbp]
\centering
\includegraphics[width=\linewidth]{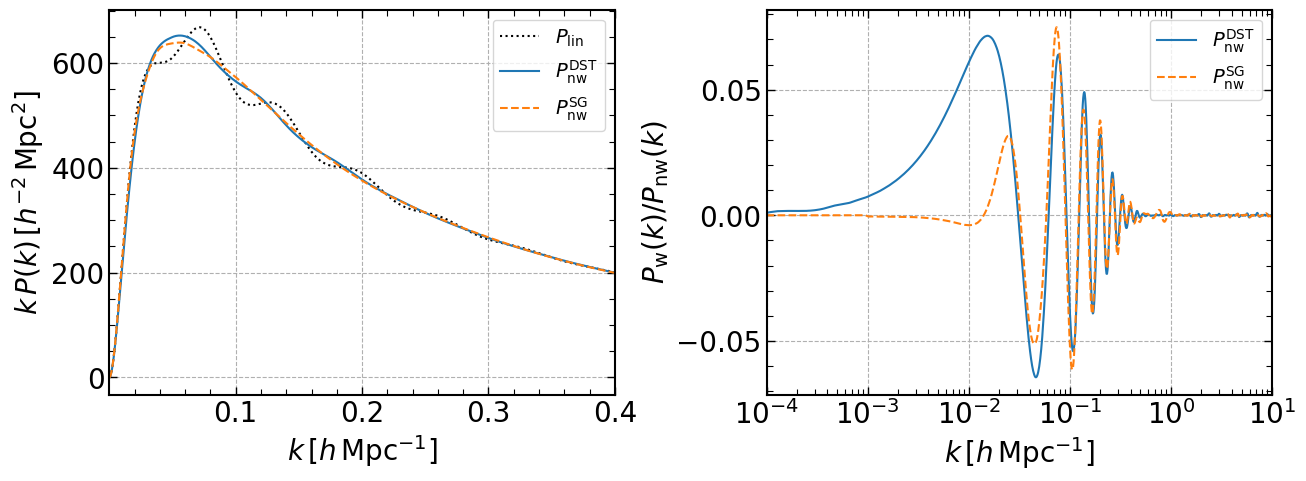}
\caption{
Comparison of the wiggle--no-wiggle splitting of the linear power spectrum using the discrete sine transform (DST) and the Savitzky--Golay (SG) filter.
Left: the linear power spectrum $P_{\lin}(k)$ and the corresponding no-wiggle spectra obtained with the two methods, $P_{\mathrm{nw}}^{\mathrm{DST}}(k)$ and $P_{\mathrm{nw}}^{\mathrm{SG}}(k)$.
Right: the corresponding wiggle components, shown as $P_{\mathrm{w}}(k)/P_{\mathrm{nw}}(k)$ with $P_{\mathrm{w}}(k) \equiv P_{\lin}(k) - P_{\mathrm{nw}}(k)$.
}
\label{fig:wiggle_split_methods}
\end{figure}

Figure~\ref{fig:wiggle_split_methods} compares the wiggle--no-wiggle decomposition of the linear power spectrum obtained with the DST and SG methods.
The resulting no-wiggle spectra are broadly consistent, while the extracted wiggle component shows noticeable differences at very low $k$.

\begin{figure}[htbp]
\centering
\includegraphics[width=1\linewidth]{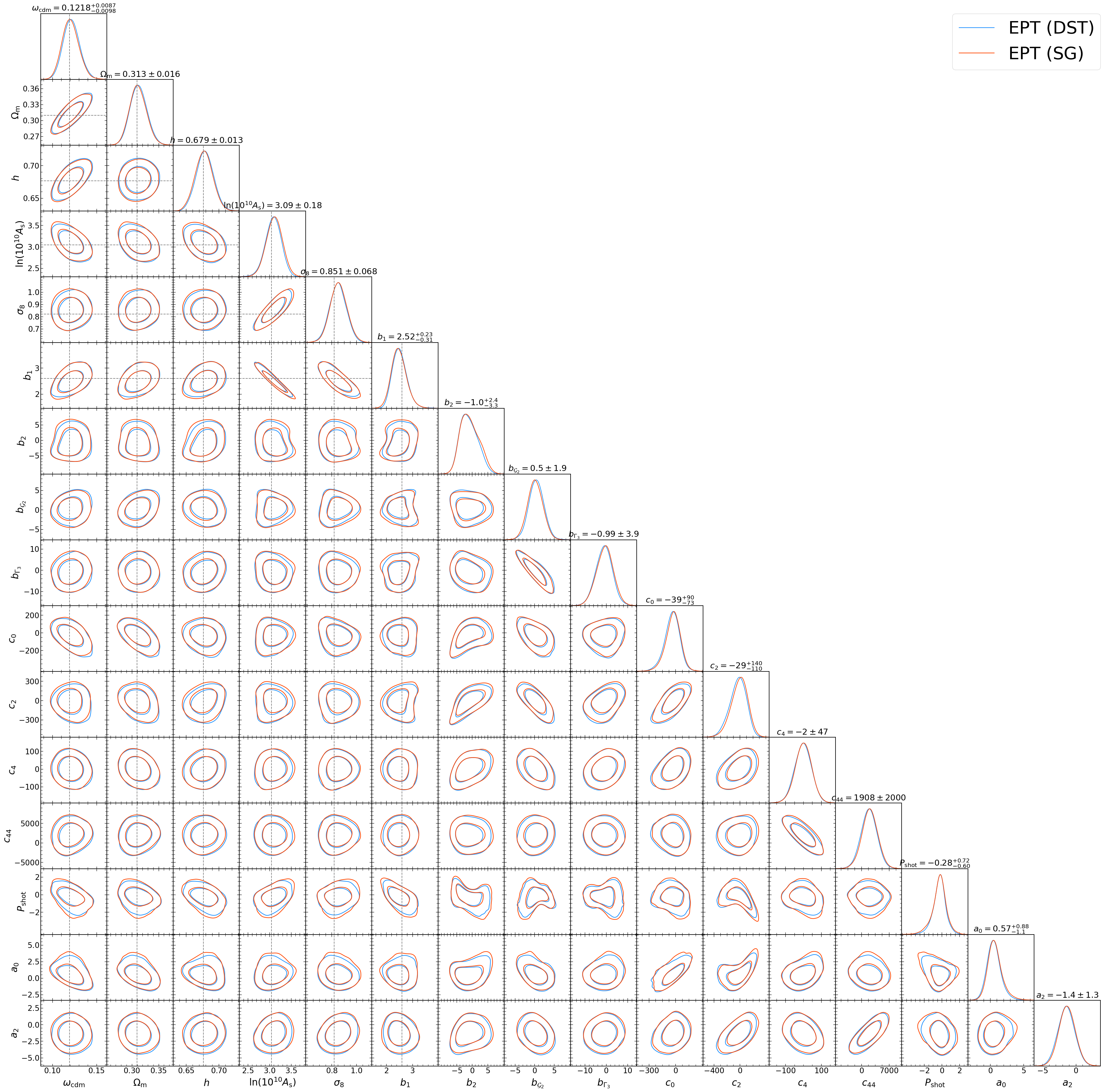}
\caption{
Same as Fig.~\ref{fig:nuts_ept_vs_lpt_halo}, but showing a comparison between two EPT models in which the IR resummation is performed using either the DST-based method (blue; identical to the blue contours in Fig.~\ref{fig:nuts_ept_vs_lpt_halo}) or the SG filter (orange).
}
\label{fig:nuts_ept_DST_vs_SG}
\end{figure}

As a validation test for parameter inference, Fig.~\ref{fig:nuts_ept_DST_vs_SG} compares the posterior distributions inferred from the halo power spectrum using the EPT model, where the IR resummation is performed either with the DST-based method (blue) or with the SG filter (orange).
The two sets of posterior distributions are nearly identical over the full parameter space, indicating that our inference results are robust to the choice of wiggle-splitting method used for IR resummation.

\section{Another option for LPT counterterms}
\label{sec:lpt_ctr}

In Sec.~\ref{sec:ctr_and_stoch}, we have described the default counterterm implementation.
Following the standard convention \cite{Chen:2019lpf,Chen:2020zjt}, we also implement another form of counterterms for the LPT that uses the Zel'dovich power spectrum in place of the linear power spectrum, \textit{i.e.},
\begin{align}
    P_{\mathrm{ctr}}^{\mathrm{LPT}}(\bk) &= - 2 \left( c_{0}^{\mathrm{L}} + c_{2}^{\mathrm{L}} f \mu^2 + c_{4}^{\mathrm{L}} f^{2} \mu^4 + c_{6}^{\mathrm{L}} f^{3} \mu^{6} \right) k^{2} P^{\mathrm{ZA}}(\bk),
    \label{eq:pkmu_lpt_ctr} \\
    P_{\mathrm{ctr, NLO}}^{\mathrm{LPT}}(\bk) &= -2 \left( c_{44}^{\mathrm{L}} + c_{46}^{\mathrm{L}}  f \mu^{2} + c_{48}^{\mathrm{L}} f^{2} \mu^{4} \right) (k \mu f)^{4} P^{\mathrm{ZA}}(\bk).
    \label{eq:pkmu_lpt_ctr_nlo}
\end{align}
where $P^{\mathrm{ZA}}(\bk)$ is the Zel'dovich power spectrum in redshift space.

However, the above formulas and Eqs.~(\ref{eq:pkmu_ept_ctr}) and (\ref{eq:pkmu_ept_ctr_nlo}) lead to a complicated correspondence between the counterterms in both models.
For example, suppose that one considers a linear mapping between $\bc = (c_{0}, c_{2}, c_{4})^{T}$ and $\bc^{\mathrm{L}} = (c_{0}^{\mathrm{L}}, c_{2}^{\mathrm{L}}, c_{4}^{\mathrm{L}})^{T}$ (fixing $c_{6} = c_{6}^{\mathrm{L}} = 0$) that ensures that the three multipoles of the counterterm $\bP_{\mathrm{ctr}} = ( P_{\mathrm{ctr}, 0}(k), P_{\mathrm{ctr}, 2}(k), P_{\mathrm{ctr}, 4}(k) )^{T}$ based on Eqs.~(\ref{eq:pkmu_ept_ctr}) and (\ref{eq:pkmu_lpt_ctr}) are similar to each other.
Since $P^{\mathrm{ZA}}(\bk) \approx (1 + f \mu^{2})^{2} P_{\lin}(k)$ on large scales, we have
\begin{align}
    \bP^{\mathrm{EPT}}_{\mathrm{ctr}} = -2 k^{2} P_{\lin}(k) \times R^{\mathrm{E}}(f) \bc \text{ and }
    \bP^{\mathrm{LPT}}_{\mathrm{ctr}} \approx -2 k^{2} P_{\lin}(k) \times R^{\mathrm{L}}(f) \bc^{\mathrm{L}},
\end{align}
where the matrices that represent the response of the power spectrum multipoles to the counterterm parameters are defined as
\begin{align}
    R^{\mathrm{E}}(f) \equiv \begin{pmatrix}
        1 & \frac{1}{3} f & \frac{1}{5} f^{2} \\
        0 & \frac{2}{3} f & \frac{4}{7} f^{2} \\
        0 & 0 & \frac{8}{35} f^{2}
    \end{pmatrix}
    \text{ and }
    R^{\mathrm{L}}(f) \equiv \begin{pmatrix}
        1 + \frac{2}{3} f + \frac{1}{5} f^{2} & \frac{1}{3} f + \frac{2}{5} f^{2} + \frac{1}{7} f^{3} & \frac{1}{5} f^{2} + \frac{2}{7} f^{3} + \frac{1}{9} f^{4} \\
        \frac{4}{3} f + \frac{4}{7} f^{2} & \frac{2}{3} f + \frac{8}{7} f^{2} + \frac{10}{21} f^{3} & \frac{4}{7} f^{2} + \frac{20}{21} f^{3} + \frac{40}{99} f^{4} \\
        \frac{8}{35} f^{2} & \frac{16}{35} f^{2} + \frac{24}{77} f^{3} & \frac{8}{35} f^{2} + \frac{48}{77} f^{3} + \frac{48}{143} f^{4}
    \end{pmatrix}.
\end{align}
Thus, by requiring that $\bP^{\mathrm{EPT}}_{\mathrm{ctr}} \approx \bP^{\mathrm{LPT}}_{\mathrm{ctr}}$ on large scales, we obtain the linear mapping between EPT and LPT counterterm parameters,
\begin{align}
    \bc^{\mathrm{L}} = \left( R^{\mathrm{L}}(f) \right)^{-1} R^{\mathrm{E}}(f) \bc.
    \label{eq:mapping_btw_ept_lpt_ctr_params}
\end{align}
Although one can employ this kind of mapping, using different counterterms between the EPT and LPT complicates parameter inference across the EPT and LPT, and hence we adopt the counterterms in Sec.~\ref{sec:ctr_and_stoch} as the default.

\section{Derivation of the closed-form expression for $G_{\ell}^{(0,0)}$}
\label{sec:G00_closed_form}

In Sec.~\ref{sec:angular_int_lpt}, we have described the implementation of the angular integrals of the LPT power spectrum in redshift space.
It is based on the analytic integral formulas $G_{\ell}^{(m,n)}$ in Eq.~(\ref{eq:Imn_formula}), which is constructed from $G_{\ell}^{(0,0)}$ through the recursion relations, Eq.~(\ref{eq:Gmn_recursion}).
In this appendix, we show that the analytic expression of $G_{\ell}^{(0,0)}$, Eq.~(\ref{eq:G00_vlah_white}), which was derived by \cite{Vlah:2018ygt}, reduces to the closed form Eq.~(\ref{eq:G00_closed_form}) as follows.

Using Euler's transformation, the hypergeometric function part of Eq.~(\ref{eq:G00_vlah_white}) becomes
\begin{align}
    {_2F_1}\left(\frac{1}{2}-n, -n; \frac{1}{2} - \ell - n; z \right) = (1-z)^{n} {_2F_1}\left(-\ell, -n; \frac{1}{2} - \ell - n; \frac{z}{z-1} \right).
\end{align}
For non-negative integers $\ell$ and $n$ with $\ell \le n$, the right-hand side reduces to a polynomial of degree $\ell$,
\begin{align}
    {_2F_1}\left(-\ell, -n; \frac{1}{2} - \ell - n; x \right) &= \sum_{k=0}^{\ell} \frac{(-\ell)_{k} (-n)_{k}}{(\frac{1}{2} - \ell - n)_{k} k!} x^{k},
\end{align}
where $(x)_{k}$ is the Pochhammer symbol.
Using these, Eq.~(\ref{eq:G00_vlah_white}) becomes
\begin{align}
    G_{\ell}^{(0,0)}(A, B, C)
    &= \sum_{k=0}^{\ell} \sum_{n=\ell}^{\infty} \frac{\Gamma(\ell + n + \frac{1}{2})}{\Gamma(\ell + 1) \Gamma(n + \frac{1}{2}) \Gamma(1-\ell+n)} \frac{(-\ell )_{k} (-n)_{k}}{(\frac{1}{2}-\ell-n)_{k} k!} \left( - B \right)^{n} \left( \frac{C^2}{\rho^2} \right)^{n-k} \nonumber\\
    &= \sum_{k=0}^{\ell} \frac{(-1)^{-k} \ell!}{[(\ell-k)!]^{2} k!}\frac{\Gamma(\frac{1}{2}+2\ell-k)}{\Gamma(\frac{1}{2}+\ell)} \pFq{2}{2}{1 + \ell, \frac{1}{2} - k + 2 \ell}{\frac{1}{2} + \ell, 1 - k + \ell}{\frac{- B C^2}{\rho^2}} \left( - B \right)^{\ell} \left( \frac{C^2}{\rho^2} \right)^{\ell - k},
    \label{eq:G00_in_terms_of_2F2}
\end{align}
where ${_2F_2}$ is the generalized hypergeometric function.
Here, for non-negative integers $\ell$ and $n$ with $\ell \le n$, the ${_2F_2}$ term in the above reduces to the following form
\begin{align}
    \pFq{2}{2}{1 + \ell, \frac{1}{2} - k + 2 \ell}{\frac{1}{2} + \ell, 1 - k + \ell}{z} &= e^{z} \sum_{i=0}^{\ell} \tilde{c}_{(\ell,k,i)} z^{i} = \sum_{\ell=0}^{\infty} \sum_{i=0}^{\ell} \frac{\tilde{c}_{(\ell,k,i)}}{\ell!} z^{i+\ell} = 
    \sum_{n=0}^{\infty} \sum_{i=0}^{\max(n,\ell)}  \frac{\tilde{c}_{(\ell,k,i)}}{(n-i)!} z^{n},
    \label{eq:2F2_series_for_intergers_ln}
\end{align}
where the coefficient $\tilde{c}_{(\ell,k,i)}$ for each $(\ell, k, i)$ is a rational number, which is determined as follows.
By the definition of the generalized hypergeometric functions, we can also write
\begin{align}
    \pFq{2}{2}{1 + \ell, \frac{1}{2} - k + 2 \ell}{\frac{1}{2} + \ell, 1 - k + \ell}{z} &= \sum_{n=0}^{\infty} \frac{ (1 + \ell)_{n} (\frac{1}{2} - k + 2 \ell)_{n} }{ (\frac{1}{2} + \ell)_{n} (1 - k + \ell)_{n} n! } z^n \equiv \sum_{n=0}^{\infty} a_{(\ell, k, n)} z^{n}.
    \label{eq:2F2_series_by_definition}
\end{align}
Comparing the coefficients between these two series expansions (\ref{eq:2F2_series_for_intergers_ln}) and (\ref{eq:2F2_series_by_definition}) leads to
\begin{align}
    \sum_{i=0}^{\max(n,\ell)}  \frac{\tilde{c}_{(\ell, k, i)}}{(n-i)!} = a_{(\ell, k, n)} 
    \Rightarrow a_{(\ell, k, i)} 
    = \sum_{j=0}^{i} \frac{1}{(i-j)!} \tilde{c}_{(\ell, k, j)},
\end{align}
We can inversely solve this to find $\tilde{c}_{(m,k,i)}$ to be
\begin{align}
    \tilde{c}_{(\ell, k, i)} = \sum_{j=0}^{i} \frac{(-1)^{i-j}}{(i-j)!} a_{(\ell, k, j)} = \frac{(-1)^{i}}{i!} \pFq{3}{2}{1 + \ell, \frac{1}{2} + 2 \ell - k, -i}{1 + \ell - k, \frac{1}{2} + \ell}{1}.
    \label{eq:2F2_coeffs}
\end{align}
Thus, using Eqs.~(\ref{eq:G00_in_terms_of_2F2}), (\ref{eq:2F2_series_for_intergers_ln}), and (\ref{eq:2F2_coeffs}), we arrive at the final expression (\ref{eq:G00_closed_form}), with the coefficients $c_{(\ell, k, i)}$ being
\begin{align}
    c_{(\ell, k, i)} 
    &= \frac{(-1)^{i-k} \ell!}{[(\ell-k)!]^2 k! i!} \frac{\Gamma(\frac{1}{2} + 2 \ell - k) }{\Gamma(\frac{1}{2} + \ell)} \pFq{3}{2}{1 + \ell, \frac{1}{2} + 2 \ell - k, -i}{1 + \ell - k, \frac{1}{2} + \ell}{1} \nonumber\\
    &= \frac{ (-1)^{i-k} }{ (\ell-k)! i! } \sum_{n=0}^{i} (-1)^{n} \binom{i}{n} \binom{\ell+n}{k} \left( \ell + n + \frac{1}{2} \right)_{\ell - k}.
\end{align}

\section{Relation between Eulerian and Lagrangian biases up to third order}
\label{sec:relation_btw_eulerian_and_largangian_biases}

In Sec.~\ref{sec:likelihood_and_priors}, we have used the relation between the Eulerian and Lagrangian biases to impose consistent priors on the bias parameters for the parameter inference.
In this appendix, we derive the relation, starting from the Lagrangian bias model (see also \textit{e.g.}, \cite{Mirbabayi:2014zca,Abidi:2018eyd} for similar calculations).
We model the density field in the initial Lagrangian space at some initial time $\tau_{0}$ that is going to collapse to the biased tracers as
\begin{align}
    \delta_{\mathrm{g}}(\bq, \tau_{0}) &= b_{1}^{\mathrm{L}}(\tau_{0}) \delta(\bq, \tau_{0}) + \frac{b_{2}^{\mathrm{L}}(\tau_{0})}{2} \left[ \delta^{2}(\bq,\tau_{0}) - \sigma^{2}(\tau_{0}) \right] + b_{\G_{2}}^{\mathrm{L}}(\tau_{0}) \G_{2}(\bq, \tau_{0}) \nonumber\\
    &\,\, + \frac{ b_{3}^{\mathrm{L}}(\tau_{0}) }{6} \left[ \delta^{3}(\bq,\tau_{0}) - 3 \sigma^{2}(\tau_{0}) \delta(\bq,\tau_{0}) \right] + b_{\G_{3}}^{\mathrm{L}}(\tau_{0}) \G_{3}(\bq, \tau_{0}) \nonumber\\
    &\,\, + b_{\delta \G_{2}}^{\mathrm{L}}(\tau_{0}) \left[ \delta(\bq, \tau_{0}) \G_{2}(\bq, \tau_{0}) + \frac{4}{3} \sigma^{2}(\tau_{0}) \delta(\bq,\tau_{0}) \right] + b_{\Gamma_{3}}^{\mathrm{L}}(\tau_{0}) \Gamma_{3}(\bq,\tau_{0}),
    \label{eq:lagrangian_bias_model}
\end{align}
up to third order of the initial linear density field $\delta(\bq, \tau_{0})$.
In the above, we construct the renormalized operators \cite{Assassi:2014fva} in the Lagrangian field.
To relate this field to the Eulerian tracer density field at later time $\tau$, we use the mass conservation, \textit{i.e.}, $\left[ 1 + \delta(\bx, \tau) \right] \mathrm{d}^{3}x = \mathrm{d}^{3}q$, as well as the conservation of the number of the (proto-)tracers, \textit{i.e.}, $\left[ 1 + \delta_{\mathrm{g}}(\bx, \tau) \right] \mathrm{d}^{3}x = \left[ 1 + \delta_{\mathrm{g}}(\bq, \tau_{0}) \right] \mathrm{d}^{3}q$.
Those two means
\begin{align}
    \delta_{\mathrm{g}}(\bx, \tau) &= \left[ 1 + \delta_{\mathrm{g}}(\bq, \tau_{0}) \right] \left[ 1 + \delta(\bx, \tau) \right] - 1.
    \label{eq:eulerian_tracer_field_in_terms_of_lagrangian_tracer_field_and_eulerian_matter_density}
\end{align}
Note that $\delta(\bx, \tau)$ is the full nonlinear matter density field at a Eulerian position $\bx$ and at time $\tau$.
To proceed with the right-hand side of Eq.~(\ref{eq:eulerian_tracer_field_in_terms_of_lagrangian_tracer_field_and_eulerian_matter_density}), we first rewrite the proto-tracer field in Lagrangian space $\delta_{\mathrm{g}}(\bq, \tau_{0})$ in terms of operators at the later time $\tau$.
The $n$th-order Lagrangian bias operator $O^{(n)}$ evolves as $O^{(n)}(\bq, \tau) = \left[ \frac{D(\tau)}{D(\tau_{0})} \right]^{n} O^{(n)}(\bq, \tau_{0})$, where $D(\tau)$ is the linear growth factor at time $\tau$, and accordingly we can write the Lagrangian bias parameters at time $\tau$ as $b_{O^{(n)}}^{\mathrm{L}}(\tau) = \left[ \frac{D(\tau_{0})}{D(\tau)} \right]^{n} b_{O^{(n)}}^{\mathrm{L}}(\tau_{0})$.
Thence we obtain
\begin{align}
    \delta_{\mathrm{g}}(\bq, \tau_{0}) &= b_{1}^{\mathrm{L}}(\tau) \delta(\bq, \tau) + \frac{b_{2}^{\mathrm{L}}(\tau)}{2} \left[ \delta^{2}(\bq,\tau) - \sigma^{2}(\tau) \right] + b_{\G_{2}}^{\mathrm{L}}(\tau) \G_{2}(\bq, \tau) \nonumber\\
    &\,\, + \frac{ b_{3}^{\mathrm{L}}(\tau) }{6} \left[ \delta^{3}(\bq, \tau) - 3 \sigma^{2}(\tau) \delta(\bq, \tau) \right] + b_{\G_{3}}^{\mathrm{L}}(\tau) \G_{3}(\bq, \tau) \nonumber\\
    &\,\, + b_{\delta \G_{2}}^{\mathrm{L}}(\tau) \left[ \delta(\bq, \tau) \G_{2}(\bq, \tau) + \frac{4}{3} \sigma^{2}(\tau) \delta(\bq, \tau) \right] + b_{\Gamma_{3}}^{\mathrm{L}}(\tau) \Gamma_{3}(\bq, \tau).
\end{align}
Second, we take into account the difference between the Lagrangian position $\bq$ and the Eulerian position $\bx$, which is described in terms of the displacement $\bPsi(\bq, \tau) = \bx - \bq$.
For the calculation up to third order, we consider the shifts in $\delta$ and $\G_{2}$,
\begin{align}
    \delta(\bq, \tau) &= \delta^{(1)}(\bx, \tau) - H_{\delta}^{(2)}(\bx, \tau) - H_{\delta}^{(3)}(\bx, \tau), \label{eq:shift_for_delta} \\
    \G_{2}(\bq, \tau) &= \G_{2}^{(2)}(\bx, \tau) - H_{\G_{2}}^{(3)}(\bx, \tau), \label{eq:shift_for_G2}
\end{align}
where we have put the superscripts on each term in the right-hand side to indicate its perturbative order, and we have defined the second- and third-order shift terms of $\delta$ and the third-order shift term of $\G_{2}$ as
\begin{align}
    H_{\delta}^{(2)}(\bx, \tau) &= \Psi^{(1)}_{i}(\bx, \tau) \nabla_{i} \delta^{(1)}(\bx, \tau), \\
    H_{\delta}^{(3)}(\bx, \tau) &= \Psi^{(2)}_{i}(\bx, \tau) \nabla_{i} \delta^{(1)}(\bx, \tau) - \Psi^{(1)}_{i}(\bx, \tau) \left[ \nabla_{i} \Psi^{(1)}_{j}(\bx,\tau) \right] \nabla_{j} \delta^{(1)}(\bx, \tau) - \frac{1}{2} \Psi^{(1)}_{i}(\bx, \tau) \Psi^{(1)}_{j}(\bx,\tau) \nabla_{i} \nabla_{j} \delta^{(1)}(\bx, \tau), \\
    H_{\G_{2}}^{(3)}(\bx, \tau) &= \Psi^{(1)}_{i}(\bx, \tau) \nabla_{i} \G_{2}^{(2)}(\bx, \tau).
\end{align}
Each of these shift terms is expressed by a linear combination of the matter density perturbations and the Eulerian bias operators at each order.
After some algebra, one can find
\begin{align}
    H_{\delta}^{(2)}(\bx) &= - \delta^{(2)}(\bx) + \left[ \delta^{(1)}(\bx) \right]^{2} + \frac{2}{7} \G_{2}^{(2)}(\bx), \\
    H_{\delta}^{(3)}(\bx) &= - \delta^{(3)}(\bx) - \left[ \delta^{(1)}(\bx) \right]^{3} + 2 \delta^{(1)}(\bx) \delta^{(2)}(\bx) + \frac{4}{7} \G_{2}^{(3)}(\bx) - \frac{2}{7} \delta^{(1)}(\bx) \G_{2}^{(2)}(\bx) + \frac{22}{63} \G_{3}^{(3)}(\bx) - \frac{23}{42} \Gamma_{3}^{(3)}(\bx), \\
    H_{\G_{2}}^{(3)}(\bx) &= - 2 \G_{2}^{(3)}(\bx) + \delta^{(1)}(\bx) \G_{2}^{(2)}(\bx) - 2 \G_{3}^{(3)}(\bx) + \frac{5}{2} \Gamma_{3}^{(3)}(\bx).
\end{align}
All in all, the tracer density field in Eulerian coordinates Eq.~(\ref{eq:eulerian_tracer_field_in_terms_of_lagrangian_tracer_field_and_eulerian_matter_density}) is perturbatively expanded as
\begin{align}
    \delta_{\mathrm{g}}(\bx, \tau) &= \delta_{\mathrm{g}}^{(1)}(\bx, \tau) + \delta_{\mathrm{g}}^{(2)}(\bx, \tau) + \delta_{\mathrm{g}}^{(3)}(\bx, \tau) + \cdots,
\end{align}
where the contributions up to third order are
\begin{align}
    \delta_{\mathrm{g}}^{(1)}(\bx,\tau) &= \left[ 1 + b_{1}^{\mathrm{L}}(\tau) \right] \delta^{(1)}(\bx, \tau), \\
    \delta_{\mathrm{g}}^{(2)}(\bx,\tau) &= \left[ 1 + b_{1}^{\mathrm{L}}(\tau) \right] \delta^{(2)}(\bx, \tau) + \frac{ b_{2}^{\mathrm{L}}(\tau) }{ 2 } \left\{ \left[ \delta^{(1)}(\bx, \tau) \right]^{2} - \sigma^{2}(\tau) \right\} + \left[ b_{\G_{2}}^{\mathrm{L}}(\tau) - \frac{2}{7} b_{1}^{\mathrm{L}}(\tau) \right] \G_{2}^{(2)}(\bx, \tau), \\
    \delta_{\mathrm{g}}^{(3)}(\bx, \tau) &= \left[ 1 + b_{1}^{\mathrm{L}}(\tau) \right] \delta^{(3)}(\bx, \tau) + \left[ \frac{1}{6} b_{3}^{\mathrm{L}}(\tau) - \frac{1}{2} b_{2}^{\mathrm{L}}(\tau) \right] \left\{ \left[ \delta^{(1)}(\bx, \tau) \right]^{3} - 3 \sigma^{2}(\tau) \delta^{(1)}(\bx,\tau) \right\} \nonumber\\
    &\,\, + b_{2}^{\mathrm{L}}(\tau) \left\{ \delta^{(1)}(\bx, \tau) \delta^{(2)}(\bx, \tau) - \frac{34}{21} \sigma^{2}(\tau) \delta^{(1)}(\bx,\tau) \right\} \nonumber\\
    &\,\, + \left[ b_{\G_{3}}^{\mathrm{L}}(\tau) - \frac{22}{63} b_{1}^{\mathrm{L}}(\tau) - b_{\G_{2}}^{\mathrm{L}}(\tau) \right] \G_{3}^{(3)}(\bx, \tau) + \left[ b_{\delta \G_{2}}^{\mathrm{L}}(\tau) - \frac{2}{7} b_{2}^{\mathrm{L}}(\tau) \right] \left\{ \delta^{(1)}(\bx, \tau) \G_{2}^{(2)}(\bx, \tau) + \frac{4}{3} \sigma^{2}(\tau) \delta^{(1)}(\bx,\tau) \right\} \nonumber\\
    &\,\, + \left[ b_{\Gamma_{3}}^{\mathrm{L}}(\tau) + \frac{23}{42} b_{1}^{\mathrm{L}}(\tau) - \frac{5}{2} b_{\G_{2}}^{\mathrm{L}}(\tau) \right] \Gamma_{3}^{(3)}(\bx, \tau) + 2 \left[ b_{\G_{2}}^{\mathrm{L}}(\tau) - \frac{2}{7} b_{1}^{\mathrm{L}}(\tau) \right] \G_{2}^{(3)}(\bx, \tau).
    \label{eq:eulerian_lagrangian_map_3rd_order}
\end{align}
From these, one can read out the Eulerian bias parameters in terms of Lagrangian bias parameters, \textit{e.g.},
\begin{align}
    &b_{1} = 1 + b_{1}^{\mathrm{L}}, 
    b_{2} = b_{2}^{\mathrm{L}}, 
    b_{\G_{2}} = b_{\G_{2}}^{\mathrm{L}} - \frac{2}{7} b_{1}^{\mathrm{L}}, 
    b_{3} = b_{3}^{\mathrm{L}} - 3 b_{2}^{\mathrm{L}}, 
    b_{\G_{3}} = b_{\G_{3}}^{\mathrm{L}} - \frac{22}{63} b_{1}^{\mathrm{L}} - b_{\G_{2}}^{\mathrm{L}}, \nonumber\\
    &b_{\delta \G_{2}} = b_{\delta \G_{2}}^{\mathrm{L}} - \frac{2}{7} b_{2}^{\mathrm{L}}, 
    b_{\Gamma_{3}} = b_{\Gamma_{3}}^{\mathrm{L}} + \frac{23}{42} b_{1}^{\mathrm{L}} - \frac{5}{2} b_{\G_{2}}^{\mathrm{L}}.
\end{align}

\section{Comparison of LPT terms to brute-force integration}
\label{sec:comarison_to_bf_lpt}

\begin{figure}[htbp]
\centering
\begin{minipage}{0.48\linewidth}
\centering
\includegraphics[width=\linewidth]{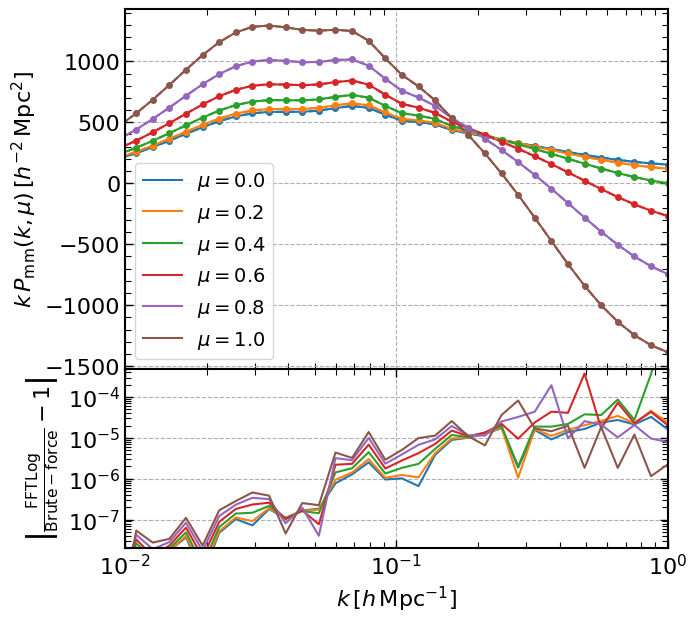}
\end{minipage}
\begin{minipage}{0.48\linewidth}
\centering
\includegraphics[width=\linewidth]{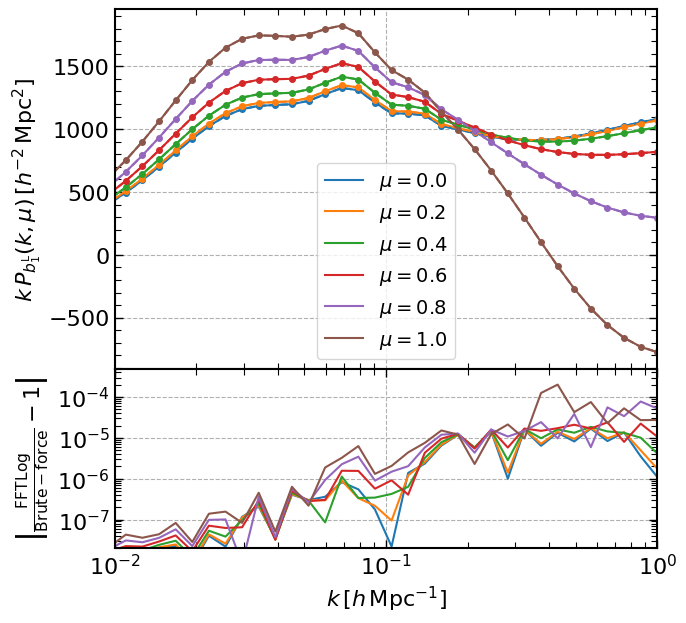}
\end{minipage}
\begin{minipage}{0.48\linewidth}
\centering
\includegraphics[width=\linewidth]{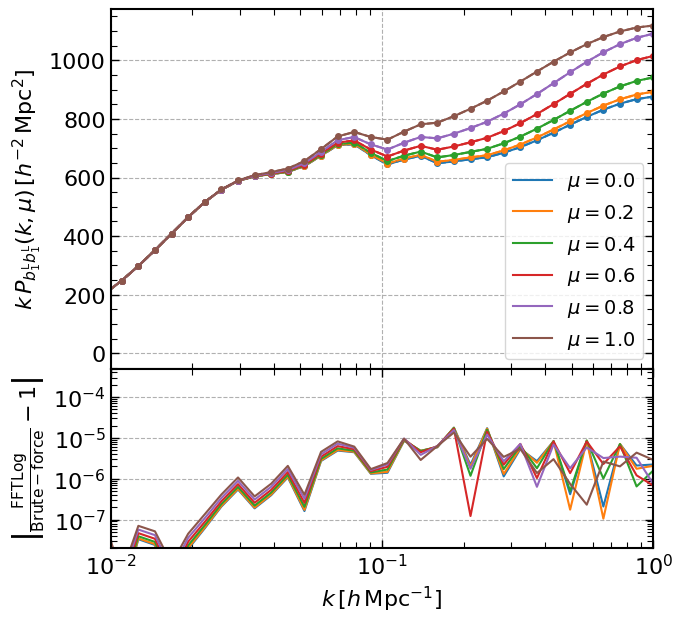}
\end{minipage}
\begin{minipage}{0.48\linewidth}
\centering
\includegraphics[width=\linewidth]{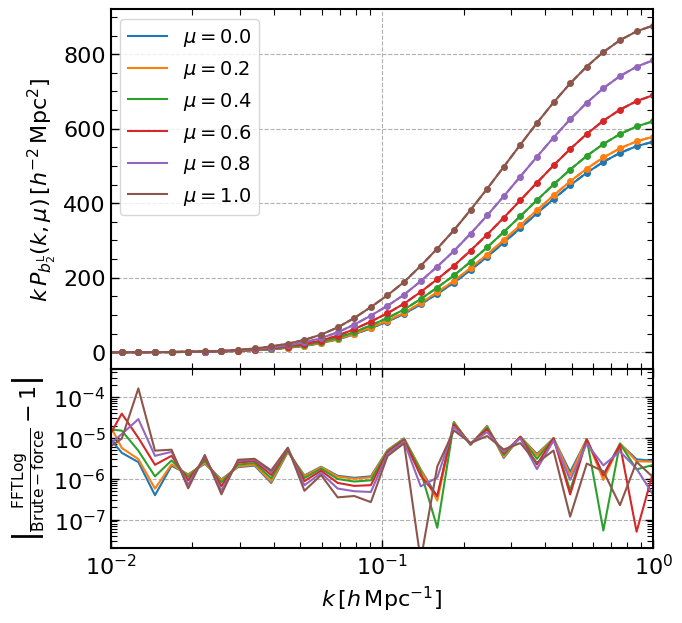}
\end{minipage}
\begin{minipage}{0.48\linewidth}
\centering
\includegraphics[width=\linewidth]{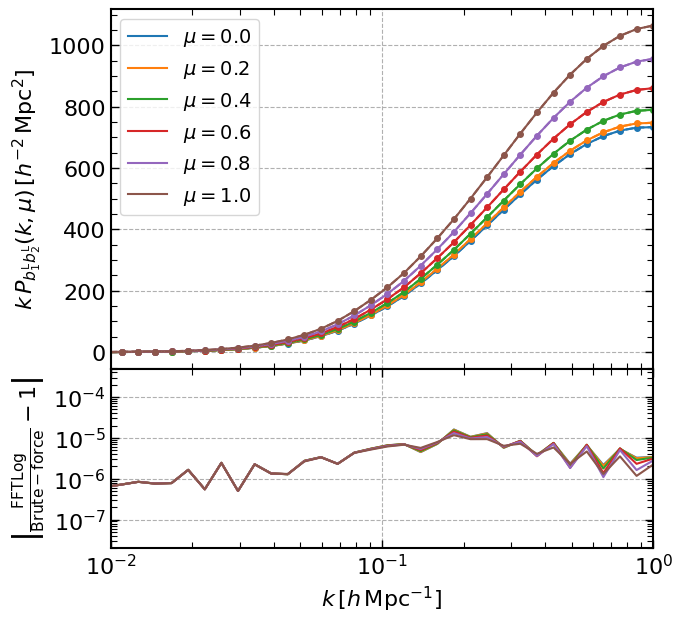}
\end{minipage}
\begin{minipage}{0.48\linewidth}
\centering
\includegraphics[width=\linewidth]{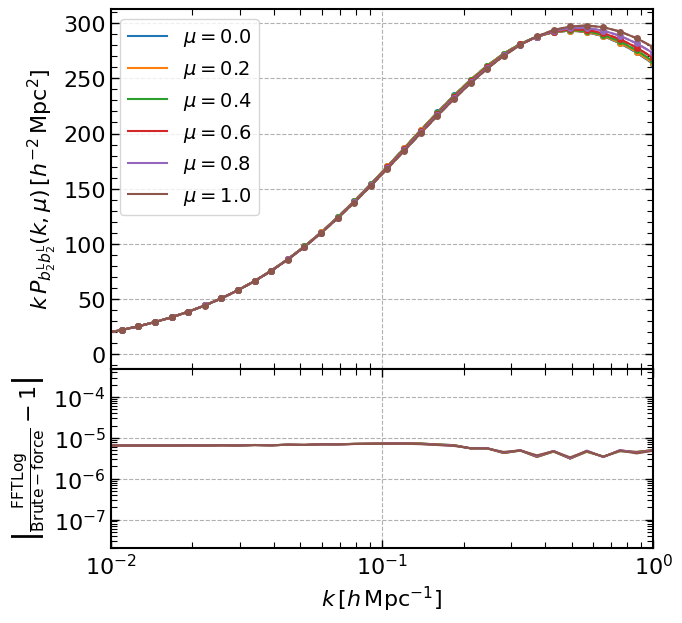}
\end{minipage}
\caption{
Comparison of the LPT terms between our JAX implementation (solid lines) and brute-force integration (dashed lines).
}
\label{fig:comparison_to_bf_lpt_terms1}
\end{figure}

\begin{figure}[htbp]
\centering
\begin{minipage}{0.48\linewidth}
\centering
\includegraphics[width=\linewidth]{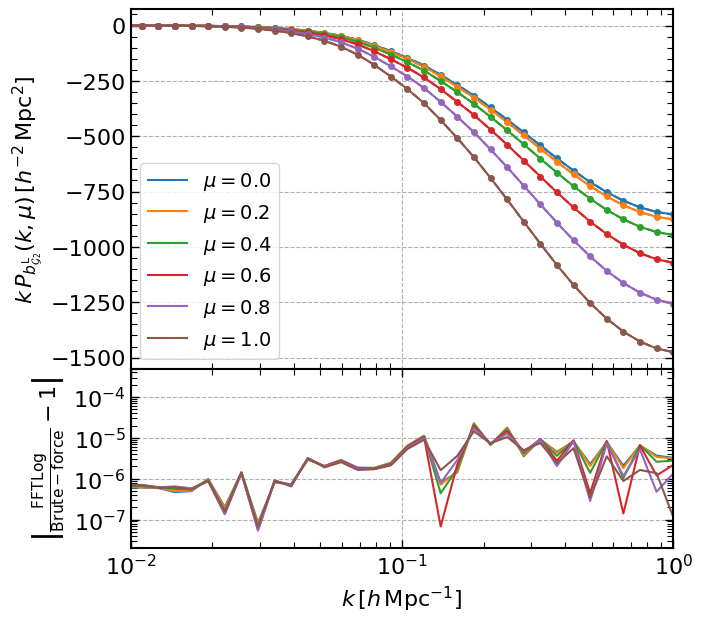}
\end{minipage}
\begin{minipage}{0.48\linewidth}
\centering
\includegraphics[width=\linewidth]{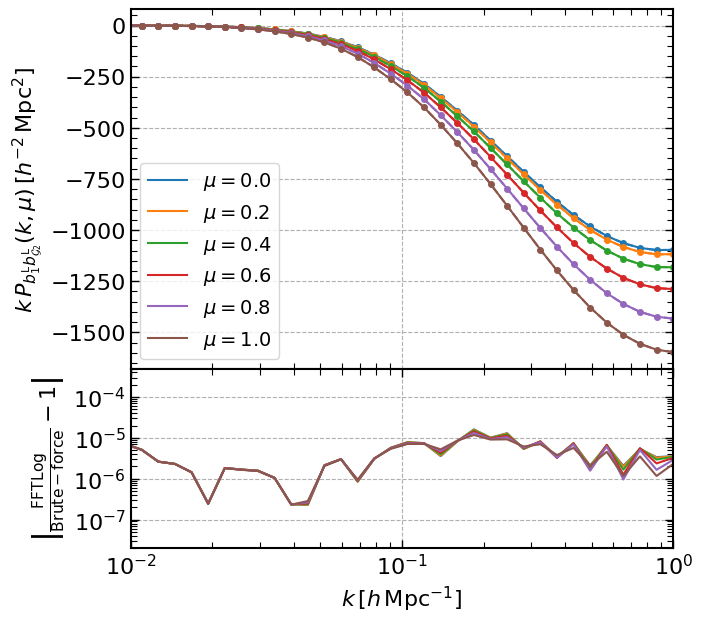}
\end{minipage}
\begin{minipage}{0.48\linewidth}
\centering
\includegraphics[width=\linewidth]{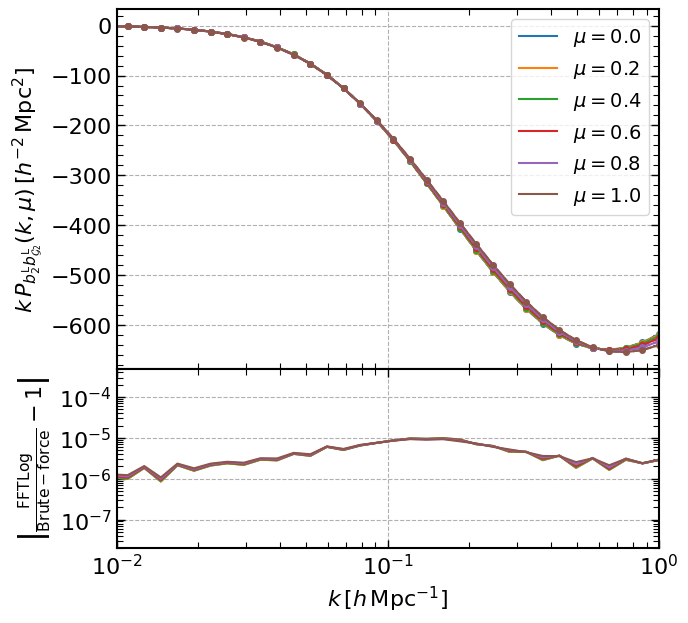}
\end{minipage}
\begin{minipage}{0.48\linewidth}
\centering
\includegraphics[width=\linewidth]{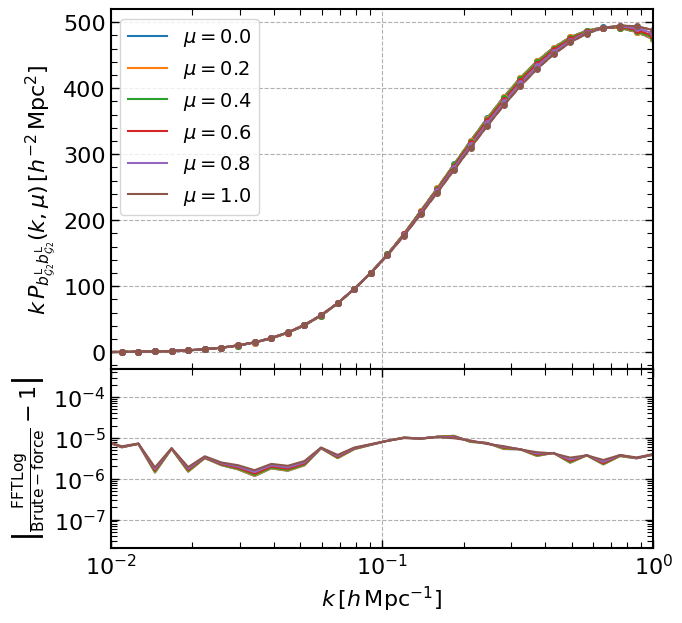}
\end{minipage}
\begin{minipage}{0.48\linewidth}
\centering
\includegraphics[width=\linewidth]{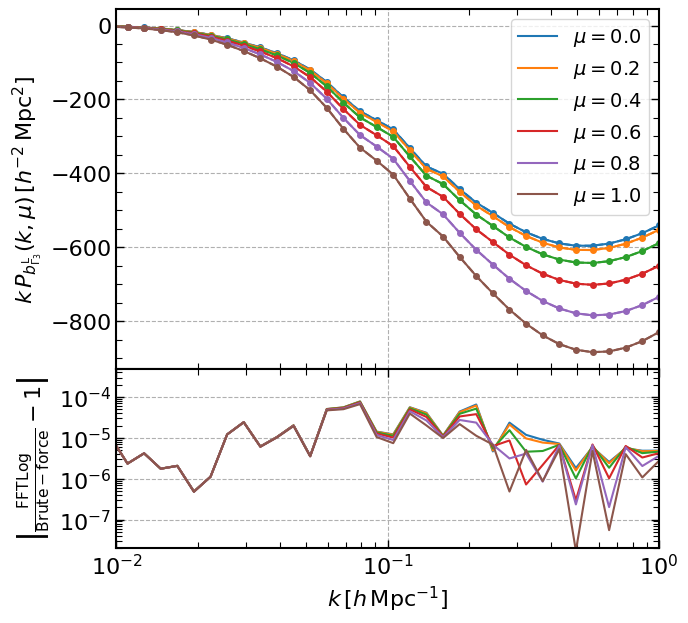}
\end{minipage}
\begin{minipage}{0.48\linewidth}
\centering
\includegraphics[width=\linewidth]{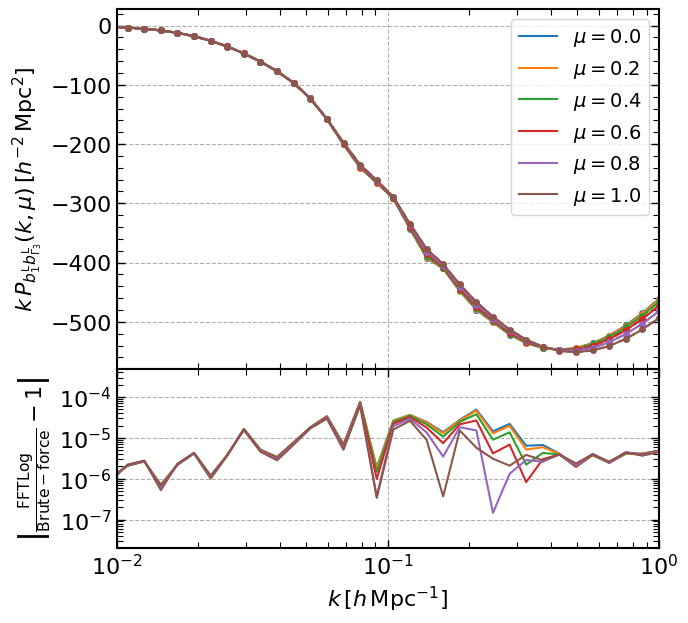}
\end{minipage}
\caption{
Comparison of the LPT terms between our JAX implementation (solid lines) and brute-force integration (dashed lines).
}
\label{fig:comparison_to_bf_lpt_terms2}
\end{figure}

In Sec.~\ref{sec:code_accuracy_lpt}, we have shown the accuracy of our JAX implementation of the LPT power spectrum by comparing the whole $P(k, \mu)$ with a brute-force implementation that replaces FFTs in all the Hankel transforms appearing in the computation with numerical integration using composite Gauss--Legendre quadrature.
For completeness, this appendix shows the same comparison for the 12 distinct contributions to the LPT power spectrum associated with different combinations of the Lagrangian bias parameters in 
Figs.~\ref{fig:comparison_to_bf_lpt_terms1} and \ref{fig:comparison_to_bf_lpt_terms2}.

\bibliographystyle{unsrt}
\bibliography{ref}

\end{document}